\documentclass[letterpaper,12pt]{article}

\usepackage[utf8]{inputenc}

\usepackage{geometry}
\usepackage{sectsty}
\allsectionsfont{\sffamily\mdseries\upshape}

\usepackage{fancyhdr}
\usepackage{graphicx}
\usepackage{float}
\usepackage{pgfplots}
\usepackage{caption}
\usepackage{subcaption}

\usepackage{hyperref}

\usepackage[numbers]{natbib}

\makeatletter

\title{Optimal Auctions with Convex Perceived Payments}

\author{%
Amy Greenwald \\
Department of Computer Science \\
Brown University \\
\texttt{amy@cs.brown.edu} \\
\and
Takehiro Oyakawa \\
Department of Computer Science \\
Brown University \\
\texttt{oyakawa@cs.brown.edu}
\and
Vasilis Syrgkanis \\
Microsoft Research, NYC \\
\texttt{vasy@microsoft.com}
}

\usepackage{amsmath}
\usepackage{amsthm}
\usepackage{amsfonts}
\usepackage{amssymb}
\usepackage{amscd}
\usepackage{eurosym}
\usepackage{mathrsfs}
\usepackage{mathtools}
\usepackage{bm} 
\usepackage{bbm}

\theoremstyle{plain}
\newtheorem{theorem}{Theorem}[section]
\newtheorem{lemma}[theorem]{Lemma}
\newtheorem{corollary}[theorem]{Corollary}

\theoremstyle{definition}

\newtheorem{example}[theorem]{Example}

\theoremstyle{remark}
\newtheorem{remark}[theorem]{Remark}

\usepackage{algorithm}
\usepackage{algpseudocode}

\DeclarePairedDelimiter{\ceil}{\lceil}{\rceil}

\DeclarePairedDelimiter{\abs}{\lvert}{\rvert}
\DeclarePairedDelimiter{\norm}{\lVert}{\rVert}

\DeclareMathOperator*{\argmax}{arg\,max}

\DeclareMathOperator*{\Exp}{\mathbb{E}}
\DeclareMathOperator*{\Reals}{\mathbb{R}}

\newcommand{\Fun}[1][]{Q_{#1}}
\newcommand{\fun}[1][]{q_{#1}}

\newcommand{\setofbidders}[1][]{N}
\newcommand{\numbidders}[1][]{n}

\newcommand{\alloc}[1][]{x_{#1}} 
\newcommand{\allocvec}[1][]{\mathbf{\alloc}_{#1}}
\newcommand{\hatalloc}[1][]{\hat{x}_{#1}}
\newcommand{\hatallocvec}[1][]{\hat{\mathbf{\alloc}}_{#1}}

\newcommand{\bid}[1][]{b_{#1}}
\newcommand{\bidvec}[1][]{\mathbf{\bid}_{#1}}

\newcommand{\x}[1][]{y_{#1}}
\newcommand{\xvec}[1][]{\mathbf{\x}_{#1}}

\newcommand{\const}[1][]{c_{#1}}
\newcommand{\cvec}[1][]{\mathbf{\const}_{#1}}

\newcommand{\payment}[1][]{p_{#1}} 
\newcommand{\paymentvec}[1][]{\mathbf{\payment}_{#1}}
\newcommand{\hatpayment}[1][]{\hat{p}_{#1}}
\newcommand{\hatpaymentvec}[1][]{\hat{\mathbf{\payment}_{#1}}}
\newcommand{\barpayment}[1][]{\bar{p}_{#1}}
\newcommand{\barpaymentvec}[1][]{\bar{\mathbf{\payment}_{#1}}}
\newcommand{\tildepayment}[1][]{\tilde{p}_{#1}}
\newcommand{\tildepaymentvec}[1][]{\tilde{\mathbf{\payment}_{#1}}}

\newcommand{\paymentq}[1][]{q_{#1}}
\newcommand{\paymentvecq}[1][]{\mathbf{\paymentq}_{#1}}
\newcommand{\hatpaymentq}[1][]{\hat{q}_{#1}}
\newcommand{\hatpaymentvecq}[1][]{\hat{\mathbf{\paymentq}_{#1}}}
\newcommand{\barpaymentq}[1][]{\bar{q}_{#1}}
\newcommand{\barpaymentvecq}[1][]{\bar{\mathbf{\paymentq}_{#1}}}
\newcommand{\tildepaymentq}[1][]{\tilde{q}_{#1}}
\newcommand{\tildepaymentvecq}[1][]{\tilde{\mathbf{\paymentq}_{#1}}}

\newcommand{\val}[1][]{v_{#1}}
\newcommand{\valvec}[1][]{\mathbf{\val}_{#1}}
\newcommand{\valspace}[1][]{T_{#1}}
\newcommand{\valspacesize}[1][]{M_{#1}}
\newcommand{\valw}[1][]{w_{#1}}
\newcommand{\valz}[1][]{z_{#1}}

\newcommand{\virvecd}[1][]{\boldsymbol\psi_{#1}}
\newcommand{\vvald}[1][]{\psi_{#1}} 

\newcommand{\util}[1][]{u_{#1}} 

\newcommand{\mydef}[1]{\textbf{#1}}

\begin{document}

\maketitle

\begin{abstract}

Myerson derived a simple and elegant solution to the single-parameter
revenue-maximization problem in his seminal work on optimal auction
design assuming the usual model of quasi-linear utilities.  In this
paper, we consider a slight generalization of this usual model---from
linear to convex ``perceived'' payments.  This more general problem
does not appear to admit a solution as simple and elegant as Myerson's.

While some of Myerson's results extend to our setting, like his
payment formula (suitably adjusted), and the observation that ``the
mechanism is incentive compatible only if the allocation rule is
monotonic,'' others do not.  For example, we observe that the
solutions to the Bayesian and the robust (i.e., non-Bayesian) optimal
auction design problems in the convex perceived payment setting do not
coincide like they do in the case of linear payments.  We therefore
study the two problems in turn.

Myerson finds an optimal robust (and Bayesian) auction by solving
pointwise: for each vector of virtual values, he finds an optimal,
ex-post feasible auction; then he plugs the resulting allocation,
which he first verifies is monotonic, into his payment formula.  This
strategy relies on a key theorem, that expected revenue equals
expected virtual surplus, which does not hold in our setting.  Still,
we derive an upper and a heuristic lower bound on expected revenue in
our setting.  These bounds are easily computed pointwise, and yield
monotonic allocation rules, so can be supported by Myerson payments
(suitably adjusted).  In this way, our bounds yield heuristics that
approximate the optimal robust auction, assuming convex perceived
payments.


In tackling the Bayesian problem, we derive a 
mathematical program that improves upon the default formulation in
that it has only polynomially-many payment variables; however, it
still has exponentially-many allocation variables and ex-post
feasibility constraints.  To address this latter issue, we also study
the ex-ante relaxation, which requires only polynomially-many
constraints, in all.  Specifically, we present a closed-form solution
to a straightforward relaxation of this relaxation.
Then, following similar logic, we present a closed-form upper bound
and a heuristic lower bound on the solution to the (ex-post) robust
problem.  As above, the resulting allocation rules can then be
supported by Myerson payment rules, yielding faster heuristics than the
greedy ones for approximating the optimal robust auction.
Interestingly, all our closed-form solutions are rather intuitive:
they allocate in proportion to values (or virtual values).

We close with experiments, the final set of which massages the output
of one of the closed-form heuristics for the robust problem into an
extremely fast, near-optimal heuristic solution to the Bayesian
optimal auction design problem.

\end{abstract}

\tableofcontents


\section{Introduction}

In a seminal paper, Myerson~\cite{Myerson:Optimal/1981} provides a simple and elegant solution to a fundamental problem in optimal auction design: the single-parameter revenue-maximization problem, assuming quasi-linear utilities with linear payments: i.e., $\util[i] = \val[i] \alloc[i] - \payment[i]$, where $\val[i] > 0$ is $i$'s private value, $\alloc[i]$ is his allocation, and $\payment[i]$ is his payment to the auctioneer.
Generally speaking, $\alloc[i]$ and $\payment[i]$ are not expressed in the same units;
hence, we can think of $\val[i]$ as a conversion factor, converting units of the good being allocated into units of payment (often, money).
In this paper, we investigate the extent to which Myerson's observations carry over to the case of convex ``perceived''%
\footnote{There are two potential payments for bidders to consider in
  auction problems: those paid to the auctioneer, and those subtracted
  from $\val[i] \alloc[i]$ in the bidders' utility functions.  In
  general, these two payments need not be equated.  While we choose to
  label the former payments (made to the auctioneer) ``actual,''
  and the latter, ``perceived,'' we could just as well have taken the
  perspective of the bidders in our naming, and referred to the former
  as ``perceived,'' and the latter as ``actual.''

  We also considered referring to perceived payments instead as
  ``costs'' instead of payments, which they rightfully are.  But as
  our results include a payment formula for these costs which closely
  relates to Myerson's original payment formula, it seemed that
  qualifying the term payment would be more illuminating.} payments.
Specifically, we consider utilities of the form $\util[i] = \val[i]
\alloc[i] - \fun[i](\payment[i])$, where $\fun[i]$ is a convex
function that describes payments that $i$ perceives, which we
distinguish from $\payment[i]$ itself ($i$'s ``actual'' payment to the
auctioneer).  Although our results apply to any convex function
$\fun[i]$, as a concrete running example throughout this paper, we
assume $\fun[i](\payment[i]) = \payment[i]^2$, for all bidders $i$.

Our problem formulation is motivated by a reverse auction setting in
which a government with a fixed budget is offering subsidies (in
euros, say) to power companies in exchange for a supply of renewable
energy (in watts, say).
We assume the power companies' utility functions take this form:
$\util[i]' = \alloc[i] - \fun[i](\payment[i]) / \val[i]$, 
where $\alloc[i]$ is some fraction of the total budget in euros, and
$\payment[i]$ is some deliverable amount of power in watts.
As already mentioned, $\val[i]$ is a conversion factor, in this case from euros to watts.
The assumption that
$\paymentq[i] (\payment[i])$ is
convex reflects the fact that energy production costs may not be linear;
for example, because of diseconomies of scale, it may be the case that
as more energy is produced, further units become more and more
expensive to produce.

Note that multiplying $\util[i]'$ by $\val[i]$ yields a more familiar utility function---that of the forward auction setting: $\util[i] = \val[i] \alloc[i] - \fun[i](\payment[i])$,
with utility measured in units of power, rather than money.
From this point of view, defining perceived payments
$\fun[i](\payment[i]) = \payment[i]^2$ can be interpreted as an assumption of risk aversion.
A power company might be risk averse because it might be concerned that it has overestimated its own value, 
or it might worry that the government won't actually deliver on the promised subsidies.

Another problem which also fits into our framework is the problem of allocating a fixed block of advertising time to retailers during the Superbowl.
In this application, an advertiser's utility is calculated by converting its allocation, in time, into dollars via its private value, and then subtracting the cost of production:
$\util[i] = \val[i] \alloc[i] - \fun[i](\payment[i])$.
Here, dividing by $\val[i]$ yields the utility function $\util[i]'$,
which, under the assumption that $\fun[i](\payment[i]) = \payment[i]^2$, again yields the interpretation that production costs are convex;
for example, it may be the case that an advertisement that is three times as long as another takes nine times as long to produce.

In both of these problems, the auctioneer's objective is to maximize its total revenue:
$\Exp_{\valvec} [ \sum_i \payment[i] ]$, where $\valvec = (\val[1], \ldots, \val[\numbidders])$ is distributed according
to some commonly known joint distribution.
Specifically, in the energy problem, the government's objective is to maximize the amount of power produced, subject to its budget constraint.
In the advertising problem, the television network is seeking to maximize its revenue for selling a fixed block of advertising time during the Superbowl (or any other television program).

\paragraph{Summary of Results}

In our setting, in which perceived payments are convex, some of
Myerson's observations, such as ``the mechanism is incentive
compatible only if the allocation rule is monotonic,'' continue to
hold.  So does his payment formula (suitably adjusted).

But others do not.  The solutions to the robust (i.e., non-Bayesian)
and Bayesian optimal auction design problems (in the latter, incentive
compatibility and individual rationality need only hold in
expectation) do not coincide in our setting like they do in the
linear-payment setting.  Moreover, bidder surplus is no longer an
upper bound on revenue, and expected revenue no longer equals expected
virtual surplus.

We do show, however, that revenue can be upper bounded by a quantity
we call \emph{pseudo-surplus}, and that expected revenue can be
heuristically lower bounded by expected virtual surplus.  
Because it is straightforward to compute these bounds (i.e., greedy
algorithms suffice),
we propose the following heuristic procedure for the robust optimal
auction design problem: 1.~solve greedily for a feasible allocation
that achieves
the upper or heuristic lower bound, 
and 2.~plug that allocation into Myerson's payment formula to ensure
incentive compatibility and individual rationality.  We show
experimentally that the performance of these heuristics can be
near-optimal, and that they are faster than solving the corresponding
revenue-maximizing mathematical programs using standard solvers.

Our heuristic that lower bounds the solution to the robust optimal auction design
problem also heuristically lower bounds the Bayesian problem, because in
the latter the constraints are strictly weaker.
Regardless, we prove a theorem for the Bayesian setting that allows us
to simplify the revenue-maximizing mathematical program from one that
has exponentially-many payment variables to one that has only
polynomially-many.


Even so, a Bayesian optimal auction still involves exponentially-many
allocation variables and requires satisfying exponentially-many
ex-post feasibility constraints.  To address this latter concern, we
also study the ex-ante relaxation, which requires only
polynomially-many constraints, in all.  Specifically, we present a
closed-form solution to a straightforward relaxation of this
relaxation, which yields a closed-form upper bound on the ex-ante
Bayesian problem.
Intuitively, this solution allocates in proportion to virtual values.

Lastly, following similar logic, we present a closed-form upper bound
and a heuristic lower bound on the solution to the (ex-post) robust
problem.  (Interestingly, these closed-form solutions allocate in
proportion to values.)  An analog of the aforementioned greedy
heuristics then applies, in which the closed-form allocation rule is
supported by Myerson payment rules, yielding closed-form heuristics that
approximate the optimal robust auction.  In our final set of
experiments, we massage the output of one of these closed-form
heuristics for the robust problem using the payment variables in the
simpler formulation of the Bayesian problem to arrive at a
near-optimal heuristic solution to the Bayesian problem, which is
faster in practice than standard solvers (and faster than our greedy
heuristic).

\paragraph{Related Work}

Vickery~\cite{Vickrey:Counterspeculation/1961} showed that 
auctions in which the highest bidder wins and pays the second-highest bid 
incentivizes bidders to bid truthfully.
Myerson~\cite{Myerson:Optimal/1981} showed that in the
single-parameter setting, with the usual utility function
involving linear payments, expected revenue is maximized by a
Vickrey auction with reserve prices.  Our setting is not captured by
Myerson's classic characterization because payments in our model are
convex.

The technical difficulties that arise in our setting are similar in
spirit to the ones faced by Pai and Vohra~\cite{Pai2014} when designing
optimal auctions for budget-constrained bidders.
If $\fun[i](\payment[i]) / \val[i] = \payment[i]^k$ for some $k \gg 0$,
then $\util[i] = \val[i] \alloc[i] - \payment[i]^k \ge 0$ when
$\left( \val[i] \alloc[i] \right)^{1/k} \ge \payment[i]$.
Additionally, if $\payment[i]$ ever exceeds $\left( \val[i] \alloc[i] \right)^{1/k}$,
utility quickly approaches $-\infty$.  Therefore, we can interpret a utility
function with convex perceived payments as a continuous approximation of that of
a budget-limited agent whose utility is $-\infty$ whenever her payment
exceeds her budget.

Our model also has strong connections with the literature on optimal auctions for risk-averse buyers (see Maskin and Riley~\cite{Maskin1984}), since the fact that utility is a concave function in terms of payments can be seen as a form of risk aversion.
In fact, our model is captured by their generic formulation if the good to be allocated is indivisible.
Optimal auctions for risk-averse bidders are notoriously hard to characterize in theory. 
Developing and testing theoretically-inspired heuristics may prove to be a fruitful alternative. 

The idea of translating a reverse auction into a direct auction by
multiplying utility by the private parameter $\val[i]$ was previously
proposed in the literature on optimal contests (see, for example,
Chawla, \emph{et al.}~\cite{Chawla2012} or DiPalantino and
Vojnovic~\cite{Dipalantino2009}).

Procuring services subject to a budget constraint is also the subject
of the literature on budget-feasible mechanisms initiated by
Singer~\cite{Singer2014}.  However, in this literature, the service of
each seller is fixed and the utility of the buyer is a combinatorial
function of the set of sellers the buyer picks.  In our setting, each
seller can produce a different level of service (i.e., amount of
energy) by incurring a different cost, so the buyer picks not only a
set of sellers, but a level of service that each seller should provide
as well.  This renders the two procurement models incomparable.

Our model is also related to the parameterized supply bidding game of Johari and Tsitsiklis~\cite{Johari2011}, 
where firms play a game in which they submit a single-parameter family of supply functions. 
There, the amount each firm is asked to produce is decided via a non-truthful mechanism, 
and efficiency at equilibrium is analyzed. 
Here, we consider the design of truthful mechanisms and we study a different objective, 
but we also restrict attention to single-parameter families of supply functions $\fun[i](\payment[i])$.

In principle, some of our problem formulations
can be solved using Border's characterization of interim feasible outcomes~\cite{Border1991} 
and an ellipsoid type algorithm with a separation oracle. 
However, such mechanisms tend to be computationally expensive.  
Here, we seek fast allocation heuristics with potential economic justification, 
such as virtual-value-based maximizations.  
Virtual-value-maximizing approximations to optimal auction design were also studied recently by 
Alaei, \emph{et al.}~\cite{alaei2013simple} in the context of multi-dimensional mechanism design, 
and from a worst-case point of view.  
Proving worst-case approximation guarantees for our heuristics is an interesting future direction.

We close this discussion of related work by pointing out that the case
of a finitely divisible good, such as euros or seconds, which
motivates the current work, lies along a continuum between an
indivisible and an infinitely divisible good.
An optimal solution that is subject to a discretization constraint
(i.e., finitely divisible) can differ from a corresponding optimal
continuous solution (i.e., infinitely divisible) by at most a
discretization term.
%
%
For a budget that is small relative to this discretization factor, the
magniture of the error is large, so the budget behaves more like an
indivisible good; but as a budget increases relative to the
discretization factor, the magnitude of the error decreases, so the
budget behaves more like an infinitely divisible good.


\section{Our Model}

Consider a reverse auction with $\numbidders$ bidders.
Each bidder $i \in \setofbidders = \{ 1, \dots, \numbidders \}$ 
has a private type $0 \le \val[i] \in \valspace[i]$ 
that is independently drawn from some distribution $F_i$.
Let $\valspace = \valspace[1] \times \cdots \times \valspace[\numbidders]$
be the set of all possible type vectors, 
and let $F = F_1 \times \cdots \times F_{\numbidders}$ be the distribution over type vectors  
$\valvec = ( \val[1], \dots, \val[\numbidders] ) \in \valspace$. 
Let $\bidvec = ( \bid[1], \dots, \bid[\numbidders] ) \in {\Reals}^{\numbidders}$ be a vector of bids, 
where the $i$th entry $\bid[i]$ is bidder $i$'s bid.
For $\xvec \in \{ \bidvec, \valvec \}$,
we use the notation $\xvec = ( \x[i], \xvec[-i] )$, 
where $\xvec[-i] = (\x[1], \ldots, \x[i-1], \x[i+1], \ldots, \x[\numbidders])$.  
Similarly, we use the notation $\valspace[-i]$, 
where $\valspace[-i] = \valspace[1] \times \cdots \times \valspace[i-i] \times \valspace[i+1] \times \cdots \times \valspace[\numbidders]$, 
and $F_{-i}$, 
where $F_{-i} = F_1 \times \cdots \times F_{i-i} \times F_{i+1} \times \cdots \times F_{\numbidders}$.

Given vector of reports $\bidvec$, a mechanism produces an allocation
rule $\allocvec (\bidvec) \in [0,1]^{\numbidders}$ that typically depends only on those
reports, together with a payment rule $\paymentvec (\bidvec, \allocvec) \in {\Reals}^{\numbidders}$,
which, in general, can depend on both the reports and the allocation rule.
Where it is clear from context, we suppress dependence of the payment
rule on $\allocvec$, and write only $\paymentvec (\bidvec)$.
We also do the same for payment terms $\payment[i] ( \bidvec, \allocvec )$,
which comprise payment rule $\paymentvec (\bidvec)$,
and refer to bidder $i$'s payment as $\payment[i] (\bid[i], \bidvec[-i])$.

\begin{figure}[h!]
\centering
\includegraphics{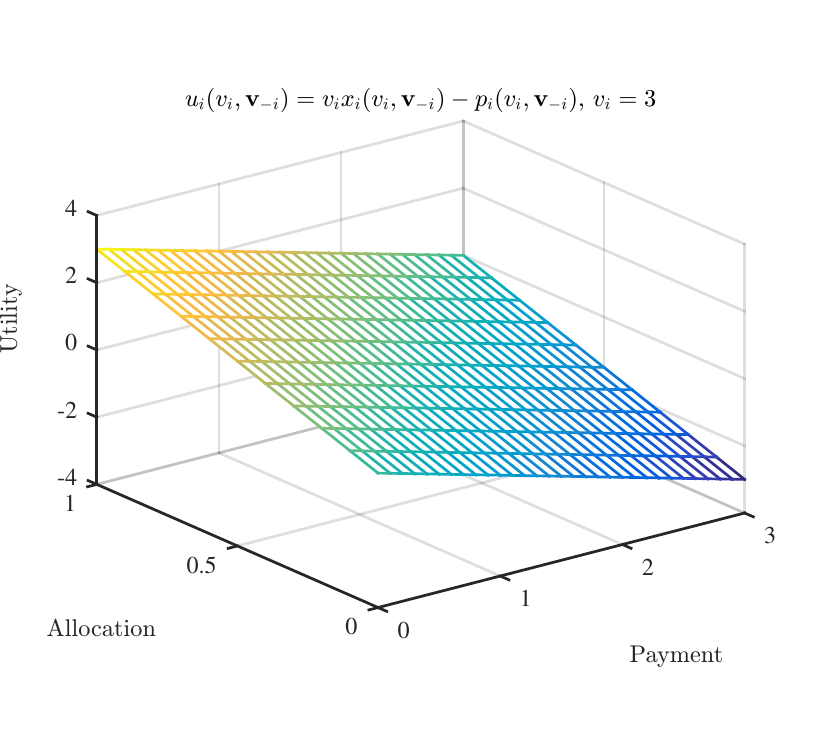}
\caption{\mydef{Linear} quasi-linear utility as a function of allocation and linear payments is shown, for 
$\util[i] (\val[i], \valvec[-i]) = 
\val[i] \alloc[i] (\val[i], \valvec[-i]) - \payment[i] (\val[i], \valvec[-i])$, 
where $\val[i] = 3$.  
For a fixed allocation $\alloc[i] (\val[i], \valvec[-i])$, 
utility scales linearly with $\payment[i] (\val[i], \valvec[-i])$, 
and for a fixed payment $\payment[i] (\val[i], \valvec[-i])$, 
utility scales linearly with $\alloc (\val[i], \valvec[-i])$.  
Utility is non-negative when 
$\val[i] \alloc[i] (\val[i], \valvec[-i]) \ge \payment[i] (\val[i], \valvec[-i])$, 
so individual rationality dictates that $\payment[i] (\val[i], \valvec[-i])$ can be at most $3$.}
\label{fig:qlUtilPlot}
\end{figure}

\begin{figure}[h!]
\centering
\includegraphics{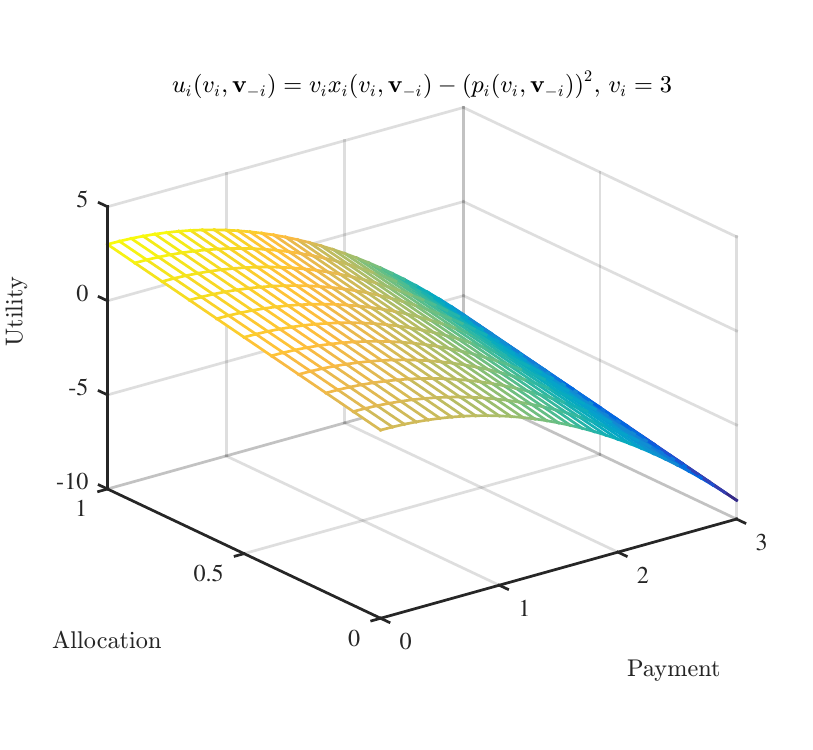}
\caption{\mydef{Concave} quasi-linear utility as a function of allocation and convex perceived payments is shown, for 
$\util[i] (\val[i], \valvec[-i]) = 
\val[i] \alloc[i] (\val[i], \valvec[-i]) - \left( \payment[i] (\val[i], \valvec[-i]) \right)^2$, 
where $\val[i] = 3$.  
For a fixed allocation $\alloc[i] (\val[i], \valvec[-i])$, 
utility scales quadratically with $\payment[i] (\val[i], \valvec[-i])$, 
and for a fixed payment $\payment[i] (\val[i], \valvec[-i])$, 
utility scales linearly with $\alloc[i] (\val[i], \valvec[-i])$.  
Utility is non-negative when 
$\sqrt{\val[i] \alloc[i] (\val[i], \valvec[-i])} \ge \payment[i] (\val[i], \valvec[-i])$, 
so individual rationality dictates that $\payment[i] (\val[i], \valvec[-i])$ can be at most $\sqrt{3}$.}
\label{fig:p2UtilPlot}
\end{figure}

We define bidder $i$'s utility function as
$\util[i]' ( \bid[i], \bidvec[-i] )
= \alloc[i] ( \bid[i], \bidvec[-i] ) - \Fun[i] ( \payment[i] ( \bid[i], \bidvec[-i] ), \val[i] )$,
where $\val[i]$ is private information, known only to bidder $i$,
and $\Fun[i](\payment[i], \val[i])$ is a function that converts payments, in units such as energy,
to the units of the good being allocated, such as euros.
If we assume $\Fun[i] ( \payment[i] ( \bid[i], \bidvec[-i] ), \val[i] ) = 
\paymentq[i] ( \payment[i] ( \bid[i], \bidvec[-i] ) ) / \val[i]$,
then maximizing the function 
$\val[i] \util[i]' ( \bid[i], \bidvec[-i] ) = \val[i] \alloc[i] ( \bid[i], \bidvec[-i] ) - \paymentq[i] ( \payment[i] ( \bid[i], \bidvec[-i] ) )$ 
also maximizes 
$\util[i] ( \bid[i], \bidvec[-i] ) = \val[i] \util[i]' ( \bid[i], \bidvec[-i] )$.
Consequently, hereafter we assume \emph{forward\/} utility functions of this form:
\begin{equation}
\label{eq:util_with_q}
\util[i] ( \bid[i], \bidvec[-i] ) = \val[i] \alloc[i] ( \bid[i], \bidvec[-i] ) - \paymentq[i] ( \payment[i] ( \bid[i], \bidvec[-i] ) )%
.\end{equation}
The shape of the utility function varies with the choice of $\paymentq[i]$: for example,
it can be linear if we choose $\paymentq[i]$ to be the identity function, or concave if we choose 
$\paymentq[i] ( \payment[i] ( \bid[i], \bidvec[-i] ) ) = ( \payment[i] ( \bid[i], \bidvec[-i] ) )^2$.
Figures~\ref{fig:qlUtilPlot} and~\ref{fig:p2UtilPlot} plot sample utility functions for these two choices of $q_i$.

For readability, we usually write $\paymentq[i] ( \bid[i], \bidvec[-i] )$
instead of $\paymentq[i] ( \payment[i] ( \bid[i], \bidvec[-i] ) )$.
Furthermore, we refer to the rule $\paymentvecq (\bidvec) \in {\Reals}^{\numbidders}$
as the \mydef{perceived} payment rule, as these are payments the bidders impose upon themselves.
Likewise, we think of $\paymentvec (\bidvec)$ as an \mydef{actual}
payment rule, as these are payments the bidders actually pay to the
auctioneer.  (But we usually omit the descriptor ``actual,'' as what it
modifies is the actual payment rule!)

With this general utility function in mind, we proceed to formulate
the optimal auction design problem: we seek the revenue-maximizing,
ex-post feasible auction in which it is optimal for bidders to bid
truthfully, and it is rational for them to participate.
Once again, Myerson solved this problem in the quasi-linear case, assuming linear payments.

\subsection{Constraints}

In this section, we formalize the constraints we will impose on an optimal auction design.
Because we restrict our attention to incentive compatible auctions,
where it is optimal to bid truthfully, we write
$\paymentq[i] ( \val[i], \valvec[-i] )$
instead of $\paymentq[i] ( \bid[i], \bidvec[-i] )$.
(As usual, the variables $\paymentq[i] ( \val[i], \valvec[-i] )$ comprise the
perceived payment rule $\paymentvecq ( \valvec ) \in \mathbb{R}^{\numbidders}$.)

A mechanism is called \mydef{incentive compatible} (IC) if each bidder maximizes her utility 
by reporting bids $\bid[i]$ that are equal to values $\val[i]$:
$\forall i \in \setofbidders$, 
$\forall \val[i], \valw[i] \in \valspace[i]$, 
and $\forall \valvec[-i] \in \valspace[-i]$, 
\begin{equation}
\label{eq:IC}
\val[i] \alloc[i] ( \val[i], \valvec[-i] ) - \paymentq[i] ( \val[i], \valvec[-i] )
\ge  \val[i] \alloc[i] ( \valw[i], \valvec[-i] ) - \paymentq[i] ( \valw[i], \valvec[-i] )
.\end{equation}

\mydef{Individual rationality} (IR) ensures that bidders have non-negative utilities:
$\forall i \in \setofbidders$, $\forall \val[i] \in \valspace[i]$, and $\forall \valvec[-i] \in \valspace[-i]$, 
\begin{equation}
\label{eq:IR}
\val[i] \alloc[i] ( \val[i], \valvec[-i] ) - \paymentq[i] ( \val[i], \valvec[-i] )
\ge  0
.\end{equation}

Next, we define IC and IR in expectation (with respect to $F_{-i}$).  
To do so, we introduce \mydef{interim allocation} and \mydef{interim perceived payment}
variables, respectively:
$\hatalloc[i] ( \val[i] ) \equiv \hatalloc[i] ( \val[i], \cdot ) =
\Exp_{\valvec[-i]} \left[ \alloc[i] ( \val[i], \valvec[-i] ) \right]$ and 
$\hatpaymentq[i] ( \val[i] ) \equiv \hatpaymentq[i] ( \val[i], \cdot ) =
\Exp_{\valvec[-i]} \left[ \paymentq[i] ( \val[i], \valvec[-i] ) \right]$.
These variables comprise the interim allocation
and perceived payment rules,
$\hatallocvec (\valvec) \in [0,1]^{\numbidders}$ and
$\hatpaymentvecq  (\valvec) \in \mathbb{R}^{\numbidders}$.

We call a mechanism \mydef{Bayesian incentive compatible} (BIC) if utility is maximized by 
truthful reports in expectation:
$\forall i \in \setofbidders$ and $\forall \val[i], \valw[i] \in \valspace[i]$,
\begin{equation}
\label{eq:BIC}
\val[i] \hatalloc[i] ( \val[i] ) - \hatpaymentq[i] ( \val[i] )
\ge  \val[i] \hatalloc[i] ( \valw[i] ) - \hatpaymentq[i] ( \valw[i] )
.\end{equation}

\mydef{Bayesian individual rationality} (BIR) insists on non-negative utilities in expectation:
$\forall i \in \setofbidders$ and $\forall \val[i] \in \valspace[i]$,
\begin{equation}
\label{eq:BIR}
\val[i] \hatalloc[i] ( \val[i] ) - \hatpaymentq[i] ( \val[i] )
\ge  0
.\end{equation}

We say a mechanism is \mydef{ex-post feasible} (XP) if it never overallocates:
$\forall \valvec \in \valspace$, 
\begin{equation}
\sum_{i=1}^{\numbidders} \alloc[i] ( \val[i], \valvec[-i] ) \le 1.
\end{equation}

\mydef{Ex-ante feasibility} (XA) is satisfied if, in expectation, the mechanism does not over-allocate:
\begin{equation}
\Exp_{\valvec} \left[ \sum_{i=1}^{\numbidders} \alloc[i] ( \valvec ) \right] \le 1.
\end{equation}

Finally, we require that
$0 \le \alloc[i] ( \val[i], \valvec[-i] ), \hatalloc[i] ( \val[i] ) \le 1$,
$\forall i \in \setofbidders$, 
$\forall \val[i] \in \valspace[i]$ and $\forall \valvec[-i] \in \valspace[-i]$.

\subsection{Revenue Maximization}

The \mydef{robust revenue-maximization problem} (RRM) is to maximize
total expected payments
(i.e., $\Exp_{\valvec} \left[ \sum_{i=1}^{n} \payment[i] ( \val[i], \valvec[-i] ) \right]$),
subject to IR, IC, and ex-post feasibility.
RRM is expressed as a mathematical program in Section~\ref{sssec:mpdesc_rrm_xp}.
A typical relaxation of this problem insists on ex-ante feasibility only.

Likewise, the \mydef{Bayesian revenue-maximization problem} (BRM)
is to maximize total expected payments
subject to BIR, BIC, and ex-post feasibility (or ex-ante feasibility).
BRM with ex-post feasibility is expressed as a mathematical program in Section~\ref{sssec:mpdesc_brm_xp_naive}.  
BRM with ex-ante feasibility is expressed as a mathematical program in Section~\ref{sssec:mpdesc_brm_xa}.

In both RRM and BRM, there are exponentially-many ex-post feasibility
constraints, and exponentially-many allocation and payment variables.
However, revenue maximization subject to BIC and BIR but only ex-ante
feasibility involves only one feasibility constraint and
polynomially-many interim allocation and payment variables.

In this paper, we assume $\paymentq[i] ( \payment[i] ( \val[i], \valvec[-i] ) ) =
(\payment[i] ( \val[i], \valvec[-i] ) )^2$ (or $\paymentq[i] = \payment[i]^2$, for short).  
This choice of perceived payments yields quadratic constraints.  
While small instances of the quadratic programs that express BRM and
RRM can be solved by standard mathematical programming solvers, this
approach does not scale.  We are interested in finding scalable
algorithms that produce approximately-optimal solutions to these
problems.


\section{Myerson's Payment Formula}
\label{sec:payments}

We start by providing a straightforward adaptation of Myerson's
payment formula \cite{Myerson:Optimal/1981} to the case where the
utilities are of the concave form of interest.

For consistency with our implementations,%
\footnote{True to our profession---computer science---we have 
implemented all mechanisms discussed in this paper.} 
we describe our contributions under the
assumption that type distributions are discrete, but our theoretical
results are in no way contingent on this assumption.  Specifically, we
assume the distribution of values is drawn from the discrete type
space $\valspace[i] = \{ \valz[i,k] : 1 \le k \le \valspacesize[i] \}$,
of cardinality $\valspacesize[i]$, 
where $\valz[i,j] < \valz[i,k]$ for $j < k$, 
and we let $\valz[i,{\valspacesize[i]}+1] = \valz[i,{\valspacesize[i]}]$.
Furthermore,
let $f_i (\val[i])$ be the probability of $\val[i] \in \valspace[i]$, 
let $f (\valvec)$ be the probability of $\valvec \in \valspace$,
and let $f_{-i} (\valvec[-i])$ be the probability of $\valvec[-i] \in \valspace[-i]$.

Myerson's payment theorem, which takes as a starting point the
bidders' utility functions (i.e., Equation~\eqref{eq:util_with_q}),
applies immediately to the perceived payment rules
$\paymentvecq$ and $\hatpaymentvecq$:

\begin{theorem}
\label{thm:payment-char}
Assume bidders' utilities take the form of Equation~\eqref{eq:util_with_q}.
A mechanism is IC and IR iff for each bidder $i \in \setofbidders$:
\begin{itemize}
\item the allocation rule $\allocvec$ is monotone, i.e., 
$\forall \val[i] \ge \valw[i] \in \valspace[i]$ and $\forall \valvec[-i] \in \valspace[-i]$,
$\alloc[i] ( \val[i], \valvec[-i] ) \ge \alloc[i] ( \valw[i], \valvec[-i] )$; and

\item the perceived payment rule $\paymentvecq$ is given by: $\forall \valz[i,\ell] \in \valspace[i]$ and $\forall \valvec[-i] \in \valspace[-i]$,
\begin{equation}
\label{eq:qPayment}
\paymentq[i] ( \valz[i,\ell], \valvec[-i] )
= \valz[i,\ell] \alloc[i] ( \valz[i,\ell], \valvec[-i] ) - 
\sum_{j=1}^{\ell - 1} ( \valz[i,j+1] - \valz[i,j] ) \alloc[i] ( \valz[i,j], \valvec[-i] )
\end{equation}
where $\paymentq[i] ( 0, \valvec[-i] ) = 0$.
\end{itemize}

Likewise, in the Bayesian setting, 
a mechanism is BIC and BIR iff for each bidder $i \in \setofbidders$:
\begin{itemize}  
\item the allocation rule $\hatallocvec$ is monotone, i.e., 
$\forall \val[i] \ge \valw[i] \in \valspace[i]$,
$\hatalloc[i] ( \val[i] ) \ge \hatalloc[i] ( \valw[i] )$; and

\item the perceived payment rule $\hatpaymentvecq$ is given by: $\forall \valz[i,\ell] \in \valspace[i]$,
\begin{equation}
\label{eq:qhatPayment}
\hatpaymentq[i] (\valz[i,\ell])
= \valz[i,\ell] \hatalloc[i] (\valz[i,\ell]) - \sum_{j=1}^{\ell - 1} (\valz[i,j+1] - \valz[i,j]) \hatalloc[i] (\valz[i,j])
\end{equation}
where $\hatpaymentq[i] ( 0 ) = 0$.
\end{itemize}
\end{theorem}

\begin{remark}
The monotonicity of the allocation rule ensures that the perceived payment rule is also monotonic.
Specifically, the perceived payment associated with a bidder of type $\valz[i,\ell+1]$ is 
at least as great as the perceived payment associated with a bidder of type $\valz[i,\ell]$:
\begin{align*}
\paymentq[i] (\valz[i,\ell+1], \valvec[-i])
&= \valz[i,\ell+1] \alloc[i] ( \valz[i,\ell+1], \valvec[-i] ) - 
\sum_{j=1}^{\ell} \left( \valz[i,j+1] - \valz[i,j] \right) \alloc[i] ( \valz[i,j], \valvec[-i] ) \\
&= \valz[i,\ell+1] \left( \alloc[i] ( \valz[i,\ell+1], \valvec[-i] )
-  \alloc[i] (\valz[i,\ell], \valvec[-i]) \right)
+ \valz[i,\ell] \alloc[i] (\valz[i,\ell], \valvec[-i]) 
- \sum_{j=1}^{\ell - 1} ( \valz[i,j+1] - \valz[i,j] ) \alloc[i] ( \valz[i,j], \valvec[-i] ) \\
&= \valz[i,\ell+1] \left( \alloc[i] ( \valz[i,\ell+1], \valvec[-i] )
-  \alloc[i] (\valz[i,\ell], \valvec[-i]) \right)
+ \paymentq[i] (\valz[i,\ell], \valvec[-i]) \\
&\ge \paymentq[i] (\valz[i,\ell], \valvec[-i])
.\end{align*}
\end{remark}

But as our goal is to maximize revenue, we are interested
not in the bidders' perceived payments, but rather 
in the actual payments to the auctioneer, namely $\paymentvec (\valvec),
\hatpaymentvec (\valvec) \in \mathbb{R}^{\numbidders}$,
the latter of which is comprised of variables 
$\hatpayment[i] ( \val[i] ) \equiv \hatpayment[i] ( \val[i], \cdot ) =
\Exp_{\valvec[-i]} \left[ \payment[i] ( \val[i], \valvec[-i] ) \right]$.
In the robust problem,
the generalization is straightforward:
since $\paymentq[i] ( \valz[i,\ell], \valvec[-i] ) = ( \payment[i] ( \valz[i,\ell], \valvec[-i] ) )^2$,
it follows that $\payment[i] ( \valz[i,\ell], \valvec[-i] )$
is simply the square root of $\paymentq[i] ( \valz[i,\ell], \valvec[-i] )$.

In the Bayesian problem, however,
$\hatpaymentq[i] ( \valz[i,\ell] )$ need not equal
$( \hatpayment[i] ( \valz[i,\ell] ) )^2$, because
$\hatpaymentq[i] ( \valz[i,\ell] ) 
= \Exp_{\valvec[-i]} \left[ \paymentq[i] ( \valz[i,\ell], \valvec[-i] ) \right] 
= \Exp_{\valvec[-i]} \left[ ( \payment[i] ( \valz[i,\ell], \valvec[-i] ) )^2 \right] 
\ne \left( \Exp_{\valvec[-i]} \left[ \payment[i] ( \valz[i,\ell], \valvec[-i] ) \right] \right)^2
\equiv ( \hatpayment[i] ( \valz[i,\ell] ) )^2$.
Nonetheless, in Section~\ref{sec:brm}, we successfully derive
an interim payment function $h_i ( \valz[i,\ell] )$ for which 
$\hatpaymentq[i] ( \valz[i,\ell] ) = ( h_i ( \valz[i,\ell] ) )^2$,
so that $h_i ( \valz[i,\ell] )$ is the square root of $\hatpaymentq[i] ( \valz[i,\ell] )$.%
\footnote{In Algorithms~\ref{alg:heurSolverBrm}) and \ref{alg:heurSolverBrmCf},
  the payment rule $\mathbf{h} (\valvec) \in \mathbb{R}^{\numbidders}$, 
  is comprised of variables $h_i (\val[i])$.}

\begin{corollary}
\label{cor:payment-char}
Under the assumptions of Theorem~\ref{thm:payment-char},
for $\paymentq[i] ( \valz[i,\ell], \valvec[-i] ) = ( \payment[i] ( \valz[i,\ell], \valvec[-i] ) )^2$,
\begin{equation}
\label{eq:pSquaredPayment}
\payment[i] ( \valz[i,\ell], \valvec[-i] )
= \sqrt{ \valz[i,\ell] \alloc[i] ( \valz[i,\ell], \valvec[-i] ) - \sum_{j=1}^{\ell - 1} ( \valz[i,j+1] - \valz[i,j] ) \alloc[i] ( \valz[ij], \valvec[-i] ) }%
.\end{equation}

Likewise, given an interim payment function $h_i: \valspace[i] \to \Reals$ such that
$\hatpaymentq[i] ( \valz[i,\ell] ) = ( h_i ( \valz[i,\ell] ) )^2$, then
\begin{equation}
\label{eq:hSquaredPayment}
h_i ( \valz[i,\ell] )
= \sqrt{ \valz[i,\ell] \hatalloc[i] ( \valz[i,\ell] ) - \sum_{j=1}^{\ell - 1} ( \valz[i,j+1] - \valz[i,j] ) \hatalloc[i] ( \valz[i,j] ) }%
.\end{equation}
\end{corollary}

Myerson's payment formula is immensely powerful.  It gives rise to a
procedure for optimal auction design that at first glance borders on
the miraculous.  The procedure is thus: First, solve for an allocation
rule that optimizes your objective (surplus, revenue, what have you),
subject only to ex-post feasibility.  Second, check for monotonicity.
If the optimal allocation rule is indeed monotonic, then you are home
free; you need only plug that allocation rule into Myerson's payment
formula and you will have imposed incentive compatibility and
individual rationality \emph{without sacrificing one ounce of
  optimality!}.  In other words, given a monotone allocation rule, IC
and IR are yours for the taking.  That is the essence of Myerson's
approach, and something we exploit heavily in this work.


\section{Robust vs.\ Bayesian Optimal Auctions}

Another of the more surprising facts that Myerson discovered about the
optimal auction design problem (assuming linear perceived payments) is
that the total expected revenue in the robust and Bayesian problems
are equivalent.  It is clear that the value of a solution to BRM must
be at least that of RRM, since the objectives are the same while BRM's
constraints are weaker.  It is the other direction, namely that the
value of a solution to BRM never exceeds that of RRM, which is
surprising.  This latter relies on the assumption that perceived
payments, and hence utilities, are linear, and therefore does not hold
in our setting with convex perceived payments, and hence concave
utilities, which we now proceed to show via counterexample.

\subsection{Robust Revenue $\ne$ Bayesian Revenue}

We begin our analysis of BRM vs.\ RRM with convex perceived payments by
demonstrating via example that solutions to BRM can strictly exceed
those of RRM when $\paymentq[i] = \payment[i]^2$.  
Therefore, an optimal solution to RRM does not generally yield an 
optimal solution to BRM, although it does yield an immediate lower bound.

\begin{example}
\label{ex:BicIcPayments}
Suppose we have $\numbidders = 2$ symmetric bidders, 
where for each bidder $i \in \setofbidders = \{ 1, 2\}$, $\valspace[i] = \{0, 100\}$,
and for each value $\val[i] \in \valspace[i]$, $f_i ( \val[i] ) = 1/2$.

\paragraph{RRM} 

Our goal is to maximize revenue, while satisfying IC, IR, and the ex-post feasibility constraints.
This mathematical program appears in Section~\ref{sssec:mpdesc_rrm_xp}.

We know from IR that $\paymentq[i] (\val[i], \valvec[-i])$ 
is upper-bounded by $\val[i] \alloc[i] (\val[i], \valvec[-i])$.
So when $\val[i] = 0$, $\paymentq[i] (\val[i], \valvec[-i]) \le 0$. 
So we can set $\alloc[i] (\val[i], \valvec[-i]) = 0$,
for all $\valvec[-i] \in \valspace[-i]$. 

When one bidder, say $i$, has type $100$, and the other bidder has type $0$, 
we maximize total payments by setting $\alloc[i] (100, 0) = 1$.  

When both bidders have type $100$, by Equation~\eqref{eq:pSquaredPayment},
$\payment[i] (100, 100) = \sqrt{100 \, \alloc[i] (100, 100)}$.
Therefore, total payments are maximized when $\alloc[i] (100, 100) = 1/2$.

Thus, in the robust problem, the following allocation is optimal:
\begin{align*}
\alloc[1] ( 0, 0 ) &= 0  &\quad  \alloc[2] ( 0, 0 ) &= 0 \\
\alloc[1] ( 100, 0 ) &= 1  &\quad  \alloc[2] ( 100, 0 ) &= 0 \\
\alloc[1] ( 0, 100 ) &= 0  &\quad  \alloc[2] ( 0, 100 ) &= 1 \\
\alloc[1] ( 100, 100 ) &= 1/2  &\quad  \alloc[2] ( 100, 100 ) &= 1/2
.\end{align*}
Computing payments according to Equation~\eqref{eq:pSquaredPayment} yields:
\begin{align*}
\payment[1] ( 0, 0 ) &= 0  &\quad  \payment[2] ( 0, 0 ) &= 0 \\
\payment[1] ( 100, 0 ) &= 10  &\quad  \payment[2] ( 100, 0 ) &= 0 \\
\payment[1] ( 0, 100 ) &= 0  &\quad  \payment[2] ( 0, 100 ) &= 10 \\
\payment[1] ( 100, 100 ) &= \sqrt{ 50 }  &\quad  \payment[2] ( 100, 100 ) &= \sqrt{ 50 }
.\end{align*}
Therefore, the total expected revenue in the robust problem is:
\begin{align*}
\sum_{i=1}^{n} \sum_{\valvec} f (\valvec) \payment[i] ( \val[i], \valvec[-i] )
&= 2 \sum_{\valvec} f (\valvec) \payment[i] ( \val[i], \valvec[-i] ) \\
&= 2 \left( \frac{1}{4} \left( 0 + 10 + 0 + \sqrt{50} \right) \right) \\
&= 5 \left( 1 + \frac{\sqrt{2}}{2} \right)
.\end{align*}

\paragraph{BRM} 

Our goal is to maximize revenue, while satisfying BIC, BIR and the ex-post feasibility constraints.
This mathematical program appears in Section~\ref{sssec:mpdesc_brm_xp_naive}.

We know from BIR that $\hatpaymentq[i] (\val[i])$ 
is upper-bounded by $\val[i] \hatalloc[i] (\val[i])$.
So when $\val[i] = 0$, $\hatpaymentq[i] (\val[i]) \le 0$. 
So we can set $\hatalloc[i] (\val[i]) = 0$, which implies that
$\alloc[i] (0, \valvec[-i]) = 0$, for all $\valvec[-i] \in \valspace[-i]$.  
Additionally, since $\hatpaymentq[i] (0) 
= \Exp_{\valvec[-i]} \left[ \left( \payment[i] (0, \valvec[-i]) \right)^2 \right]$,
this also means that $\payment[i] (0, \valvec[-i]) = 0$,
for all $\valvec[-i] \in \valspace[-i]$.

When $\val[i] = 100$, we want to maximize $\hatalloc[i] (\val[i])$ 
so that we can maximize $\hatpaymentq[i] (\val[i])$, 
as this will maximize the payments the auctioneer collects when a bidder has type $100$.
For bidder 1:
\begin{align*}
\hatalloc[1] (100) 
&= f_{-1} (0) \alloc[1] (100, 0) + f_{-1} (100) \alloc[1] (100, 100) \\
&= \frac{1}{2} \left( \alloc[1] (100, 0) + \alloc[1] (100, 100) \right)
.\end{align*}
We can set $\alloc[1] (100, 0)$ to $1$, which
leaves only $\alloc[1] (100, 100)$ to be determined.  
Since the setting is symmetric, we can also set $\alloc[2] (0, 100) = 1$, 
leaving $\alloc[2] (100, 100)$ also to be determined.  

Rewriting the BIR constraints using payment terms explicitly, we have
\begin{align*}
100 \left( \frac{1}{2} \alloc[1] (100, 0) + \frac{1}{2} \alloc[1] (100, 100) \right) 
\ge \frac{1}{2} \left( \left( \payment[1] (100, 0) \right)^2 + \left( \payment[1] (100, 100) \right)^2 \right) \\
100 \left( \frac{1}{2} \alloc[2] (0, 100) + \frac{1}{2} \alloc[2] (100, 100) \right) 
\ge \frac{1}{2} \left( \left( \payment[2] (0, 100) \right)^2 + \left( \payment[2] (100, 100) \right)^2 \right)
\end{align*}
Equivalently,
\begin{align*}
100 \left( \frac{1}{2} + \frac{1}{2} \alloc[1] (100, 100) \right) 
\ge \frac{1}{2} \left( \left( \payment[1] (100, 0) \right)^2 + \left( \payment[1] (100, 100) \right)^2 \right) \\
100 \left( \frac{1}{2} + \frac{1}{2} \left( 1 - \alloc[1] (100, 100) \right) \right) 
\ge \frac{1}{2} \left( \left( \payment[2] (0, 100) \right)^2 + \left( \payment[2] (100, 100) \right)^2 \right)
\end{align*}
Combining the inequalities yields
\begin{equation*}
150 \ge \frac{1}{2} \bigg(
\left( \payment[1] (100, 0) \right)^2 + \left( \payment[1] (100, 100) \right)^2 
+ \left( \payment[2] (0, 100) \right)^2 + \left( \payment[2] (100, 100) \right)^2
\bigg)
\end{equation*}
Subject to this constraint,
we can maximize total payments,
\begin{align*}
\payment[1] (100, 0) + \payment[1] (100, 100) 
+ \payment[2] (0, 100) + \payment[2] (100, 100) 
,\end{align*}
by equating each of the four payment terms, so that $\payment[i] (100, \valvec[-i]) = 5 \sqrt{3}$.

In summary, the following payment rule maximizes revenue:
\begin{align*}
\payment[i] (0, \valvec[-i]) &= 0,
&& \forall i \in \setofbidders, \forall \valvec[-i] \in \valspace[-i] \\
\payment[i] (100, \valvec[-i]) &= 5 \sqrt{3},
&& \forall i \in \setofbidders, \forall \valvec[-i] \in \valspace[-i] 
.\end{align*}

This payment rule implies a symmetric allocation in which $\alloc[1] (100, 0) = \alloc[2] (0, 100) = \frac{1}{2}$.
Therefore, this BRM problem can be solved using the same allocation rule as the corresponding RRM problem:
\begin{align*}
\alloc[1] ( 0, 0 ) &= 0  &\quad  \alloc[2] ( 0, 0 ) &= 0 \\
\alloc[1] ( 100, 0 ) &= 1  &\quad  \alloc[2] ( 100, 0 ) &= 0 \\
\alloc[1] ( 0, 100 ) &= 0  &\quad  \alloc[2] ( 0, 100 ) &= 1 \\
\alloc[1] ( 100, 100 ) &= 1/2  &\quad  \alloc[2] ( 100, 100 ) &= 1/2
.\end{align*}
Furthermore, the interim allocation rule is given by:
\begin{align*}
\hatalloc[1] (0) &= 0 &\quad \hatalloc[2] (0) = 0 \\
\hatalloc[1] (100) &= \frac{3}{4} &\quad \hatalloc[2] (100) = \frac{3}{4}
.\end{align*}
From this interim allocation rule, we derive the interim payment rule according to 
Equation~\eqref{eq:hSquaredPayment}:
\begin{align*}
h_i (0) &= \sqrt{0 \, \hatalloc[i](0)} = 0 \\
h_i (100) &= \sqrt{100 \, \hatalloc[i] (100) - (100 - 0) \, \hatalloc[i] (0)} = 5 \sqrt{3}
.\end{align*}
Observe that $h_i (0)
= \payment[i] (0, \valvec[-i])
= \Exp_{\valvec[-i]} \left[ \payment[i] (0, \valvec[-i]) \right]
= \hatpayment[i] (0)$, for all $\valvec[-i] \in \valspace[-i]$;
likewise for 100.


Finally, the total expected revenue in the Bayesian problem 
(which we can compute using either the $h_i (\val[i])'s$ or the
$\payment[i] (\val[i], \valvec[-i])'s$) is:
\begin{align*}
\sum_{i=1}^{n} \sum_{\val[i] \in \valspace[i]} f_i (\val[i]) h_i ( \val[i] )
&= 2 \sum_{\val[i] \in \valspace[i]} f_i (\val[i]) h_i ( \val[i] ) \\
&= 2 \left( \frac{1}{2} \left( 0 + 5 \sqrt{3} \right) \right) \\
&= 5 \sqrt{3}
.\end{align*}

Since $\sqrt{3} \approx 1.732 > 1.707 \approx (1 + \sqrt{2}/2)$,
it follows that $5 \sqrt{3} > 5 (1 + \sqrt{2}/2)$.  Thus, we conclude
that the value of the optimal solution to a Bayesian problem can
exceed the value of the optimal solution to the corresponding robust
problem, assuming convex perceived payments, and hence concave utilities.  \qed
\end{example}

\begin{remark}
  More generally, with two types, $0$ and $v$, the total expected
  robust revenue is $\frac{1}{2} \left( \sqrt{v} + \sqrt{\frac{v}{2}} \right)$,
  while the total expected Bayesian revenue is $\frac{\sqrt{3}}{2}
  \sqrt{v}$, making BRM greater than RRM by a factor of
    \[
        \frac{\frac{\sqrt{3}}{2} \sqrt{v}}
        {\frac{1}{2} \left( \sqrt{v} + \sqrt{\frac{v}{2}} \right)}
        = \frac{\sqrt{6}}{\sqrt{2} + 1}
        \approx 1.015
    .\]
An open question at present is how bad this ratio becomes as the
number of types increases.
\end{remark}

\begin{remark}
Until now,
we have not discussed the possibility of randomized auctions.
Randomization is not necessary in the linear-payment setting because
there always exists an optimal allocation rule that is integral.  But
in our examples, the optimal allocation rule is fractional.  While a
fractional allocation rule does not pose a problem in the case of an
infinitely divisible good, in the case of an indivisible good,
strictly more revenue can be obtained by interpreting the allocations
$\allocvec (\valvec)$ as probabilities, and then
allocating that one good at random based on these probabilities: i.e.,
\emph{randomization adds power assuming convex perceived payments}.

Likewise, in our case---the case of a finitely divisible good (i.e., a
budget)---randomization again adds power, assuming convex perceived
payments.  However, that power decreases as the discretization factor
decreases with respect to the budget.  Specifically, for a
discretization factor of $\Delta_B$ (say \euro0.01), given a budget of
$B$ (say 1 million euro), the potential gain due to randomization is
at most $O(\Delta_B / B)$ (which is \euro$10^{-8}$, in our example).
(We prove this claim in Appendix~\ref{sec:discretization}.)
\end{remark}

Having established that these two problems---BRM and RRM---are
distinct, we proceed to study them in turn (see Sections~\ref{sec:rrm}
and~\ref{sec:brm}).
But first a word about surplus.

\subsection{Robust Pseudo-surplus $\ne$ Bayesian Pseudo-surplus}

In addition to noting the equivalence of revenue in the robust and
Bayesian (linear perceived payment) problems, Myerson also noted the equivalence
of bidder surplus in these two problems.  Bidder surplus is a quantity
of interest in the usual setting with linear perceived payments, even when
studying revenue maximization, because bidder surplus upper bounds
revenue.  In our convex perceived payment setting, however, bidder surplus does
not upper bound revenue (see Example~\ref{ex:tight}).  Nonetheless, we
introduce a quantity we call pseudo-surplus, which does upper
bound revenue (in both the robust and the Bayesian problems).


When perceived payments are linear, so that $\paymentq[i] = \payment[i]$, IR implies that
$\val[i] \alloc[i] (\val[i], \valvec[-i]) \ge \payment[i] (\val[i], \valvec[-i])$.
The quantity on the left-hand side of this inequality is called bidder $i$'s \mydef{surplus}.
Taking expectations and summing over all bidders yields 
expected bidder surplus as an upper bound on 
expected revenue:
\begin{equation}
\label{eq:naiveUpperBound_q}
\sum_{i=1}^{\numbidders} \Exp_{\valvec} \left[ \val[i] \alloc[i] (\val[i], \valvec[-i]) \right] \ge
\sum_{i=1}^{\numbidders} \Exp_{\valvec} \left[ \payment[i] (\val[i], \valvec[-i]) \right]%
.\end{equation}

When perceived payments are quadratic, so that $\paymentq[i] = \payment[i]^2$, IR implies that
$\sqrt{\val[i] \alloc[i] (\val[i], \valvec[-i])} \ge \payment[i] (\val[i], \valvec[-i])$.
We call the quantity on the left-hand side of this inequality bidder $i$'s \mydef{(robust) pseudo-surplus}.
After taking expectations and summing over all bidders, as above, we find that 
expected revenue is upper bounded by
expected bidder pseudo-surplus:
\begin{equation}
\label{eq:naiveUpperBound_p2}
\sum_{i=1}^{n} \Exp_{\valvec} \left[ \sqrt{\val[i] \alloc[i] (\val[i], \valvec[-i])} \right]
\ge \sum_{i=1}^{n} \Exp_{\valvec} \left[ \payment[i] (\val[i], \valvec[-i]) \right]%
.\end{equation}

Likewise, in the Bayesian problem, by BIR,
$\forall i \in \setofbidders$ and $\forall \val[i] \in \valspace[i]$,
$\val[i] \hatalloc[i] ( \val[i] ) \ge \hatpaymentq[i] ( \val[i] )$.
%
When perceived payments are quadratic,
if there exists an interim payment function $h_i: \valspace[i] \to \Reals$ such that
$\hatpaymentq[i] ( \val[i] ) = ( h_i ( \val[i] ) )^2$,
as in Corollary~\ref{cor:payment-char}, then
$\val[i] \hatalloc[i] ( \val[i] ) \ge ( h_i ( \val[i] ) )^2$, so that
$\sqrt{\val[i] \hatalloc[i] ( \val[i] )} \ge h_i ( \val[i] )$.
We call the quantity on the left-hand side of this inequality bidder $i$'s \mydef{Bayesian pseudo-surplus}.
After taking expectations and summing over all bidders, as usual,
we find that 
expected revenue is upper-bounded by
expected bidder Bayesian pseudo-surplus:
\begin{equation}
\label{eq:naiveUpperBound_p2_Bayesian}
\sum_{i=1}^{n} \Exp_{\val[i]} \left[ \sqrt{\val[i] \hatalloc[i] (\val[i])} \right]
\ge \sum_{i=1}^{n} \Exp_{\val[i]} \left[ h_i (\val[i]) \right]%
.\end{equation}

\begin{remark}
Bayesian pseudo-surplus is at least as great as robust pseudo-surplus:
\begin{align*}
\Exp_{\val[i]} \left[ \sqrt{\val[i] \hatalloc[i] ( \val[i] )} \right]
&= \Exp_{\val[i]} \left[ \sqrt{\val[i] \Exp_{\valvec[-i]} \left[ \alloc[i] ( \val[i], \valvec[-i] )\right]} \right] \\
&\ge \Exp_{\val[i]} \left[ \Exp_{\valvec[-i]} \left[ \sqrt{\val[i] \alloc[i] ( \val[i], \valvec[-i] )} \right] \right] \\
&= \Exp_{\valvec} \left[ \sqrt{\val[i] \alloc[i] ( \val[i], \valvec[-i] )} \right]
.\end{align*}
%
\end{remark}

The following example, which builds on Example~\ref{ex:BicIcPayments},
shows that the pseudo-surplus bounds
(Equations~\eqref{eq:naiveUpperBound_p2}
and~\eqref{eq:naiveUpperBound_p2_Bayesian}) are tight, meaning revenue
can equal pseudo-surplus in both the robust and Bayesian problem settings.

\begin{example}
\label{ex:PW}
We continue using the framework of Example~\ref{ex:BicIcPayments}.

In the robust problem, revenue equals pseudo-surplus:
\begin{align*}
&\sum_{i=1}^{2} \Exp_{\valvec} \left[ \sqrt{\val[i] \alloc[i] (\val[i], \valvec[-i])} \right] \\
&= 2 \sum_{\valvec \in \valspace} f (\valvec) \left[ \sqrt{\val[1] \alloc[1] (\val[1], \valvec[-1])} \right] \\
&= 2 \left( \frac{1}{4} \left( \sqrt{0 \, \alloc[1] (0, 0)} + \sqrt{0 \, \alloc[1] (0, 100)} + 
\sqrt{100 \, \alloc[1] (100, 0)} + \sqrt{100 \, \alloc[1] (100, 100)} \right) \right) \\
&= 2 \left( \frac{1}{4} \left( \sqrt{0 \cdot 0} + \sqrt{0 \cdot 0} + 
\sqrt{100 \cdot 1} + \sqrt{100 \cdot \frac{1}{2}} \right) \right) \\
&= \frac{5}{2} \left( 2 + \sqrt{2} \right)
.\end{align*}

Similarly, in the Bayesian problem, revenue equals pseudo-surplus:
\begin{align*}
&\sum_{i=1}^{2} \Exp_{\val[i]} \left[ \sqrt{\val[i] \hatalloc[i] (\val[i])} \right] \\
&= 2 \sum_{\val[1] \in \valspace[1]} f_1 (\val[1]) \sqrt{\val[1] \hatalloc[1] (\val[1])} \\
&= 2 \left( \frac{1}{2} \left( \sqrt{0 \, \hatalloc[1] (0)} + \sqrt{100 \, \hatalloc[1] (100)} \right) \right) \\
&= 2 \left( \frac{1}{2} \left( \sqrt{0 \cdot 0} + \sqrt{100 \cdot \frac{3}{4}} \right) \right) \\
&= 5 \sqrt{3}
.\end{align*}
\qed
\end{example}

In Example~\ref{ex:BicIcPayments}, revenue in the Bayesian problem exceeds revenue in the robust problem.  
But in Example~\ref{ex:PW}, revenue equals pseudo-surplus in both the robust and Bayesian problems.
Therefore, Bayesian pseudo-surplus exceeds robust pseudo-surplus.
In other words, while pseudo-surplus in a Bayesian problem is always
at least that of pseudo-surplus in the corresponding robust problem
(potentially greater objective function; weaker constraints), 
Bayesian pseudo-surplus can strictly exceed robust pseudo-surplus.

In summary, both revenue and pseudo-surplus in the Bayesian and the
corresponding robust problems are not generally equal in the
convex perceived payment setting as they are in the linear perceived payment setting.
(In the linear perceived payment setting, because of linearity, there is no
distinction between Bayesian and robust surplus; there is only
surplus.)  Consequently, we are unable to proceed as Myerson did, and
solve for an optimal auction in the robust problem formulation as a
means of finding an optimal auction in the corresponding Bayesian
setting.  Instead, we are forced to address these two problems
separately.


\section{Robust Revenue Maximization}
\label{sec:rrm}

The first problem we tackle is robust revenue maximization (RRM).
Recall the power of Myerson's payment characterization
(Theorem~\ref{thm:payment-char}): the problem of optimal auction
design can be reduced to the problem of finding an optimal feasible
allocation, where feasible here means only ex-post feasible; then
if the resulting allocation is monotonic, IC and IR can be had for
free, by assigning the appropriate payments, thereby resulting in an
optimal auction.
This is precisely the approach we take here.

We have already established an upper bound on revenue (namely,
pseudo-surplus).  The present exercise will lead us to an algorithm
for finding an ex-post feasible allocation that optimizes that upper
bound, which, when saddled with Myerson payment rule, yields a heuristic
procedure that approximates RRM from above.  Likewise, this exercise
will also lead us to a heuristic lower bound, along with an analogous
algorithm for computing an ex-post feasible allocation that optimizes
that heuristic lower bound, which, along with Myerson payment rule, yields
a heuristic procedure that approximates RRM from below.

In Appendix~\ref{sec:upperBound}, we also derive an alternative,
as-of-yet non-operational, (tight) upper bound on expected revenue.

\subsection{Pseudo-Surplus Maximization}
\label{sec:surplus}

Although Myerson ultimately applied his payment formula to compute
payments in a revenue-maximizing auction, his formula applies equally
well to computing payments in a surplus-maximizing auction.  While
maximizing surplus is not the eventual goal of this work (recall our
example in which the government wished to maximize power: i.e.,
revenue), we can nonetheless use Myerson's approach to find a
pseudo-surplus-maximizing auction in our setting, in a manner
analogous to finding a surplus-maximizing auction in the usual
quasi-linear setting with linear perceived payments.

Our present objective is surplus, in the usual quasi-linear setting.
Observe the following:
\begin{equation}
\label{eq:pointwise}
\max_{\allocvec} \, \Exp_{\valvec} \left[ \sum_{i=1}^{\numbidders} \val[i] \, \alloc[i] ( \val[i], \valvec[-i] ) \right] =
\Exp_{\valvec} \left[ \max_{\allocvec ( \valvec )} \, \sum_{i=1}^{\numbidders} \val[i] \, \alloc[i] ( \val[i], \valvec[-i] ) \right]
.\end{equation}
This equality holds because $\allocvec(\valvec)$ is independent of any
other $\valvec' \ne \valvec \in \valspace$.  Consequently, surplus
(the left-hand side) can be maximized by proceeding
\mydef{pointwise\/} (the right-hand side), in which an optimal
allocation is determined for each $\valvec$ in turn.

Recall that Myerson reduced the problem of optimal auction design to
the problem of finding an optimal ex-post feasible allocation
(assuming monotonicity).
It is straightforward to enforce ex-post feasibility and
simultaneously maximize the objective function pointwise: given
$\valvec$, simply allocate
to $i^* \in \arg \max_i \{ \const[i] \}$, breaking ties randomly.
(Pseudocode for this pointwise approach is presented in Algorithm~\ref{alg:pointwise}.%
\footnote{In the case of an indivisible good, this pseudocode can be
  interpreted as defining a randomized mechanism that allocates
  uniformly at random to exactly one of the highest bidders.  In the
  case of an infinitely divisible good,
  fractionally, in proportion to the number of bidders tied for the
  highest bid.  In the case of interest, namely a finitely divisible
  budget $B$, any error in a discrete approximation as compared to the
  continuous (i.e., infinitely divisible) case, assuming
  discretization factor $\Delta_B$, is upper-bounded by $O (|M|\Delta_B/B)$.})
Since the resulting allocation is monotone (higher values are
allocated more), it can be plugged in to Myerson's payment rule to
arrive at an IC, IR, ex-post feasible surplus-maximizing auction (see
Algorithm~\ref{alg:surplus}).
%
Doing so yields Vickrey's famous second-price
auction~\cite{Vickrey:Counterspeculation/1961}.


\begin{algorithm}
\caption{Pointwise Maximization}
\label{alg:pointwise}
\begin{algorithmic}[1]
\Procedure{Pointwise\_Maximization}{$\cvec$}
\For{$i = 1$ to $\numbidders$}
    \State $\alloc[i] ( \const[i], \cvec[-i] ) \gets 0$
\EndFor
\If{any of the $\const[i]$'s are positive} \label{alg:positive}
    \State $M \gets \argmax_i \left\{ \const[i] \right\}$
    \ForAll{$i^* \in M$}
        \State $\alloc[i^*] ( \const[i^*], \cvec[-i^*] ) \gets 1 / \abs{M}$
    \EndFor
\EndIf
\State \textbf{return} $\allocvec (\cvec)$
\EndProcedure
\end{algorithmic}
\end{algorithm}

\begin{algorithm}[h]
\caption{Surplus Maximizer}
\label{alg:surplus}
\begin{algorithmic}[1]
\ForAll{$\valvec \in \valspace$}
\State $\allocvec (\valvec) \gets \Call{Pointwise\_Maximization}{\valvec}$
\EndFor
\State $W \gets \Exp_{\valvec} \left[ \sum_{i=1}^{\numbidders} \val[i] \alloc[i] ( \val[i], \valvec[-i] ) \right]$ \Comment{Total expected surplus}
\State Calculate the payment rule $\paymentvec$ using Equation~\eqref{eq:qPayment} with $\paymentq = \payment$
\State \textbf{return} $W$, $\allocvec$, $\paymentvec$
\end{algorithmic}
\end{algorithm}

Like surplus, pseudo-surplus can be maximized pointwise, because once
again
\begin{equation}
\label{eq:pseudo-pointwise}
\max_{\allocvec} \, \Exp_{\valvec} \left[ \sum_{i=1}^{\numbidders} \sqrt{\val[i] \, \alloc[i] ( \val[i], \valvec[-i] )} \right] =
\Exp_{\valvec} \left[ \max_{\allocvec ( \valvec )} \, \sum_{i=1}^{\numbidders} \sqrt{\val[i] \, \alloc[i] ( \val[i], \valvec[-i] )} \right]
.\end{equation}
Given $\valvec$, the function 
$\sqrt{\val[i] \alloc[i] (\val[i], \valvec[-i])}$ is non-decreasing
and concave.  Consequently, we can find an ex-post feasible allocation that
maximizes $\sum_{i=1}^{\numbidders} \sqrt{\val[i] \alloc[i] (\val[i], \valvec[-i])}$ 
by invoking the equi-marginal principle~\cite{gossen1854entwickelung},
which states that it is optimal (up to discretization error) to
allocate greedily until supply is exhausted.
That is, assuming $\val[i] \ge 0$, we calculate
$$\delta_i ( \val[i], \valvec[-i] ) =
\sqrt{ \val[i] } \Big( \sqrt{ \alloc[i] ( \val[i], \valvec[-i] ) + \epsilon } - 
\sqrt{ \alloc[i] ( \val[i], \valvec[-i] ) } \Big),$$
and then allocate
to $i^* \in \argmax_i \{ \delta_i ( \val[i], \valvec[-i] ) \}$,
breaking ties randomly.
Since the resulting allocation rule is monotone---higher values are
more likely to be allocated---it can be plugged into Myerson's payment
rule (as we have extended it to the convex perceived payment setting) to
obtain an optimal IC, IR, and ex-post feasible
pseudo-surplus-maximizing auction.
%
This heuristic procedure---1.~greedily solve for an allocation rule
that optimizes pseudo-surplus (an upper bound), and 2.~support that
allocation rule with Myerson's payment rule---approximates RRM from above.

Generalizing slightly, Algorithm~\ref{alg:equimarginalSolver}
solves (up to some set discretization factor $\epsilon$, and for
$\alpha = 1/2$) the following mathematical program, which we call
Program $C$, for ``concave'': given a vector of constants 
$\cvec \in {\Reals}^{\numbidders}$ and $\alpha \in (0,1)$,
\begin{align}
\max_{\allocvec (\cvec)} \,
& \sum_{i=1}^{\numbidders} \const[i]^{\alpha} \left( \alloc[i] (\const[i], \cvec[-i]) \right)^{\alpha}
&& \label{eq:obj} \\
\text{subject to }
& \sum_{i=1}^{\numbidders} \alloc[i] (\const[i], \cvec[-i]) \le 1
&& \label{eq:sum-constraint} \\
& 0 \le \alloc[i] (\const[i], \cvec[-i]) \le 1,
&& \forall i \in \setofbidders
\label{eq:zero-one-constraint}
.\end{align}
Using Algorithm~\ref{alg:equimarginalSolver},
Algorithm~\ref{alg:pseudo-surplus} produces
a pseudo-surplus maximizing IC, IR, XP auction.

\begin{algorithm}[h]
\caption{Equi-marginal Principle Solver}
\label{alg:equimarginalSolver}
\begin{algorithmic}[1]
\Procedure{eqp\_solver}{$\cvec$}
\For{$i = 1$ to $\numbidders$}
    \State $\alloc[i] (\cvec) \gets 0$
\EndFor
\For{$i = 1$ to $\numbidders$}
    \State $\const[i]^+ \gets \max \{ \const[i], 0\}$
\EndFor
\If{any of the $\const[i]$'s are positive}
    \While{$\sum_{i=1}^{\numbidders} \alloc[i] (\cvec) < 1$}
        \State $M \gets \argmax_i \left\{ \sqrt{ \const[i]^+ } \Big( \sqrt{ \alloc[i] (\const[i], \cvec[-i]) + \epsilon } - \sqrt{ \alloc[i] (\const[i], \cvec[-i])} \Big) \right\}$
        \ForAll{$i^* \in M$}
            \State $\alloc[i^*] (\const[i^*], \cvec[-i^*]) \gets \alloc[i^*] (\const[i^*], \cvec[-i^*]) + \epsilon / \abs{M}$
        \EndFor
    \EndWhile
\EndIf
\State \textbf{return} $\allocvec (\cvec)$
\EndProcedure
\end{algorithmic}
\end{algorithm}

\begin{algorithm}[h]
\caption{Pseudo-Surplus Maximizer}
\label{alg:pseudo-surplus}
\begin{algorithmic}[1]
\ForAll{$\valvec \in \valspace$}
\State $\allocvec (\valvec) \gets \Call{eqp\_solver}{\valvec}$
\EndFor
\State $W \gets \Exp_{\valvec} \left[ \sum_{i=1}^{\numbidders} \sqrt{\val[i] \alloc[i] ( \val[i], \valvec[-i] )} \right]$ \Comment{Total expected pseudo-surplus}
\State Calculate the payment rule $\paymentvec$ using Equation~\eqref{eq:pSquaredPayment}
\State \textbf{return} $W$, $\allocvec$, $\paymentvec$
\end{algorithmic}
\end{algorithm}

\begin{example}
\label{ex:tight}
Suppose there are $\numbidders$ bidders,
with $\valspace[i] = \{1\}$, for all bidders $i \in \setofbidders = \{ 1, \ldots, n \}$.
When bidder utilities are quasi-linear, surplus maximization 
prescribes that we assign $\alloc[i] (\valvec) = 1$ to a bidder $i$ whose type is maximal,
and $\alloc[j] (\valvec) = 0$ to all other bidders $j \in \setofbidders \setminus i$.
By Myerson's payment rule (Equation~\eqref{eq:qPayment} with $\paymentq[i] = \payment[i]$), 
this allocation rule yields payments
$\payment[i] ( \val[i], \valvec[-i] ) = 1$ and
$\payment[j] ( \val[j], \valvec[-j] ) = 0$
for $j \in \setofbidders \setminus i$.
Therefore, surplus and revenue are both $1$,
making surplus a tight upper bound on revenue.

When perceived payments are convex with 
$\paymentq[i] = \payment[i]^2$,
an integral allocation may not be optimal.
Pseudo-surplus and revenue are both maximized
when $\alloc[i] ( \val[i], \valvec[-i] ) = 1 / \numbidders$
for all bidders $i \in \setofbidders$,
and by Myerson's payment rule (Equation~\eqref{eq:pSquaredPayment}),
all bidders pay $\payment[i] ( \val[i], \valvec[-i] ) = \sqrt{1 / \numbidders}$.
Pseudo-surplus and revenue are both $\sqrt{\numbidders}$,
so analogous to the quasi-linear case,
pseudo-surplus is a tight upper bound on revenue
in our convex perceived payment setting.

Interestingly, in this example, bidder surplus, 
is not an upper bound on revenue, because
$\sum_{i=1}^{\numbidders} \val[i] \alloc[i] (\val[i], \valvec[-i]) = 1 \le
\sqrt{\numbidders} = \sum_{i=1}^{\numbidders} \payment[i] (\val[i], \valvec[-i])$.
Thus, unlike in the usual setting with linear perceived payments,
where bidder surplus upper bounds revenue, 
revenue in the convex perceived payment setting 
can actually exceed bidder surplus.
\qed
\end{example}



Algorithm~\ref{alg:equimarginalSolver} applies the equi-marginal
principle specifically to the sum of square root functions, subject to
ex-post feasibility.  But this algorithm works equally well when
applied to the sum of any concave functions.  In fact, when the
concave functions are differentiable, as in Equations~\eqref{eq:obj},
\eqref{eq:sum-constraint}, and~\eqref{eq:zero-one-constraint}, this
program can be solved in closed form (see
Section~\ref{sec:closedForm}).  But
Algorithm~\ref{alg:equimarginalSolver} remains applicable and
Algorithm~\ref{alg:pseudo-surplus} finds a pseudo-surplus maximizing
IC, IR, XP auction, even in cases where the concave functions are not
differentiable.

The two subroutines presented above---pointwise maximization
and the equi-marginal principle---are both used in the next
section when we revisit Myerson's virtual values.

\subsection{Myerson's Virtual Values}
\label{sec:virtual}

In his seminal work on optimal auction design, Myerson proved that
expected revenue can be expressed as something he called expected
virtual surplus, which he defined in terms of virtual values.
Myerson used this theorem to reduce the problem of finding an optimal
robust auction to pointwise optimization in virtual value space.
In this section, we restate Myerson's theorem and algorithm for
discrete, rather than continuous, types.
In the next section, we go on to use virtual values to heuristically
lower bound revenue when perceived payments are convex,
and utilities hence concave.


Following Section~\ref{sec:payments},
we assume bidder $i$'s type space is
$\valspace[i] = \{ \valz[i,k] : 1 \le k \le \valspacesize[i] \}$,
of cardinality $\valspacesize[i]$, 
where $\valz[i,j] < \valz[i,k]$ for $j < k$, 
and we let $\valz[i,{\valspacesize[i]}+1] = \valz[i,{\valspacesize[i]}]$.
We also assume the probability of type $\valz[i,k] \in \valspace[i]$ 
is given by cumulative distribution function $F_{i,k} \equiv F_i ( \valz[i,k] )$
and corresponding probability mass function $f_{i,k} \equiv f_i ( \valz[i,k] )$.

Using this notation, here is the definition of \mydef{virtual values} for discrete distributions:
\begin{equation}
\label{eq:discvvf}
\vvald[i] ( \valz[i,k], \valz[i,k+1] ) = \valz[i,k] - ( \valz[i,k+1] - \valz[i,k] ) 
\left( \frac{1 - F_{i,k}}{f_{i,k}} \right)
.\end{equation}
We abbreviate the virtual value $\vvald[i] ( \valz[i,k], \valz[i,k+1] )$ by $\vvald[i,k]$.  
We also use the shorthand 
$\vvald[i] (\val[i]) \equiv \vvald[i]( \valz[i,k], \valz[i,k+1] )$,
when $\val[i] = \valz[i,k]$ for some $1 \le k \le \valspacesize[i]$.

Using the discrete version of Myerson's payment formula (Theorem~\ref{thm:payment-char}), 
and following a similar analysis to that of Myerson~\cite{Myerson:Optimal/1981},
we arrive at the following theorem.

\begin{theorem}
\label{thm:ExpRevVirVal}
Assume bidders' utilities take the form of Equation~\eqref{eq:util_with_q}.
If a mechanism is IC and IR, then for all bidders $i$,
\begin{equation}
\label{eq:ExpRevVirVal}
\Exp_{\valz[i,k] \sim F_i} \left[ \paymentq[i] ( \valz[i,k], \valvec[-i] ) \right] = 
\Exp_{\valz[i,k] \sim F_i} \left[ \vvald[i,k] \, \alloc[i] ( \valz[i,k], \valvec[-i] ) \right]
.\end{equation}
(The right-hand side of Equation~\eqref{eq:totalExpRevExpVvQl} is
called bidder $i$'s expected \mydef{virtual surplus}, because it is
surplus in virtual value space.)
\end{theorem}

By Theorem~\ref{thm:ExpRevVirVal},
when perceived payments are linear (i.e., $\paymentq[i] = \payment[i]$), 
each bidder's expected (actual) payment is equal to his expected virtual surplus.
Furthermore, total expected revenue is equal to total expected virtual surplus:
\begin{equation}
\label{eq:totalExpRevExpVvQl}
\Exp_{\valvec} \left[ \sum_{i=1}^{\numbidders} \payment[i] ( \val[i], \valvec[-i] ) \right] 
= \Exp_{\valvec} \left[ \sum_{i=1}^{\numbidders} \vvald[i] (\val[i]) \, \alloc[i] ( \val[i], \valvec[-i] ) \right]
.\end{equation}

\noindent
Mimicking Equations~\eqref{eq:pointwise}
and~\eqref{eq:pseudo-pointwise}, virtual surplus (the right-hand side
of Equation~\eqref{eq:totalExpRevExpVvQl}) can be maximized in the
same pointwise fashion as surplus and pseudo-surplus:
\begin{equation}
\label{eq:pointwise2}
\max_{\allocvec} \Exp_{\valvec} \left[ \sum_{i=1}^{n} \vvald[i] (\val[i]) \, \alloc[i] ( \val[i], \valvec[-i] ) \right] =
\Exp_{\valvec} \left[ \max_{\allocvec ( \valvec )} \sum_{i=1}^{n} \vvald[i] (\val[i]) \, \alloc[i] ( \val[i], \valvec[-i] ) \right]
.\end{equation}

\noindent
Furthermore, by Equation~\eqref{eq:totalExpRevExpVvQl}, maximizing
virtual surplus also maximizes revenue.
So, in fact, Myerson's virtual value theorem
(Theorem~\ref{thm:ExpRevVirVal}) gives rise to a pointwise method
(Algorithm~\ref{alg:virtual_surplus}) for finding a revenue-maximizing
ex-post feasible allocation rule: for each $\valvec$, simply allocate
to one of the bidders with the highest \emph{virtual\/} value.

The next question, then, is: is this allocation rule monotone? If so,
we can apply Myerson's payment formula to arrive at a mechanism that
is also IC and IR.  But there is nothing that guarantees that the
resulting allocation rule is monotone; the rule depends on virtual
values, while monotonicity is a constraint on values.
So, we need one further assumption on the bidders' type distributions
to ensure monotonicity of virtual values in values:
i.e., for all bidders $i \in \setofbidders$,
$\vvald[i,k+1] \ge \vvald[i,k]$
whenever $\valz[i,k+1] \ge \valz[i,k]$.
This assumption is called \mydef{regularity}.  Assuming regularity,
pointwise optimization of virtual surplus yields a monotonic
allocation rule,
and Algorithm~\ref{alg:virtual_surplus}
yields a solution to RRM in the linear perceived payment case.

\begin{algorithm}[h]
\caption{Virtual Surplus Maximizer}
\label{alg:virtual_surplus}
\begin{algorithmic}[1]
\ForAll{$\valvec \in \valspace$}
\ForAll{$i \in \setofbidders$}
\State $\virvecd[i] \gets \vvald[i] (\val[i])$ \Comment{Calculate virtual values using Equation~\eqref{eq:discvvf}.}
\EndFor
    \State $\allocvec ( \valvec ) \gets \Call{Pointwise\_Maximization}{\virvecd}$
\EndFor
\State Calculate the payment rule $\paymentvec$ using Equation~\eqref{eq:qPayment} with $\paymentq = \payment$
\State $R \gets \Exp_{\valvec} \left[ \sum_{i=1}^{\numbidders} \payment[i] ( \val[i], \valvec[-i] ) \right]$ 
\Comment{Total expected revenue/virtual surplus}
\State \textbf{return} $R$, $\allocvec$, $\paymentvec$
\end{algorithmic}
\end{algorithm}

\begin{remark}
  The auction that falls out of Algorithm~\ref{alg:virtual_surplus}
  can be interpreted as one with per-bidder reserve prices.
Restricting the search to positive
virtual values
(Algorithm~\ref{alg:pointwise}, Line~\ref{alg:positive})
ensures that the winning bidder's bid exceeds not only
the second-highest bid, but in addition the smallest value for which
his inverse virtual value function is zero.  Plugging the resulting
allocation into Myerson's payment formula then dictates that the
winning bidder pay the maximum of the second-highest bid and the
inverse of his virtual value function at zero,%
\footnote{When the inverse is not a singleton, he pays the minimum value in this set.}
the latter of which functions as his reserve price.
\end{remark}

\subsection{Heuristic Lower Bound}
\label{sec:lowerBound}

In the more general case of convex perceived payments,
expected revenue does not equal expected virtual surplus. 
Nonetheless, for $\paymentq[i] = \payment[i]^2$,
we can derive bounds on expected revenue in terms of expected virtual surplus.  
One operational upper bound is pseudo-surplus; we present a second,
non-operational upper bound expressed in terms of virtual surplus in
Appendix~\ref{sec:upperBound}.
Here, we present a heuristic lower bound, also in terms of virtual surplus,
which lends itself to a heuristic procedure that approximates RRM.

\begin{theorem}
\label{thm:payment-lower-bound}
If bidders' utilities take the form of Equation~\eqref{eq:util_with_q}
and $\paymentq[i] (\payment[i])$ is quadratic
(i.e., $\paymentq[i] ( \payment[i] ( \valz[i,k], \valvec[-i] ) ) = 
\left( \payment[i] ( \valz[i,k], \valvec[-i] ) \right)^2$),
then expected payments can be lower-bounded as follows:
\begin{equation}
\label{eq:expPayLb}
\Exp_{\valz[i,k] \sim F_i} \left[ \payment[i]( \valz[i,k], \valvec[-i] ) \right]
\ge \Exp_{\valz[i,k] \sim F_i} \left[ \frac{\vvald[i,k] \, \alloc[i](\valz[i,k], \valvec[-i])}
  {\payment[i] ( \valz[i,k], \valvec[-i])} \right]
.\end{equation}
\end{theorem}

\begin{proof}
First, observe the following:
\begin{align}
  \Exp_{\valz[i,k] \sim F_i} \left[ \payment[i](\valz[i,k], \valvec[-i]) \right]
  &= \Exp_{\valz[i,k] \sim F_i} \left[ \frac{\payment[i](\valz[i,k], \valvec[-i])^2}
    {\payment[i](\valz[i,k], \valvec[-i])} \right] \\
  &= \sum_{k=1}^{\valspacesize[i]} f_{i,k} \left[ \frac{\valz[i,k] \, \alloc[i] ( \valz[i,k], \valvec[-i] ) - 
\sum_{j=1}^{k - 1} ( \valz[i,j+1] - \valz[i,j] ) \, \alloc[i] ( \valz[i,j], \valvec[-i] )}
    {\payment[i](\valz[i,k], \valvec[-i])} \right] \\ 
  &= \sum_{k=1}^{\valspacesize[i]} \frac{f_{i,k} \, \valz[i,k] \, \alloc[i] ( \valz[i,k], \valvec[-i])}{\payment[i] ( \valz[i,k], \valvec[-i])} - \sum_{k=1}^{\valspacesize[i]} \sum_{j=1}^{k - 1} \frac{f_{i,k} \, (\valz[i,j+1] - \valz[i,j]) \, \alloc[i] (\valz[i,j], \valvec[-i])}{\payment[i]( \valz[i,k], \valvec[-i])} \label{eq:neg-part}
.\end{align}

We now work only with the negative part of Equation~\eqref{eq:neg-part}.
By summation by parts, this expression equals:
\begin{align}
  & \sum_{j=1}^{\valspacesize[i]} \sum_{k=j+1}^{\valspacesize[i]} \frac{f_{i,k} \, (\valz[i,j+1] - \valz[i,j]) \, \alloc[i] (\valz[i,j], \valvec[-i])}{\payment[i](\valz[i,k], \valvec[-i])} \\
  &\le \sum_{j=1}^{\valspacesize[i]} (\valz[i,j+1] - \valz[i,j]) \, \alloc[i] (\valz[i,j], \valvec[-i])
  \left( \sum_{k=j+1}^{\valspacesize[i]} \frac{f_{i,k}}{\payment[i](\valz[i,j], \valvec[-i])} \right) \label{eq:ub} \\
  &= \sum_{j=1}^{\valspacesize[i]} (\valz[i,j+1] - \valz[i,j]) \, \alloc[i] (\valz[i,j], \valvec[-i])
  \left( \frac{1 - F_{i,j}}{\payment[i](\valz[i,j], \valvec[-i])} \right) \label{eq:cdf}
.\end{align}
Equation~\eqref{eq:ub} follows from the fact that
$\payment[i] ( \valz[i,j+1], \valvec[-i] ) \ge \payment[i] ( \valz[i,j], \valvec[-i] )$,
which follows in turn from the fact that
$\paymentq[i] ( \valz[i,j+1], \valvec[-i] ) \ge \paymentq[i] ( \valz[i,j], \valvec[-i] )$.
Equation~\eqref{eq:cdf} is the result of substituting
$\sum_{k=j + 1}^{\valspacesize[i]} f_{i,k} = 1 - F_{i,j}$.

Merging back in the positive part of Equation~\eqref{eq:neg-part} and re-arranging yields:
\begin{align}
  \Exp_{\valz[i,k] \sim F_i} \left[ \payment[i](\valz[i,k], \valvec[-i]) \right]
  &\ge \sum_{k=1}^{\valspacesize[i]} \frac{f_{i,k} \, \valz[i,k] \, \alloc[i] ( \valz[i,k], \valvec[-i])}{\payment[i] ( \valz[i,k], \valvec[-i])} - \sum_{j=1}^{\valspacesize[i]} (\valz[i,j+1] - \valz[i,j]) \, \alloc[i] (\valz[i,j], \valvec[-i]) \left( \frac{1 - F_{i,j}}{\payment[i](\valz[i,j], \valvec[-i])} \right) \\  
  &= \sum_{k=1}^{\valspacesize[i]} \frac{f_{i,k} \, \alloc[i](\valz[i,k], \valvec[-i])}{\payment[i] ( \valz[i,k], \valvec[-i] )}\left( \valz[i,k] -  (z_{k+1} - \valz[i,k]) \left( \frac{1 - F_{i,k}}{f_{i,k}} \right) \right) \\
  &= \sum_{k=1}^{\valspacesize[i]} f_{i,k} \left( \frac{\vvald[i,k] \, \alloc[i](\valz[i,k], \valvec[-i])}
  {\payment[i] ( \valz[i,k], \valvec[-i])} \right) \\
  &= \Exp_{\valz[i,k] \sim F_i} \left[ \frac{\vvald[i,k] \, \alloc[i](\valz[i,k], \valvec[-i])}
  {\payment[i] ( \valz[i,k], \valvec[-i])} \right]
.\end{align}
\end{proof}

\begin{example}
\label{ex:tight3}
Following up on Example~\ref{ex:tight}, in which $\valspace[i] = \{ 1 \}$,
we find that $\vvald[i,1] = 1$, for all bidders $i \in \setofbidders$.
In this case, the lower bound (Equation~\eqref{eq:expPayLb}) is tight:
\begin{equation*}
\frac{\vvald[i,1] \, \alloc[i] (\val[i], \valvec[-i])}{\payment[i] (\valz[i], \valvec[-i])}
= \frac{(1) \left( \frac{1}{\numbidders} \right)}{\sqrt{\frac{1}{\numbidders}}}
= \frac{\sqrt{\numbidders}}{\numbidders}
= \sqrt{\frac{1}{\numbidders}}
= \payment[i] (\val[i], \valvec[-i])%
.\end{equation*}
\qed
\end{example}

Although Theorem~\ref{thm:payment-lower-bound} is actually a claim about expectations,
let us make the stronger assumption that for all $\val[i] \in \valspace[i]$
and $\valvec[-i] \in \valspace[-i]$,
\begin{equation}
\payment[i] (\val[i], \valvec[-i])
\ge \frac{\vvald[i] (\val[i]) \, \alloc[i](\val[i], \valvec[-i])}{\payment[i] ( \val[i], \valvec[-i])}
;\end{equation}
equivalently,
$\left( \payment[i] (\val[i], \valvec[-i]) \right)^2 \ge
\vvald[i] (\val[i]) \, \alloc[i](\val[i], \valvec[-i])$.
Now, letting $\vvald[i]^{+} (\val[i]) = \max \{ \vvald[i] (\val[i]), 0 \}$,
it also holds that $\left( \payment[i] (\val[i], \valvec[-i]) \right)^2
\ge \vvald[i]^{+} (\val[i]) \, \alloc[i](\val[i], \valvec[-i])$;
equivalently, $\payment[i] (\val[i], \valvec[-i])
\ge \sqrt{\vvald[i]^{+} (\val[i]) \, \alloc[i](\val[i], \valvec[-i])}$.
Summing over all bidders and then taking expectations yields:
\begin{equation}
\label{eq:p2ExpRevLowerBound}
\Exp_{\valvec} \left[ \sum_{i=1}^{\numbidders} \payment[i] ( \val[i], \valvec[-i] ) \right]
\ge \Exp_{\valvec} \left[ \sum_{i=1}^{\numbidders} \sqrt{\vvald[i]^{+} (\val[i]) \, \alloc[i](\val[i], \valvec[-i])} \right]
,\end{equation}
which gives us a new \emph{heuristic\/} objective of maximizing the
right-hand side of Equation~\eqref{eq:p2ExpRevLowerBound}.  As usual,
we optimize this expression in a pointwise fashion subject to ex-post
feasibility.

Given $\valvec$, the function $\sqrt{\vvald[i]^{+} (\val[i]) \alloc[i](\val[i], \valvec[-i])}$
is non-decreasing and concave.  Consequently, we can find an ex-post feasible allocation that maximizes
$\sum_{i=1}^{\numbidders} \sqrt{\vvald[i]^{+} (\val[i]) \alloc[i](\val[i], \valvec[-i])}$
by invoking the equi-marginal principle~\cite{gossen1854entwickelung}.  That is,
we calculate $$\delta_i ( \val[i], \valvec[-i] ) =
\sqrt{ \vvald[i,k]^{+} } \left( \sqrt{ \alloc[i] ( \val[i], \valvec[-i] ) + \epsilon } - 
\sqrt{ \alloc[i] ( \val[i], \valvec[-i] ) } \right),$$
and then allocate $\epsilon$ to $\argmax_i \{ \delta_i ( \val[i], \valvec[-i] ) \}$.  
Assuming regularity, the resulting allocation rule is monotone (higher
values are allocated more), so it can be plugged into Myerson's payment
rule (as we have extended it to the convex perceived payment setting) to obtain
an optimal IC, IR, and ex-post feasible revenue-maximizing auction.
%
This heuristic procedure---1.~greedily solve for an allocation rule
that optimizes the heuristic lower bound, and 2.~support that
allocation rule with Myerson's payment rule---approximates RRM from
below.  (See Algorithm~\ref{alg:heurSolver}.)

\begin{algorithm}
\caption{Heuristic Lower Bound Maximizer}
\label{alg:heurSolver}
\begin{algorithmic}[1]
\ForAll{$\valvec \in \valspace$}
\ForAll{$i \in \setofbidders$}
\State $\virvecd[i] \gets \vvald[i] (\val[i])$ \Comment{Calculate virtual values using Equation~\eqref{eq:discvvf}}
\EndFor
\State $\allocvec (\valvec) \gets \Call{eqp\_solver}{\virvecd}$
\EndFor
\State Calculate the payment rule $\paymentvec$ using Equation~\eqref{eq:pSquaredPayment}
\State $R \gets \Exp_{\valvec} \left[ \sum_{i=1}^{\numbidders} \payment[i] ( \val[i], \valvec[-i] ) \right]$ \Comment{Total expected revenue}
\State \textbf{return} $R$, $\allocvec$, $\paymentvec$
\end{algorithmic}
\end{algorithm}

\begin{example}[Reserve bids]

Suppose the bidders have values drawn from a uniform distribution, where
$\valz[i,k] \in \valspace[i] 
= \{ j / \valspacesize[i] : j \in \mathbb{N}, 1 \le j \le \valspacesize[i] \}$, 
$f_{i,k} = 1 / \valspacesize[i]$ and $F_{i,k} = k / \valspacesize[i]$,
for each bidder $i \in \setofbidders$ and $1 \le j \le \valspacesize[i]$.  
In an optimal auction, values that have negative corresponding virtual values are rejected.
Since $\vvald[i] (\valz[i,k]) = 2 k / \valspacesize[i] - 1$,
the minimum $k$ for which $\vvald[i] (\valz[i,k]) \ge 0$ is $k^* = \ceil{\valspacesize[i] / 2}$.
So, given $\valvec[-i]$, the allocation variable $\alloc[i] (\val[i], \valvec[-i])$ 
can only be positive if $\val[i] \ge \valz[i,k^*]
= k^* / \valspacesize[i] = \ceil{\valspacesize[i] / 2} / \valspacesize[i]$.

If we insist upon an integral allocation rule (as in the case of an indivisible good, for example), 
so that exactly one bidder $i$ is allocated with $\alloc[i] (\val[i], \valvec[-i]) = 1$, 
$i$ must bid at least $\valz[i,k^*]$, in which case $i$ pays
$\paymentq[i] (\valz[i,k^*], \valvec[-i]) = \valz[i,k^*]$.
When $\valspacesize[i]$ is even, $\valz[i,k^*] = 1/2$, so
$\paymentq[i] (\valz[i,k^*], \valvec[-i]) = 1/2$.
Assuming linear perceived payments, so that $\paymentq[i] = \payment[i]$, it follows that
$\payment[i] (\valz[i,k^*], \valvec[-i]) = 1/2$;
assuming convex perceived payments, so that $\paymentq[i] = \payment[i]^2$, it follows that
$\payment[i] (\valz[i,k^*], \valvec[-i]) = \sqrt{1/2} = \sqrt{2}/2$.


When we allow for a fractional allocation rule, 
$i$ must again bid at least $\valz[i,k^*]$ to be allocated.
If, for example, $\alloc[i] (\valz[i,k^*], \valvec[-i]) = 1 / 2$, then
$\paymentq[i](\valz[i,k^*], \valvec[-i]) = \valz[i,k^*] / 2$.
When $\valspacesize[i]$ is even, $\valz[i,k^*] = 1/2$, so
$\paymentq[i] (\valz[i,k^*], \valvec[-i]) = 1/4$.
Assuming linear perceived payments, so that $\paymentq[i] = \payment[i]$, it follows that
$\payment[i] (\valz[i,k^*], \valvec[-i]) = 1/4$;
assuming convex perceived payments, so that $\paymentq[i] = \payment[i]^2$, it follows that
$\payment[i] (\valz[i,k^*], \valvec[-i]) = \sqrt{1/4} = 1/2$.

N.B.\ In this example, if, when allowing for a fractional (or
randomized) allocation rule, two bidders are both allocated $1/2$, they
both pay $1/2$, and total revenue ($1$) exceeds total revenue in the
case of an integral allocation rule only ($\sqrt{2}/2$).
\qed
\end{example}


\section{Bayesian Revenue Maximization}
\label{sec:brm}

We now turn our attention to the problem of Bayesian revenue
maximization in our convex perceived payment setting.  With a bit more effort,
we are again able to apply Myerson's payment rule, but unlike in the
robust case, the logic employed here is not merely a straightforward
generalization or application of Myerson's original reasoning.
Specifically, we produce a smaller mathematical program than the
default revenue-maximizing one: whereas the default program has
exponentially-many payment variables, ours has only polynomially-many;
but it still has exponentially-many allocation variables and ex-post
feasibility constraints.

Our strategy is as follows:
First we show that in our search for an optimal Bayesian auction, it
suffices to restrict our attention to auctions in which each bidder's
payment is a (deterministic) function $h_i: \valspace[i] \to \Reals$
of his type alone, irrespective of other bidders' types.  Second,
since by design this interim payment function $h_i$ is such that
$\hatpaymentq[i] ( \val[i] ) = ( h_i ( \val[i] ) )^2$, by
Corollary~\ref{cor:payment-char}, $h_i$ is given by
Equation~\eqref{eq:hSquaredPayment}.

Consider a (possibly randomized) auction, Auction $A$, where $\payment[i]^A
(\val[i], \valvec[-i], r)$ denotes bidder $i$'s payment in auction
$A$.  (Here, $r$ is the outcome of some randomization device.)  We
define another (deterministic) auction, Auction $B$, with payment rule
$\payment[i]^B (\val[i], \valvec[-i]) = h_i (\val[i])$ for some
function $h_i(\val[i])$ that depends only on $\val[i]$.  More
specifically,
\begin{equation}
h_i (\val[i]) 
= \sqrt{\Exp_{\valvec[-i], r} \left[ 
\left( \payment[i]^A (\val[i], \valvec[-i], r) \right)^2 \right]}
.\end{equation}

\begin{lemma}
\label{lem:feasibleAllocAB} 
An arbitrary allocation $\allocvec \in [0,1]^{\numbidders}$,
together with the corresponding payment rule $\hatpaymentvec^A$ or $\hatpaymentvec^B$,
satisfies BIC, BIR, and ex-post feasibility for Auction $A$ if and only if 
it satisfies BIC, BIR, and ex-post feasibility for Auction $B$.
\end{lemma}

\begin{proof}
An arbitrary allocation $\allocvec \in [0,1]^{\numbidders}$ satisfies ex-post
feasibility for Auction $A$ if and only if satisfies ex-post feasibility for
$B$, as these set of constraints are identical in both auctions.  The
BIC and BIR constraints, however, can differ across the two auctions
because payments can differ.  Nonetheless, we now proceed to show
that an arbitrary allocation $\allocvec$ satisfies BIC and BIR for Auction $A$
if and only if it satisfies these properties for Auction $B$.

Since
\begin{align}
\Exp_{\valvec[-i]} \left[ \left( \payment[i]^B (\val[i], \valvec[-i]) \right)^2 \right]
&= \Exp_{\valvec[-i]} \left[ (h_i (\val[i]))^2 \right] \\
&= \left( h_i (\val[i]) \right)^2 \\
&= \left( \sqrt{\Exp_{\valvec[-i], r} \left[ \left( \payment[i]^A (\val[i], \valvec[-i], r) \right)^2 \right]} \right)^2 \\
&= \Exp_{\valvec[-i], r} \left[ \left( \payment[i]^A (\val[i], \valvec[-i], r) \right)^2 \right]
,\end{align}
it follows that 
\begin{align}
\hatpaymentq[i]^B (\val[i])
&= \Exp_{\valvec[-i]} \left[ \paymentq[i] ( \payment[i]^B (\val[i], \valvec[-i]) ) \right] \\
&= \Exp_{\valvec[-i]} \left[ \left( \payment[i]^B (\val[i], \valvec[-i]) \right)^2 \right] \\
&= \Exp_{\valvec[-i], r} \left[ \left( \payment[i]^A (\val[i], \valvec[-i], r) \right)^2 \right] \\
&= \Exp_{\valvec[-i], r} \left[ \paymentq[i] ( \payment[i]^A (\val[i], \valvec[-i], r) ) \right] \\ 
&= \hatpaymentq[i]^A (\val[i]).
\end{align}
Therefore, for all bidders $i \in \setofbidders$ and values $\val[i], \valw[i] \in \valspace[i]$, 
\begin{equation}
\val[i] \hatalloc[i] ( \val[i] ) - \hatpaymentq[i]^B (\val[i])
\ge 
\val[i] \hatalloc[i] ( \valw[i] ) - \hatpaymentq[i]^B (\valw[i])
\ \text{if and only if} \
\val[i] \hatalloc[i] ( \val[i] ) - \hatpaymentq[i]^A (\val[i])
\ge 
\val[i] \hatalloc[i] ( \valw[i] ) - \hatpaymentq[i]^A (\valw[i])
\end{equation}
and
\begin{equation}
\val[i] \hatalloc[i] ( \val[i] ) - \hatpaymentq[i]^B (\val[i]) \ge 0
\quad \text{if and only if} \quad
\val[i] \hatalloc[i] ( \val[i] ) - \hatpaymentq[i]^A (\val[i]) \ge 0.
\end{equation}
In other words, $\allocvec$ (and the corresponding payment rule $\hatpaymentvec$)
is BIC and BIR for Auction $A$ if and only if it is BIC and BIR for Auction $B$.
\end{proof}

\begin{lemma}
\label{lem:revAB}
The total expected revenue of Auction $B$ is at least that of Auction $A$.
\end{lemma}

\begin{proof}
Let $R_B = \sum_{i=1}^{\numbidders} 
\Exp_{\valvec} \left[ \payment[i]^B (\val[i], \valvec[-i]) \right]$ 
denote the expected revenue of Auction $B$, and 
let $R_A = \sum_{i=1}^{\numbidders} 
\Exp_{\valvec} \left[ \payment[i]^A (\val[i], \valvec[-i]) \right]$ 
denote the expected revenue of Auction $A$. 
By Jensen's inequality, since the square root is a concave function,
\begin{align}
h_i (\val[i]) 
&= \sqrt{\Exp_{\valvec[-i], r} \left[ \left( \payment[i]^A (\val[i], \valvec[-i], r) \right)^2 \right] } \\
&\geq \Exp_{\valvec[-i], r} \left[ \sqrt{\left( \payment[i]^A (\val[i], \valvec[-i], r) \right)^2} \right] \\
&= \Exp_{\valvec[-i], r} \left[ \payment[i]^A (\val[i], \valvec[-i], r) \right]
,\end{align}
it follows that
\begin{align}
R_B 
&= \sum_{i=1}^{\numbidders} \Exp_{\valvec} \left[ \payment[i]^B (\val[i], \valvec[-i]) \right] \\
&= \sum_{i=1}^{\numbidders} \Exp_{\val[i]} \left[ h_i (\val[i]) \right] \\
&\geq \sum_{i=1}^{\numbidders} \Exp_{\val[i]} \left[ \Exp_{\valvec[-i], r} 
\left[ \payment[i]^A (\val[i], \valvec[-i], r) \right] \right] \\
&= \sum_{i=1}^{\numbidders} \Exp_{\valvec, r} \left[ \payment[i]^A (\val[i], \valvec[-i], r) \right] \\
&= R_A
.\end{align}
In other words, the expected revenue of Auction $B$ is at least that of Auction $A$.
\end{proof}

These two lemmas establish that in our search for an optimal auction,
it suffices to restrict our attention to auctions like Auction $B$ in
which each bidder's payment is a (deterministic) function $h_i$ of his
type alone. Furthermore, since
$\hatpaymentq[i]^B (\val[i])
= \Exp_{\valvec[-i]} \left[ \paymentq[i] ( \payment[i]^B (\val[i], \valvec[-i]) ) \right]
= \Exp_{\valvec[-i]} \left[ \left( \payment[i]^B (\val[i], \valvec[-i]) \right)^2 \right]
= \Exp_{\valvec[-i]} \left[ (h_i (\val[i]))^2 \right]
= \left( h_i (\val[i]) \right)^2$,
it follows by
Corollary~\ref{cor:payment-char} that $h_i$ is given by
Equation~\eqref{eq:hSquaredPayment}: i.e.,
\begin{equation}
h_i ( \valz[i,\ell] )
=  \sqrt{ \valz[i,\ell] \hatalloc[i] ( \valz[i,\ell] ) - \sum_{j=1}^{\ell - 1} ( \valz[i,j+1] - \valz[i,j] ) \hatalloc[i] ( \valz[i,j] ) }%
.\end{equation}
These two lemmas together with this final observation establish the follow theorem:

\begin{theorem}
\label{thm:deterministic-payments} 
The mathematical program given in Section~\ref{sssec:mpdesc_brm_xp},
with only polynomially-many payment variables, gives a solution to
the BRM problem, which when stated naively involves
exponentially-many such variables
(Section~\ref{sssec:mpdesc_brm_xp_naive}).
\end{theorem}

\begin{remark}
Theorem~\ref{thm:deterministic-payments} holds not only for
$\paymentq[i] = \payment[i]^2$, but for any
$\paymentq[i] = C_i(\payment[i])$,
where $C_i(\cdot)$ is an invertible convex function.
The only necessary modification in the proof of 
Theorem~\ref{thm:deterministic-payments} 
is that we now consider deterministic payments of the form 
\begin{equation}
h_i (\val[i]) = C_i^{-1} \left( \Exp_{\valvec[-i]} 
\left[ C_i \left( \payment[i]^A (\val[i], \valvec[-i]) \right) \right] \right) %
.\end{equation}
The theorem then follows by the concavity of $C_i^{-1}(\cdot)$.  
\end{remark}

We close this section with a heuristic for approximating BRM.
Analogous to Algorithm~\ref{alg:heurSolver}, which describes a method
to approximate RRM from below, Algorithm~\ref{alg:heurSolverBrm}
approximates BRM from below.  The only difference between these
two algorithms is that Algorithm~\ref{alg:heurSolver} uses
Equation~\eqref{eq:pSquaredPayment} to determine payments, whereas
this new heuristic uses Equation~\eqref{eq:hSquaredPayment}.

\begin{algorithm}
\caption{Bayesian Heuristic Lower Bound Maximizer}
\label{alg:heurSolverBrm}
\begin{algorithmic}[1]
\ForAll{$\valvec \in \valspace$}
\ForAll{$i \in \setofbidders$}
\State $\virvecd[i] \gets \vvald[i] (\val[i])$ \Comment{Calculate virtual values using Equation~\eqref{eq:discvvf}}
\EndFor
\State $\allocvec (\valvec) \gets \Call{eqp\_solver}{\virvecd}$
\EndFor
\State Calculate the interim allocation rule $\hatallocvec$
\State Calculate the payment rule $\mathbf{h}$ using Equation~\eqref{eq:hSquaredPayment}.
\State $R \gets \Exp_{\valvec} \left[ \sum_{i=1}^{\numbidders} h_i ( \val[i] ) \right]$ \Comment{Total expected revenue}
\State \textbf{return} $R$, $\allocvec$, $\mathbf{h}$
\end{algorithmic}
\end{algorithm}


\section{Closed-Form Solutions}

Next we present closed-form solutions to three mathematical programs
of interest.  The first problem we study is the ex-ante relaxation of
BRM, which requires only polynomially-many constraints.  We present an
intuitive, closed-form solution to a relaxation of this relaxation,
which yields a closed-form upper bound on the ex-ante
Bayesian problem.

This solution also upper bounds RRM.  Nonetheless, we derive a tighter
upper bound in closed form.  Specifically, the second two programs we
discuss in this section optimize the upper and heuristic lower bounds
we derived for RRM, subject to the usual ex-post feasibility
constraints.  While both these programs can both be solved greedily
(Algorithm~\ref{alg:equimarginalSolver}), we derive closed-form
solutions, assuming the payment function is differentiable.

In searching for a near-optimal auction, all three of these problems
optimize allocation variables.  Interestingly and intuitively, all our
closed-form solutions allocate in proportion to value (or virtual
value).  Consequently, the resulting allocation rules are all
monotone,\footnote{assuming regularity} which means that they support
Myerson payments, and yield approximately-optimal auctions.

\subsection{A Relaxation of the Ex-Ante Relaxation of BRM}
\label{sec:ex-ante}

Recall Theorem~\ref{thm:payment-lower-bound}:
Expected payments, when bidders have quasi-linear utility functions as described 
by Equation~\eqref{eq:util_with_q} and
$\paymentq[i] ( \payment[i] ( \val[i], \valvec[-i] ) ) = ( \payment[i] ( \val[i], \valvec[-i] ) )^2$,
can be lower-bounded as follows:
\begin{equation}
\label{eq:expPayLb1}
\Exp_{\val[i] \sim F_i} \left[ \payment[i]( \val[i], \valvec[-i] ) \right]
\ge  \Exp_{\val[i] \sim F_i} \left[ \frac{\vvald[i] (\val[i]) \, \alloc[i](\val[i], \valvec[-i])}{\payment[i] ( \val[i], \valvec[-i])} \right]
.\end{equation}
Following similar logic, and taking expectations throughout,
yields a similar lower bound:
\begin{equation}
\label{eq:expPayLbHat}
\Exp_{\val[i] \sim F_{i}} \left[ \hatpayment[i] (\val[i]) \right]
\ge 
\Exp_{\val[i] \sim F_{i}} \left[ \frac{\vvald[i] (\val[i]) \, \hatalloc[i] (\val[i])}{\hatpayment[i] (\val[i])} \right]%
.\end{equation}
Like Equation~\eqref{eq:expPayLb1}, Equation~\eqref{eq:expPayLbHat}
is a claim about expectations.
Still, let us make the stronger assumption that for all $\val[i] \in \valspace[i]$,
\begin{equation}
\hatpayment[i] (\val[i]) \ge \frac{\vvald[i] (\val[i]) \, \hatalloc[i] (\val[i])}{\hatpayment[i] (\val[i])} %
;\end{equation}
equivalently,
$\left( \hatpayment[i] (\val[i]) \right)^2 \ge \vvald[i] (\val[i]) \, \hatalloc[i] (\val[i])$.
Now, letting $\vvald[i]^{+} (\val[i]) = \max \{ \vvald[i] (\val[i]), 0 \}$,
it also holds that $\left( \hatpayment[i] (\val[i]) \right)^2
\ge \vvald[i]^{+} (\val[i]) \, \hatalloc[i] (\val[i])$;
equivalently, $\hatpayment[i] (\val[i]) \ge \sqrt{\vvald[i]^{+} (\val[i]) \, \hatalloc[i] (\val[i])}$.
Taking expectations and then summing over all bidders yields:
\begin{equation}
\label{eq:xaRelLb}
\sum_{i=1}^{\numbidders}   \Exp_{\val[i]} \left[ \hatpayment[i] (\val[i]) \right]
\ge \sum_{i=1}^{\numbidders} \Exp_{\val[i]} \left[ \sqrt{\vvald[i]^{+} (\val[i]) \, \hatalloc[i] (\val[i])} \right]
,\end{equation}
which gives us a new \emph{heuristic\/} objective of maximizing the
right-hand side of Equation~\eqref{eq:xaRelLb}. 
Note that this new objective \emph{cannot\/} be maximized in a pointwise fashion,
because the various value vectors $\valvec$ cannot be treated
independently in an ex-ante problem.  On the contrary, all the various allocations
$\alloc[i] (\val[i], \valvec[-i])$ interact through the one ex-ante feasibility constraint.

Although we cannot maximize pointwise, we can still approximate a
solution to the ex-ante relaxation by solving a mathematical program
with this heuristic objective function and the relevant feasibility
constraints.
We go one step further and propose the following heuristic program as
a means of approximating an optimal allocation:
%
\begin{align}
\label{rel:prog-1}
\max_{\hatallocvec \ge 0} \,
& \sum_{i=1}^{\numbidders} \sum_{k=1}^{\valspacesize[i]} f_{i,k} \sqrt{\vvald[i,k]^+ \, \hatalloc[i] (\valz[i,k])} \\
\text{subject to} \,
& \sum_{i=1}^{\numbidders} \sum_{k=1}^{\valspacesize[i]} f_{i,k} \, \hatalloc[i] (\valz[i,k]) \le 1 
.\end{align}
This heuristic program ensures only ex-ante feasibility, but drops the
constraints that the interim allocation variables lie in the range
$[0, 1]$ (and indeed, they need not in the ensuing solution).
Interestingly, we can solve this program in closed form.

\begin{theorem}
\label{thm:ex-ante}
The optimal solution to this heuristic program (call it Program $A$)
is to allocate in proportion to virtual values:
\begin{equation}
\hatalloc[i](\valz[i,k]) = 
\frac{\vvald[i,k]^+}{\sum_{i=1}^{\numbidders} \Exp_{z_{i,j} \sim F_i} \left[ \vvald[i,j]^+ \right]} %
.\end{equation}
Under the regularity assumption, this proportional allocation is monotone, 
so can be supported by Bayesian (i.e., BIC and BIR) payments.
\end{theorem}

\begin{proof}
Let $y_{i,k} = \sqrt{\hatalloc[i] (\valz[i,k])}$, and rewrite Program $A$ as follows: 
\begin{align}
\max_{y} \,
& \sum_{i=1}^{\numbidders} \sum_{k=1}^{\valspacesize[i]} f_{i,k} \sqrt{\vvald[i,k]^+} \, y_{i,k} \label{rel:prog} \\
\text{subject to} \,
& \sum_{i}^{\numbidders} \sum_{k=1}^{\valspacesize[i]} f_{i,k} \, y_{i,k}^2 \leq 1 \label{constr:upper}
&& \\
& y_{i,k} \geq 0,
&& \forall i \in \setofbidders, k \in \{1, \dots, \valspacesize[i]\} %
.\end{align}
Call this new program Program $B$.

Now consider the Lagrangian of Program $B$,
dropping the positivity constraints as they are redundant:
\begin{equation}
L(\hatalloc, \mu) = 
\sum_{i=1}^{\numbidders} \sum_{k=1}^{\valspacesize[i]} 
f_{i,k} \sqrt{\vvald[i,k]^+} \, y_{i,k} + 
\mu \left( 1 - \sum_{i=1}^{\numbidders} \sum_{k=1}^{\valspacesize[i]} f_{i,k} \, y_{i,k}^2 \right) %
.\end{equation}
By the Karush-Kuhn-Tucker conditions, 
the partial derivative of the Lagrangian 
with respect to each $y_{i,k}$ must equal $0$:
\begin{equation}
\label{eq:prop:sol}
f_{i,k} \sqrt{\vvald[i,k]^+} - \mu \left( 2 \, f_{i,k} \, y_{i,k} \right) = 0 
.\end{equation}
In other words, $y_{i,k} = \sqrt{\vvald[i,k]}/{2 \mu}$.

At any optimal solution, 
Constraint~\eqref{constr:upper} will be tight, 
since otherwise we could increase the objective value by increasing some 
$y_{i,j}$ for which $f_{i,j} \sqrt{\vvald[i,j]} > 0$ by an infinitesimal amount.  
So:
\begin{equation}
\label{eq:tight}
\sum_{i=1}^{\numbidders} \sum_{j=1}^{\valspacesize[i]} f_{i,j} \, y_{i,j}^2 = 1 %
.\end{equation}
Replacing $y_{i,j}$ with $\sqrt{\vvald[i,j]^+} / 2 \mu$ in Equation~\eqref{eq:tight} yields:
\begin{equation}
\sum_{i=1}^{\numbidders} \sum_{j=1}^{\valspacesize[i]} f_{i,j} \left( \frac{\vvald[i,j]^+}{{(2 \mu)}^2} \right) = 1 
\end{equation}
so that
\begin{equation}
2 \mu = \sqrt{\sum_{i=1}^{\numbidders} \sum_{j=1}^{\valspacesize[i]} f_{i,j} \, \vvald[i,j]^+} %
.\end{equation}
Now substituting $\mu$ into Equation~\eqref{eq:prop:sol}, 
we arrive at a closed-form solution to Program $B$:
\begin{equation}
y_{i,k} = \sqrt{\frac{\vvald[i,k]^+}{\sum_{i=1}^{\numbidders} \sum_{j=1}^{\valspacesize[i]} 
f_{i,j} \, \vvald[i,j]^+}} %
.\end{equation}
Equivalently, the optimal solution to Program $A$ takes the form:
\begin{equation}
\hatalloc[i] (\valz[i,k]) = y_{i,k}^2 = \frac{\vvald[i,k]^+}{\sum_{i=1}^{\numbidders} \sum_{j=1}^{\valspacesize[i]} f_{i,j} \, \vvald[i,j]^+} %
.\end{equation}
Finally, observe that for all bidders $i$,
\begin{equation}
\sum_{j=1}^{\valspacesize[i]} f_{i,j} \, \vvald[i,j]^+ = \Exp_{\valz[i,j] \sim F_j} \left[ \vvald[i,j]^+ \right] %
.\end{equation}
Therefore, the closed-form solution to Program $A$ takes the intuitive form:
\begin{equation}
\hatalloc[i] (\valz[i,k]) = \frac{\vvald[i,k]^+}{\sum_{i=1}^{\numbidders} \Exp_{\valz[i,j] \sim F_i} \left[ \vvald[i,j]^+ \right]} %
.\end{equation}
\end{proof}

\subsection{A Closed-form Upper and Heuristic Lower Bound on RRM}
\label{sec:closedForm}

In this section, we present closed-form solutions to two mathematical
programs that upper and (heuristically) lower bound RRM.  
In the first, the objective function is pseudo-surplus, an upper bound
on revenue.  The second uses as an objective function the heuristic
lower bound we derived for RRM based on virtual values
(Equation~\eqref{eq:p2ExpRevLowerBound}).  
Solving these programs in closed form yields solutions that allocate
in proportion to value.  Such monotonic allocation rules can be
supported via Myerson's payment rule.

\begin{theorem}
\label{thm:rrm_pseudowelfare_alloc_closedform}
The optimal solution to Program $C$ is
\begin{equation}
\alloc[j] (\const[j], \cvec[-j]) 
= \frac{\left(\const[j]^+\right)^{\alpha / (1 - \alpha)}}{\sum_{i=1}^{\numbidders} \left(\const[i]^+\right)^{\alpha / (1 - \alpha)}},
\quad \forall j \in \setofbidders, \forall \cvec \in {\Reals}^n
,\end{equation}
where $\const[i]^+ = \max \{ 0, \const[i] \}$,
whenever there exists at least one positive entry in $\cvec$.
\end{theorem}


\begin{proof}
Bidders whose constants $\const[i]$ are negative are not allocated,
and bidders whose constants $\const[i]$ are zero have no impact, so we
restrict our attention to bidders with positive constants.  That is,
it suffices to allocate assuming we are given $\cvec^+$ instead of
$\cvec$.

The derivative of the contribution of bidder $i$ is
\begin{equation}
\label{eq:pseudoWelfareDerivativeOfBidder}
\frac{\partial \left( \const[i]^+ \right)^{\alpha} \left( \alloc[i] (\const[i], \cvec[-i]) \right)^{\alpha}}
{\partial \alloc[i] (\const[i], \cvec[-i])}
= \frac{1}{\alpha} \left( \const[i]^+ \right)^{\alpha} \left( \alloc[i] (\const[i], \cvec[-i]) \right)^{\alpha - 1}
.\end{equation}
Equating derivatives for bidders $i$ and $j$ (as per the equi-marginal principle), we get
\begin{equation}
\frac{1}{\alpha} \left( \const[i]^+ \right)^{\alpha} \left( \alloc[i] (\const[i], \cvec[-i]) \right)^{\alpha - 1}
= \frac{1}{\alpha} \left( \const[j]^+ \right)^{\alpha} \left( \alloc[j] (\const[j], \cvec[-j]) \right)^{\alpha - 1}
.\end{equation}
The common terms can be removed, and after raising the expressions to the $1 / (1 - \alpha)$ power, 
we can simplify to
\begin{equation}
\frac{ \left( \const[i]^+ \right)^{\alpha / (1 - \alpha)} }{\alloc[i] (\const[i], \cvec[-i])}
= \frac{ \left( \const[i]^+ \right)^{\alpha / (1 - \alpha)} }{\alloc[j] (\const[j], \cvec[-j])}
.\end{equation}
Therefore, bidder $i$'s allocation in terms of bidder $j$ is
\begin{equation}
\alloc[i] (\const[i], \cvec[-i]) = \left( \frac{\const[i]^+}{\const[j]^+} \right)^{\alpha / (1 - \alpha)} \alloc[j] (\const[j], \cvec[-j])
.\end{equation}

Plugging this expression into the ex-post feasibility condition yields:
\begin{equation}
\sum_{i=1}^{\numbidders} \alloc[i] (\const[i], \cvec[-i])
= \sum_{i=1}^{\numbidders} \left( \frac{\const[i]^+}{\const[j]^+} \right)^{\alpha / (1 - \alpha)} \alloc[j] (\const[j], \cvec[-j])
.\end{equation}
It is always optimal to allocate until 
$\sum_{i=1}^{\numbidders} \alloc[i] (\const[i], \cvec[-i]) = 1$
when there is at least one bidder with a positive constant,
so the ex-post feasibility constraint can be written as
\begin{equation}
\sum_{i=1}^{\numbidders} \left( \frac{\const[i]^+}{\const[j]^+} \right)^{\alpha / (1 - \alpha)} \alloc[j] (\const[j], \cvec[-j])
= 1
.\end{equation}
Therefore, bidder $j$ is allocated as follows:
\begin{equation}
\alloc[j] (\const[j], \cvec[-j]) = \frac{\left(\const[j]^+\right)^{\alpha / (1 - \alpha)}}{\sum_{i=1}^{\numbidders} \left(\const[i]^+\right)^{\alpha / (1 - \alpha)}}
.\end{equation}
\end{proof}

The following corollary is immediate:

\begin{corollary}
\label{cor:rrm_pseudowelfare_alloc_closedform}
The optimal solution to Program $C$ when $\alpha = 1/2$ is
\begin{equation}
\alloc[j] (\const[j], \cvec[-j]) = \frac{\const[j]^+}{\sum_{i=1}^{\numbidders} \const[i]^+},
\quad \forall j \in \setofbidders, \forall \cvec \in \valspace
,\end{equation}
where $\const[i]^+ = \max \{ 0, \const[i] \}$,
whenever there exists at least one positive entry in $\cvec$.
\end{corollary}

When the constants $\cvec$ are values, the objective function in
program $C$ is pseudo-surplus.  When the constants $\cvec$ are virtual
values, the objective function in program $C$ is our heuristic lower
bound.  Consequently,
Theorem~\ref{thm:rrm_pseudowelfare_alloc_closedform} immediately
gives rise to closed-form solutions to these two mathematical
programs of interest.

\begin{corollary}
\label{cor:rrm_ub_alloc_closedform}
The following allocation is optimal when $\alpha = 1/2$ and $\cvec = \valvec$
(i.e., when the objective function is pseudo-surplus):
\begin{equation}
\alloc[j] (\val[j], \valvec[-j]) = \frac{\val[j]^+}{\sum_{i=1}^{\numbidders} \val[i]^+},
\quad \forall j \in \setofbidders, \forall \valvec \in \valspace
,\end{equation}
where $\val[i]^+ = \max \{ 0, \val[i] \}$,
when there exists at least one positive entry in $\valvec$.
\end{corollary}

\begin{corollary}
\label{cor:rrm_lb_alloc_closedform}
The following allocation is optimal when $\alpha = 1/2$ and $\cvec = \virvecd$
(i.e., when the objective function is our heuristic lower bound):
\begin{equation}
\alloc[j] (\val[j], \valvec[-j]) = \frac{\vvald[j]^+ (\val[j])}{\sum_{i=1}^{\numbidders} \vvald[i]^+(\val[i])},
\quad \forall j \in \setofbidders, \forall \valvec \in \valspace
,\end{equation}
where $\vvald[i]^+ (\val[i]) = \max \{ 0, \vvald[i] (\val[i]) \}$,
when there exists at least one positive entry in $\virvecd$.
\end{corollary}

Corollaries~\ref{cor:rrm_ub_alloc_closedform}
and~\ref{cor:rrm_lb_alloc_closedform} gives rise to respective
variants of Algorithm~\ref{alg:pseudo-surplus},
Algorithm~\ref{alg:heurSolver}, and Algorithm~\ref{alg:heurSolverBrm}:
in each algorithm, simply replace the call to the equi-marginal
principle solver with the corresponding closed form.  Like their
counterparts, these heuristics, which employ closed-form solutions
instead of greedy ones, and then plug the resulting allocation rules
into Myerson's payment formula, approximate optimal auctions.

\begin{algorithm}[h!]
\caption{Pseudo-Surplus Maximizer (Closed-form)}
\label{alg:pseudo-surplusCf}
\begin{algorithmic}[1]
\ForAll{$\valvec \in \valspace$}
\State Calculate the allocation rule $\allocvec (\valvec)$ using Corollary~\ref{cor:rrm_ub_alloc_closedform}
\EndFor
\State $W \gets \Exp_{\valvec} \left[ \sum_{i=1}^{\numbidders} \sqrt{\val[i] \alloc[i] ( \val[i], \valvec[-i] )} \right]$ \Comment{Total expected pseudo-surplus}
\State Calculate the payment rule $\paymentvec$ using Equation~\eqref{eq:pSquaredPayment}.
\State \textbf{return} $W$, $\allocvec$, $\paymentvec$
\end{algorithmic}
\end{algorithm}

\begin{algorithm}[h!]
\caption{Heuristic Lower Bound Maximizer (Closed-form)}
\label{alg:heurSolverCf}
\begin{algorithmic}[1]
\ForAll{$\valvec \in \valspace$}
\ForAll{$i \in \setofbidders$}
\State $\virvecd[i] \gets \vvald[i] (\val[i])$ \Comment{Calculate virtual values using Equation~\eqref{eq:discvvf}}
\EndFor
\State Calculate the allocation rule $\allocvec (\valvec)$ using Corollary~\ref{cor:rrm_lb_alloc_closedform}
\EndFor
\State Calculate the payment rule $\paymentvec$ using Equation~\eqref{eq:pSquaredPayment}.
\State $R \gets \Exp_{\valvec} \left[ \sum_{i=1}^{\numbidders} \payment[i] ( \val[i], \valvec[-i] ) \right]$ \Comment{Total expected revenue}
\State \textbf{return} $R$, $\allocvec$, $\paymentvec$
\end{algorithmic}
\end{algorithm}

\begin{algorithm}[h!]
\caption{Bayesian Heuristic Lower Bound Maximizer (Closed-form)}
\label{alg:heurSolverBrmCf}
\begin{algorithmic}[1]
\ForAll{$\valvec \in \valspace$}
\ForAll{$i \in \setofbidders$}
\State $\virvecd[i] \gets \vvald[i] (\val[i])$ \Comment{Calculate virtual values using Equation~\eqref{eq:discvvf}}
\EndFor
\State Calculate the allocation rule $\allocvec (\valvec)$ using Corollary~\ref{cor:rrm_lb_alloc_closedform}
\EndFor
\State Calculate the interim allocation rule $\hatallocvec$
\State Calculate the payment rule $\mathbf{h}$ using Equation~\eqref{eq:hSquaredPayment}.
\State $R \gets \Exp_{\valvec} \left[ \sum_{i=1}^{\numbidders} h_i ( \val[i] ) \right]$ \Comment{Total expected revenue}
\State \textbf{return} $R$, $\allocvec$, $\mathbf{h}$
\end{algorithmic}
\end{algorithm}


\begin{remark}
If values are continuous as in Myerson's original paper, 
then Equation~\eqref{eq:qPayment} is
\begin{equation}
\label{eq:qPaymentCont}
\paymentq[i] ( \val[i], \valvec[-i] )
= \val[i] \alloc[i] (\val[i], \valvec[-i]) - \int_{0}^{\val[i]} \alloc[i] (z, \valvec[-i]) \, \mathrm{d}z
,\end{equation}
which, letting $S = \sum_{j \ne i} \val[j]^+$, simplifies to
\begin{align}
\paymentq[i] ( \val[i], \valvec[-i] )
&= \val[i] \left( \frac{\val[i]}{\val[i] + S} \right) - \int_{0}^{\val[i]} \frac{\valz}{\valz + S} \, \mathrm{d}z \\
&= \val[i] \left( \frac{\val[i]}{\val[i] + S} \right) - \left[ \valz - S \ln \left( \valz + S \right) \right]_{0}^{\val[i]} \\
&= \val[i] \left( \frac{\val[i]}{\val[i] + S} - 1 \right) + S \left( \ln \left( \val[i] + S \right) - \ln S \right) \\
&\le S \left( \ln \left( \val[i] + S \right) - \ln S \right) \\
&= S \ln \left( \frac{\val[i] + S}{S} \right) \label{eq:intuition}
.\end{align}
%
%
Since Equation~\eqref{eq:intuition} gives an upper bound that is
proportional to the marginal gain in log-surplus, we expect that
extracting more revenue from bidder $i$ will become more difficult as
bidder $i$'s type increases.  We see this explicitly when computing
derivatives.
Specifically, perceived payment $\paymentq[i] ( \val[i], \valvec[-i] )$ changes by
\begin{align}
\frac{\mathrm{d} \paymentq[i] ( \val[i], \valvec[-i] )}{\mathrm{d} \val[i]}
= \frac{\val[i] S}{\left( \val[i] + S \right)^2}
\propto \frac{1}{\val[i]}
,\end{align}
while actual payment
$\payment[i] ( \val[i], \valvec[-i] ) = \sqrt{\paymentq[i] ( \val[i], \valvec[-i] )}$
changes by
\begin{align}
\frac{\mathrm{d} \payment[i] ( \val[i], \valvec[-i] )}{\mathrm{d} \val[i]}
= \frac{\val[i] S}
{2 \left(\val[i] + S \right)^2 \payment[i] ( \val[i], \valvec[-i] )}
\propto \frac{1}{\val[i] \, \payment[i] ( \val[i], \valvec[-i] )} 
.\end{align}
\end{remark}


\section{Experiments}
\label{sec:experiments}

We close this paper with some experimental results implemented in
MATLAB demonstrating the performance of our methods in different
settings.  As a baseline, we used IBM's ILOG CPLEX Optimization Studio
to solve RRM and BRM optimally.  In comparison, we used our heuristic
procedures to generate heuristic lower bounds on total expected
revenue, for auctions that satisfy all the relevant constraints.  We
also computed pseudo-surplus (ensuring ex-post feasibility only) to
give upper bounds on total expected revenue; but we did not compute
accompanying payments, so we did not ensure IC and IR or BIC and
BIR.  Finally, we also solved the ex-ante relaxation of BRM to obtain
a nearly instantaneous upper bound on solutions to all our problems.

All programs were run on a system with an 
Intel Core i5 3.5 GHz processor and 8 GB of RAM.
We compared total expected revenue and run time for different numbers of bidders.
The number of bidders we used was limited by processing time and memory constraints.

In all simulations, bidders were symmetric, meaning $F_i = F_j$ for all bidders $i$ and $j$.
We studied three different bidder distributions: categorical, uniform, and binomial.



While most of our heuristics are powerful enough to apply to any
convex function (not only $\paymentq[i] = \payment[i]^2$), our
experiments are restricted to this case because, to our knowledge,
CPLEX (for MATLAB) can only handle objective functions and constraints
that are linear, quadratic, or integer.


\paragraph{Categorical Distribution}

\begin{figure}[H]
    \centering
    \begin{subfigure}{.32\textwidth}
        \includegraphics[width=\textwidth]{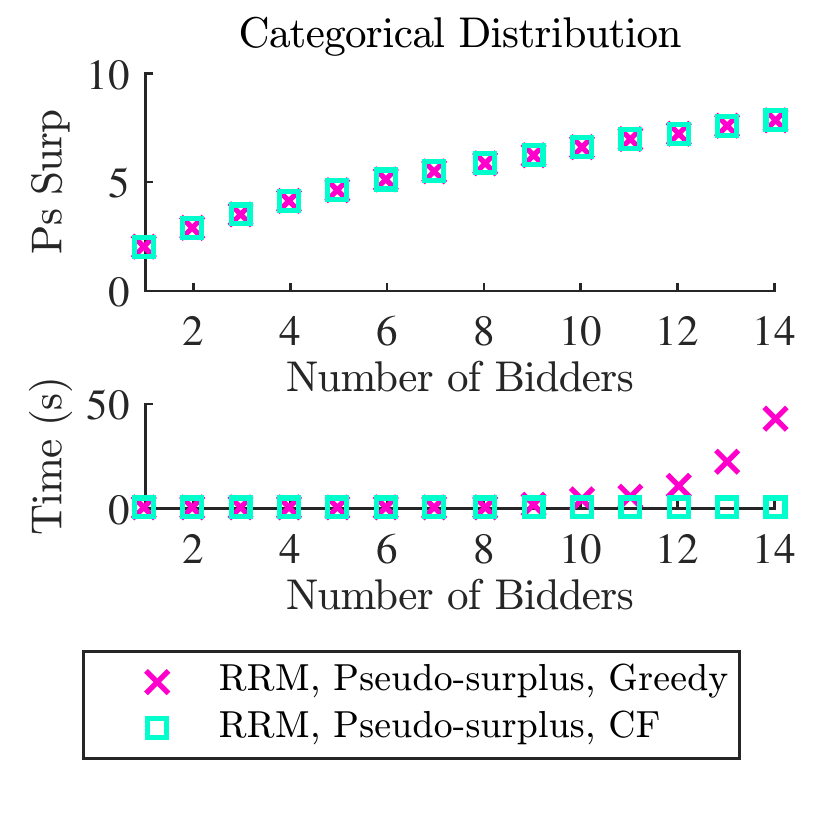}
    \caption{Categorical Distribution}
\label{fig:catPlot_rrm_ub_bef_aft_thm}
    \end{subfigure}
    \begin{subfigure}{.32\textwidth}
    \includegraphics[width=\textwidth]{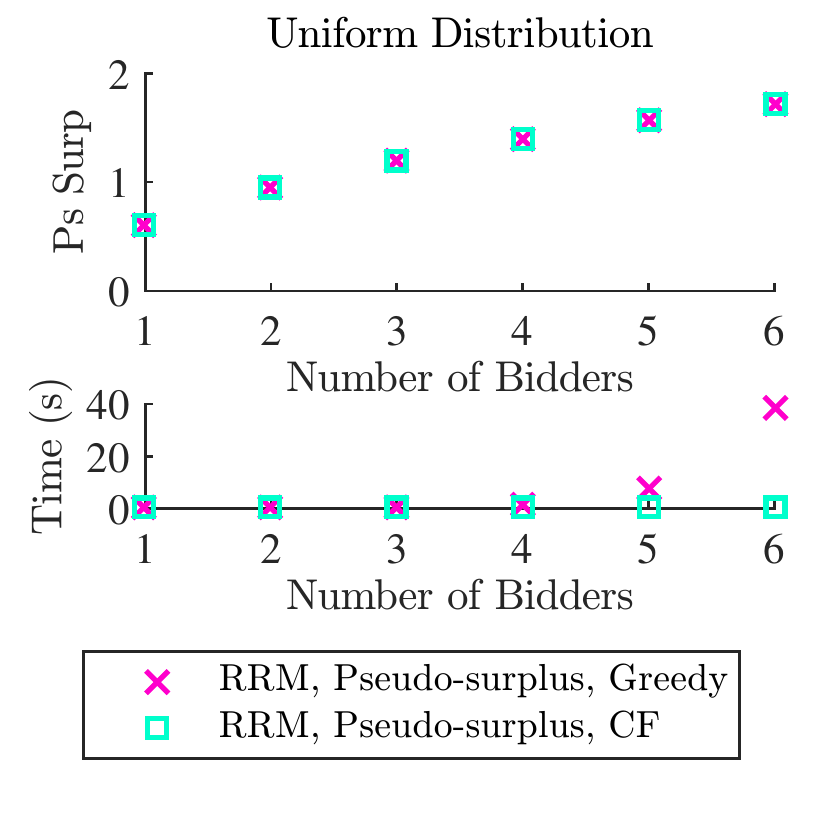}
    \caption{Uniform Distribution}
\label{fig:unifPlot_rrm_ub_bef_aft_thm}
    \end{subfigure}
    \begin{subfigure}{.32\textwidth}
        \includegraphics[width=\textwidth]{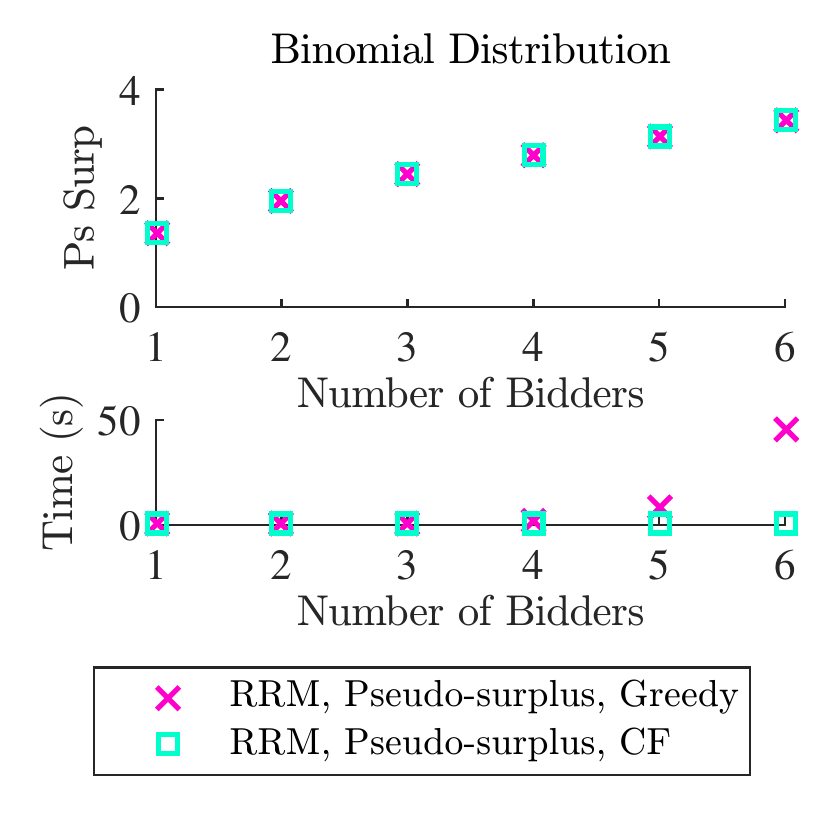}
    \caption{Binomial Distribution}
\label{fig:binomPlot_rrm_ub_bef_aft_thm}
    \end{subfigure}
    \caption{Pseudo-surplus in the Robust Problem}
\end{figure}

\begin{figure}[H]
    \centering
    \begin{subfigure}{.32\textwidth}
        \includegraphics[width=\textwidth]{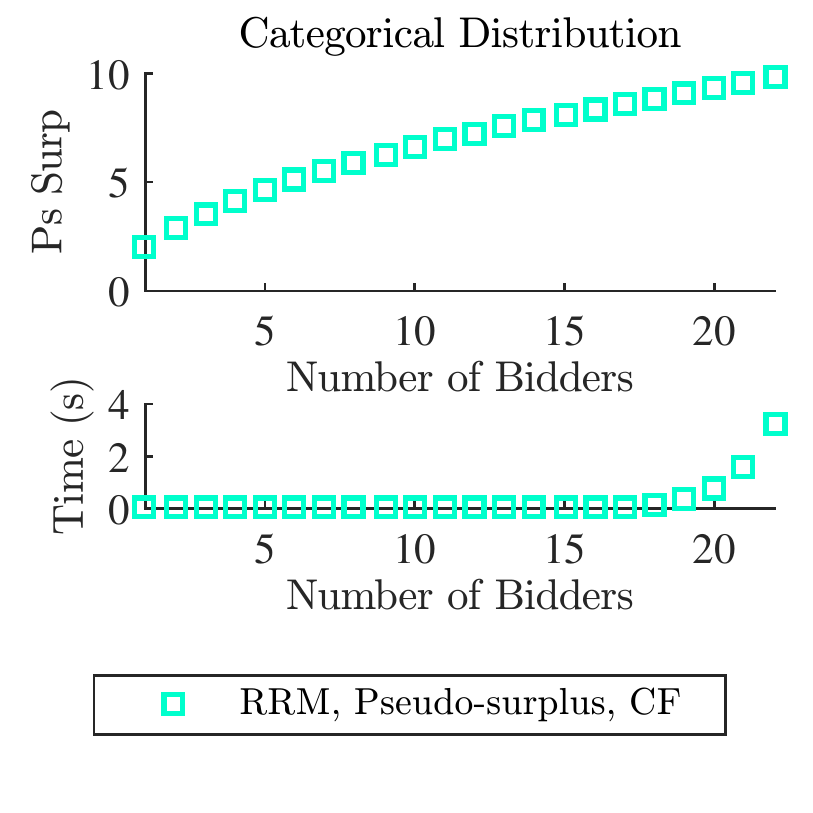}
    \caption{Categorical Distribution}
\label{fig:catPlot_pseudowelfare_scaling}
    \end{subfigure}
    \begin{subfigure}{.32\textwidth}
    \includegraphics[width=\textwidth]{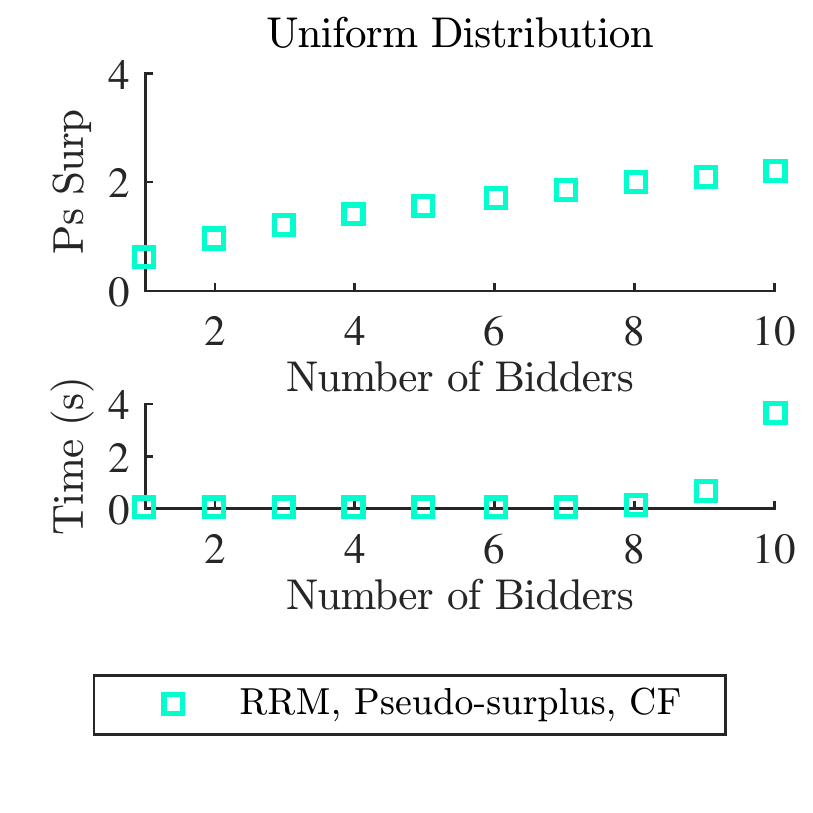}
    \caption{Uniform Distribution}
\label{fig:unifPlot_pseudowelfare_scaling}
    \end{subfigure}
    \begin{subfigure}{.32\textwidth}
        \includegraphics[width=\textwidth]{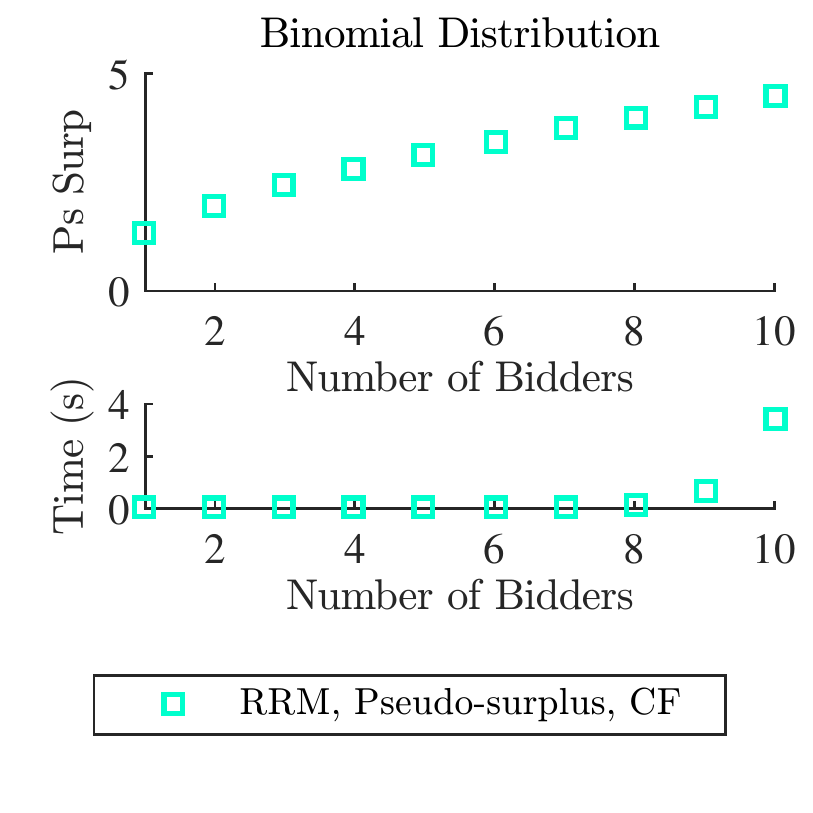}
    \caption{Binomial Distribution}
\label{fig:binomPlot_pseudowelfare_scaling}
    \end{subfigure}
    \caption{Pseudo-surplus Scaling in the Robust Problem}
\end{figure}

\begin{figure}[H]
    \centering
    \begin{subfigure}{.32\textwidth}
        \includegraphics[width=\textwidth]{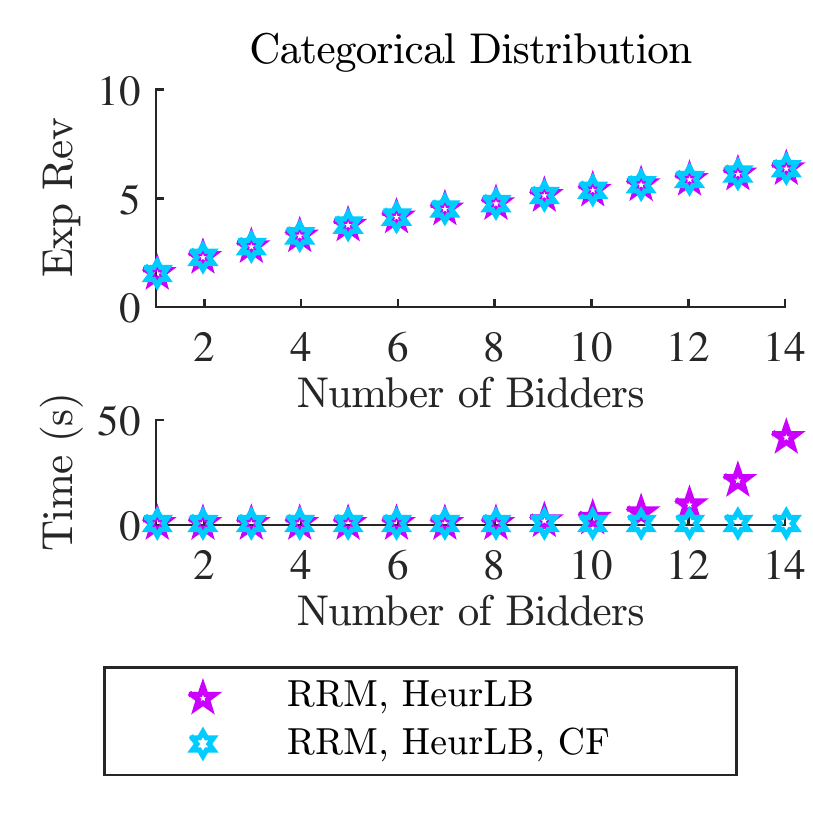}
    \caption{Categorical Distribution}
\label{fig:catPlot_rrm_heurlb_greedy_thm}
    \end{subfigure}
    \begin{subfigure}{.32\textwidth}
    \includegraphics[width=\textwidth]{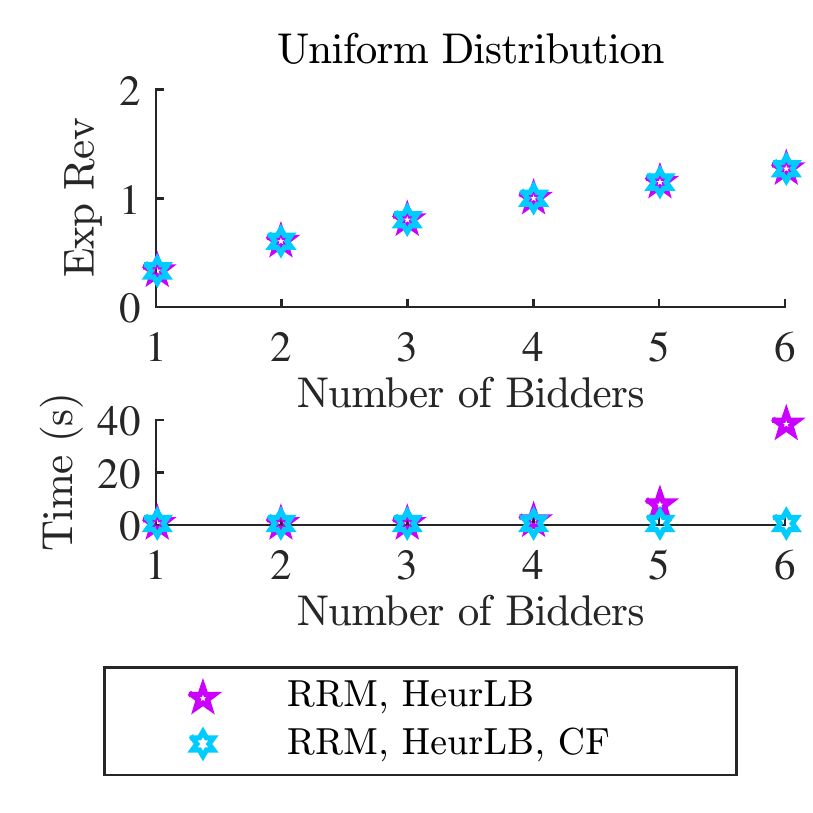}
    \caption{Uniform Distribution}
\label{fig:unifPlot_rrm_heurlb_greedy_thm}
    \end{subfigure}
    \begin{subfigure}{.32\textwidth}
        \includegraphics[width=\textwidth]{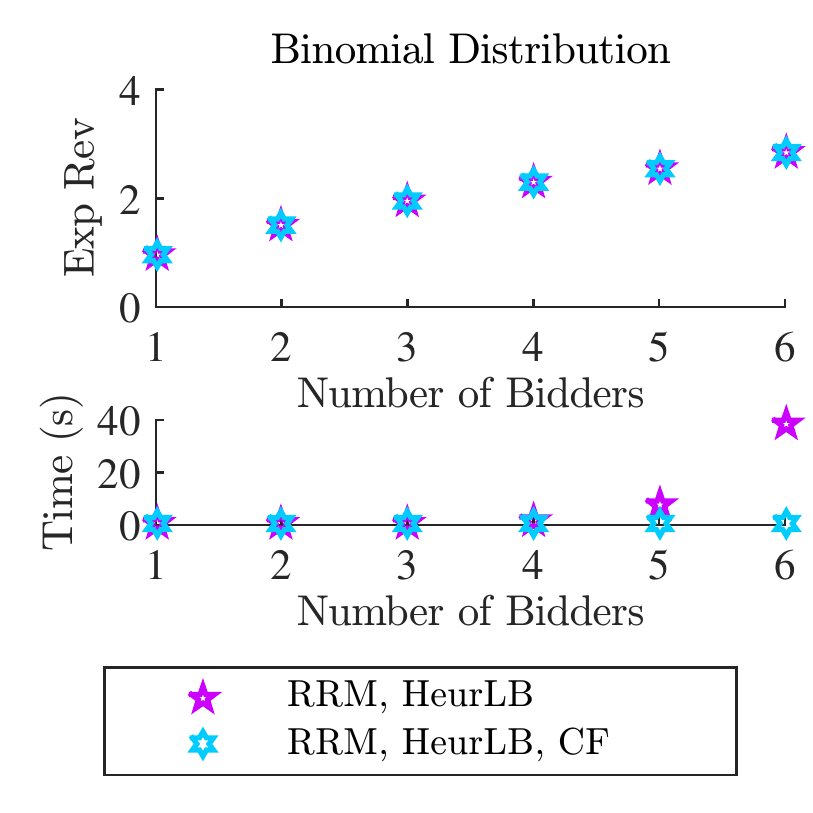}
    \caption{Binomial Distribution}
\label{fig:binomPlot_rrm_heurlb_greedy_thm}
    \end{subfigure}
    \caption{Heuristic Lower Bound in the Robust Problem}
\end{figure}

\begin{figure}[H]
    \centering
    \begin{subfigure}{.32\textwidth}
        \includegraphics[width=\textwidth]{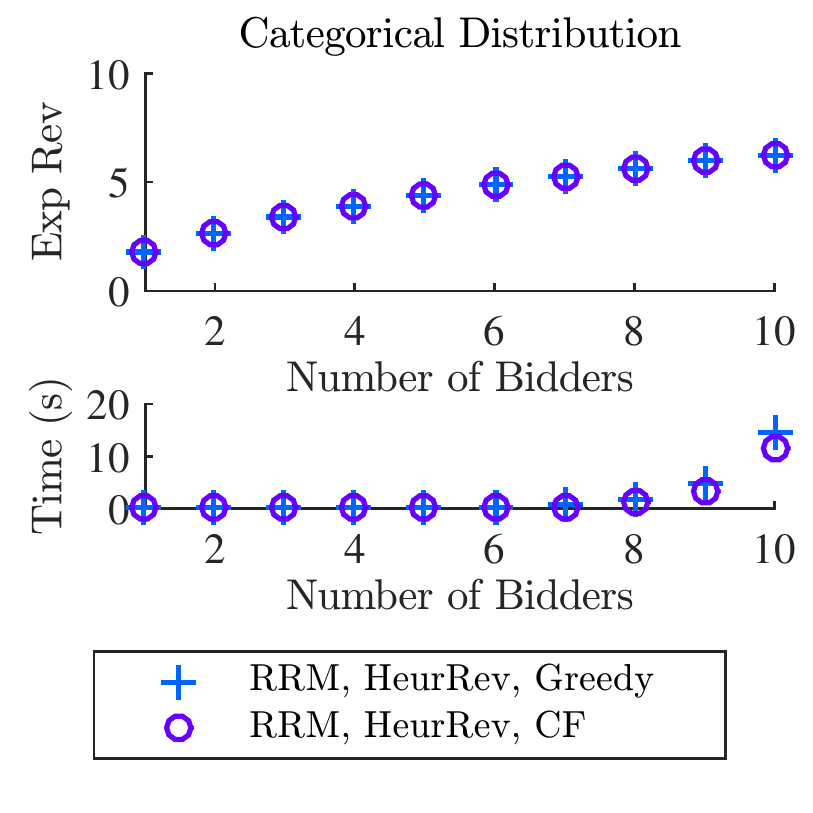}
    \caption{Categorical Distribution}
\label{fig:catPlot_rrm_lb_bef_aft_thm}
    \end{subfigure}
    \begin{subfigure}{.32\textwidth}
    \includegraphics[width=\textwidth]{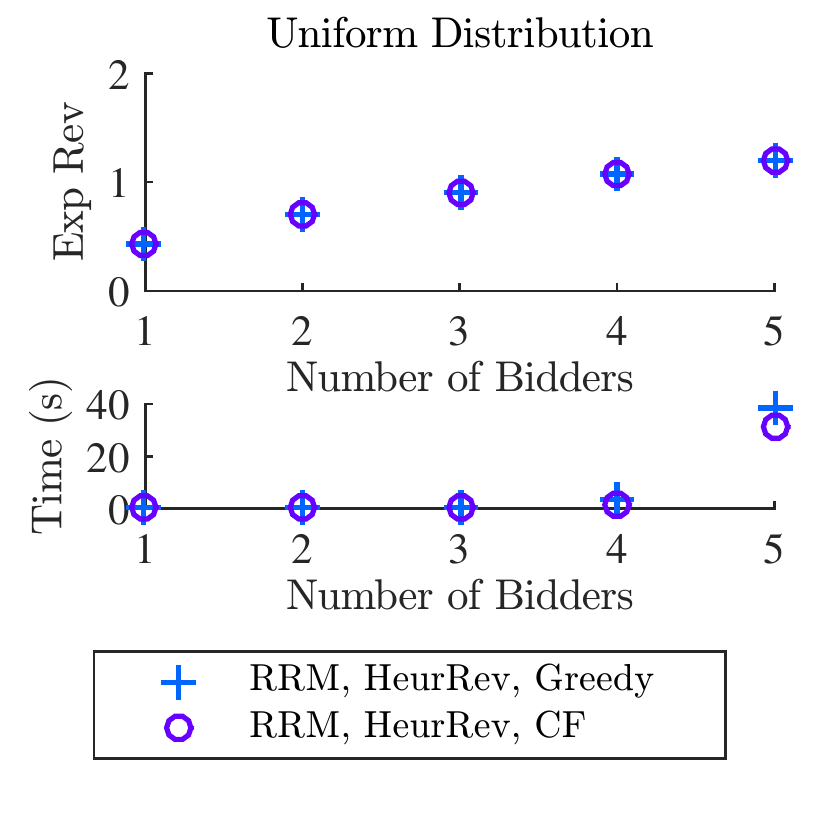}
    \caption{Uniform Distribution}
\label{fig:unifPlot_rrm_lb_bef_aft_thm}
    \end{subfigure}
    \begin{subfigure}{.32\textwidth}
        \includegraphics[width=\textwidth]{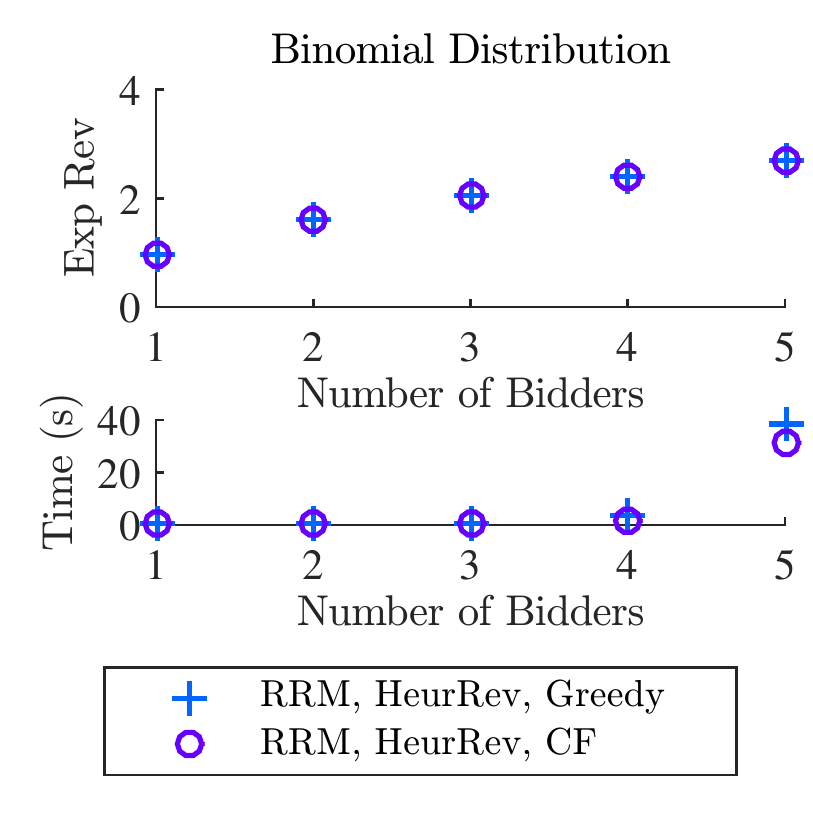}
    \caption{Binomial Distribution}
\label{fig:binomPlot_rrm_lb_bef_aft_thm}
    \end{subfigure}
    \caption{Heuristic Revenue in the Robust Problem}
\end{figure}

\begin{figure}[H]
    \centering
    \begin{subfigure}{.32\textwidth}
        \includegraphics[width=\textwidth]{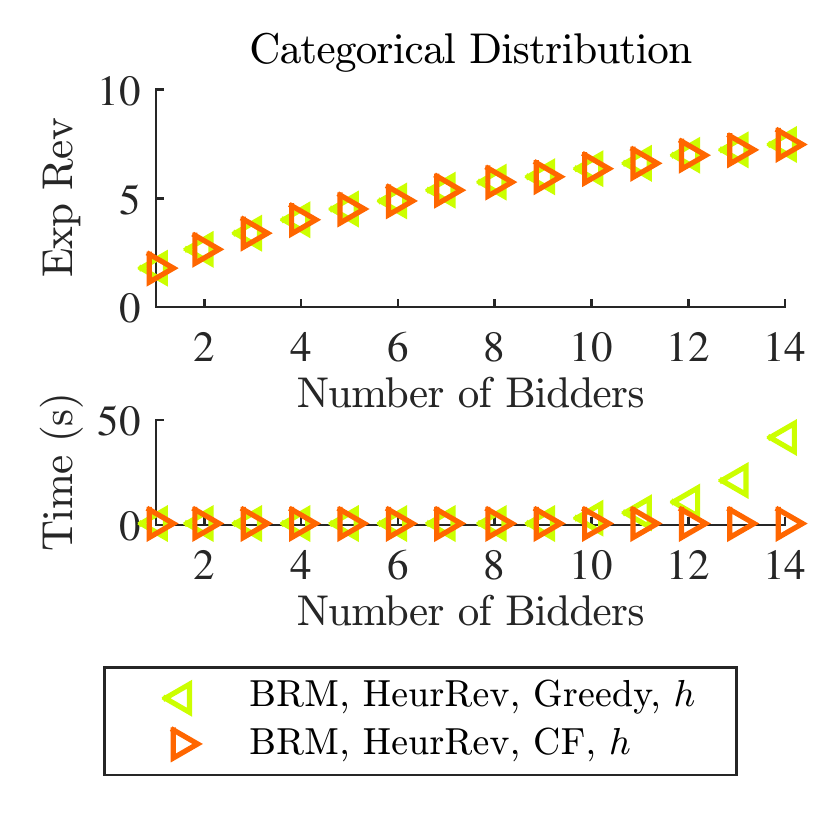}
    \caption{Categorical Distribution}
\label{fig:catPlot_brm_lb_bef_aft_thm}
    \end{subfigure}
    \begin{subfigure}{.32\textwidth}
    \includegraphics[width=\textwidth]{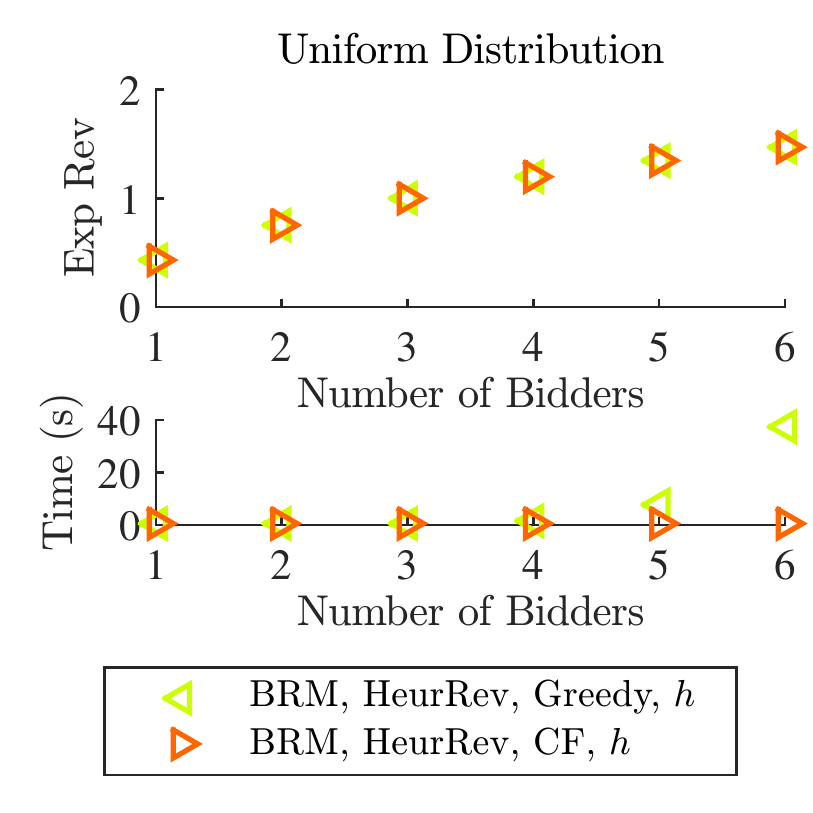}
    \caption{Uniform Distribution}
\label{fig:unifPlot_brm_lb_bef_aft_thm}
    \end{subfigure}
    \begin{subfigure}{.32\textwidth}
        \includegraphics[width=\textwidth]{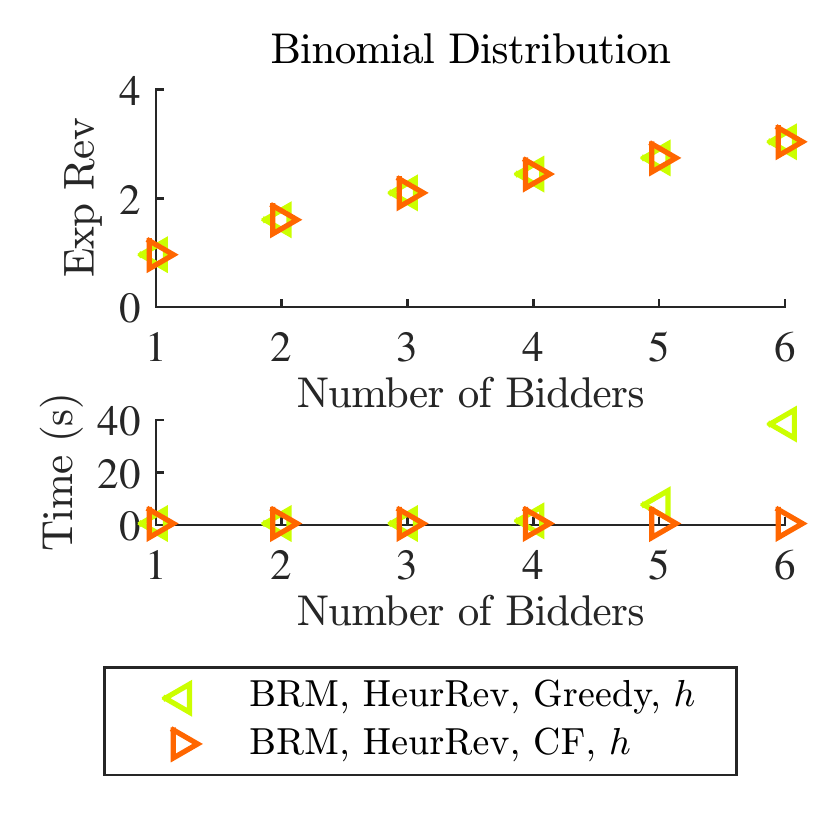}
    \caption{Binomial Distribution}
\label{fig:binomPlot_brm_lb_bef_aft_thm}
    \end{subfigure}
    \caption{Heuristic Revenue in the Bayesian Problem}
\end{figure}

\begin{figure}[H]
    \centering
    \begin{subfigure}{.32\textwidth}
        \includegraphics[width=\textwidth]{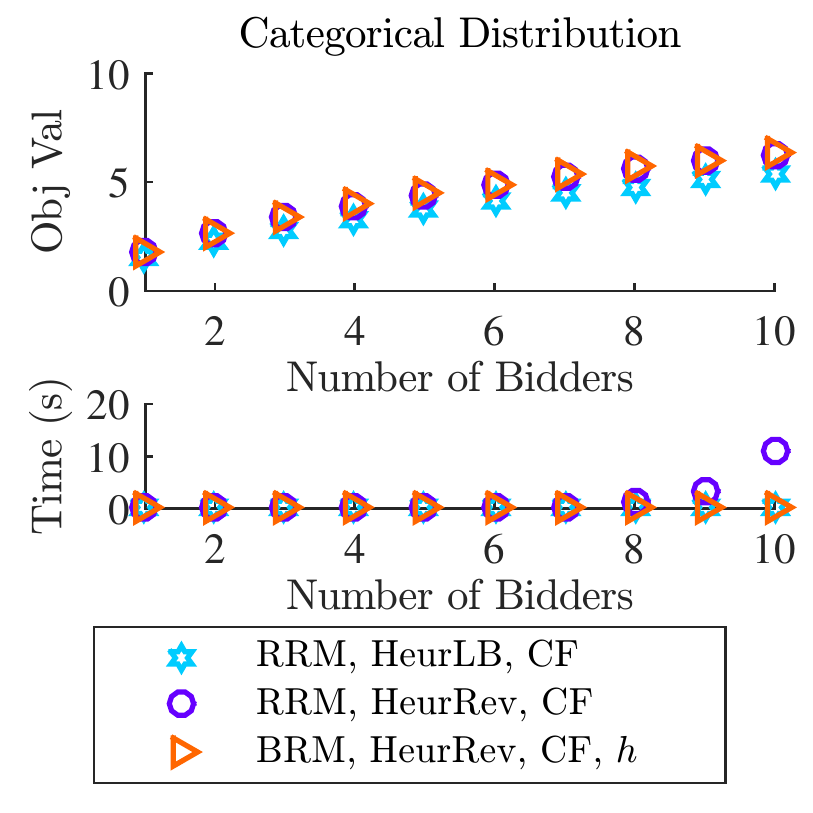}
    \caption{Categorical Distribution}
\label{fig:catPlot_rrm_brm_heur}
    \end{subfigure}
    \begin{subfigure}{.32\textwidth}
    \includegraphics[width=\textwidth]{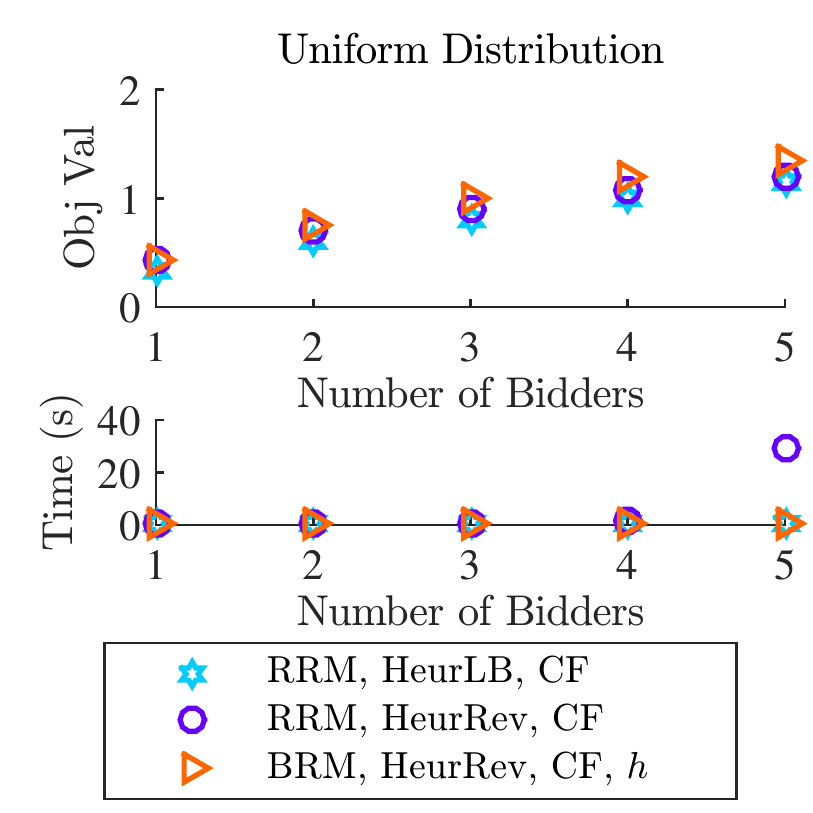}
    \caption{Uniform Distribution}
\label{fig:unifPlot_rrm_brm_heur}
    \end{subfigure}
    \begin{subfigure}{.32\textwidth}
        \includegraphics[width=\textwidth]{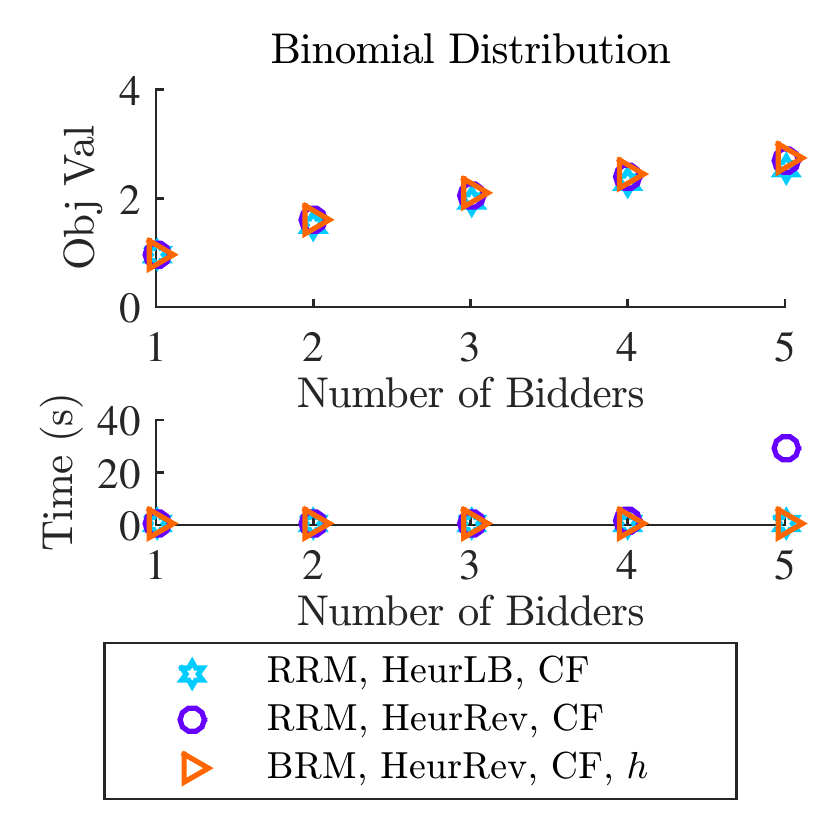}
    \caption{Binomial Distribution}
\label{fig:binomPlot_rrm_brm_heur}
    \end{subfigure}
    \caption{Robust vs.\ Bayesian Heuristic Revenue}
\end{figure}

\begin{figure}[H]
    \centering
    \begin{subfigure}{.32\textwidth}
    \includegraphics[width=\textwidth]{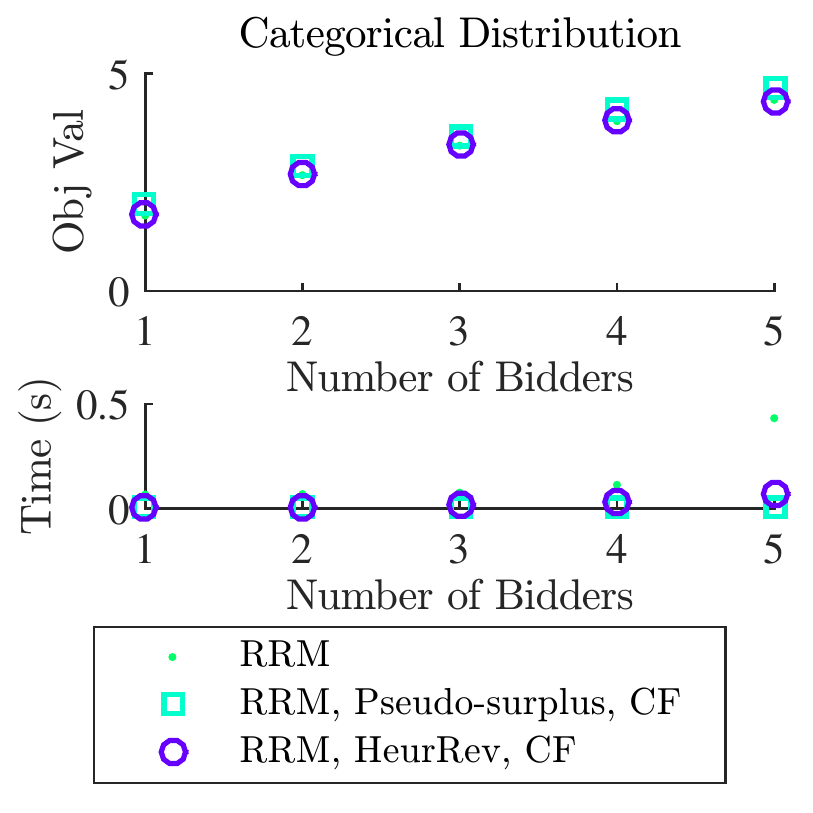}
    \caption{Categorical Distribution}
\label{fig:catPlot_rrm_heur}
    \end{subfigure}
    \begin{subfigure}{.32\textwidth}
    \includegraphics[width=\textwidth]{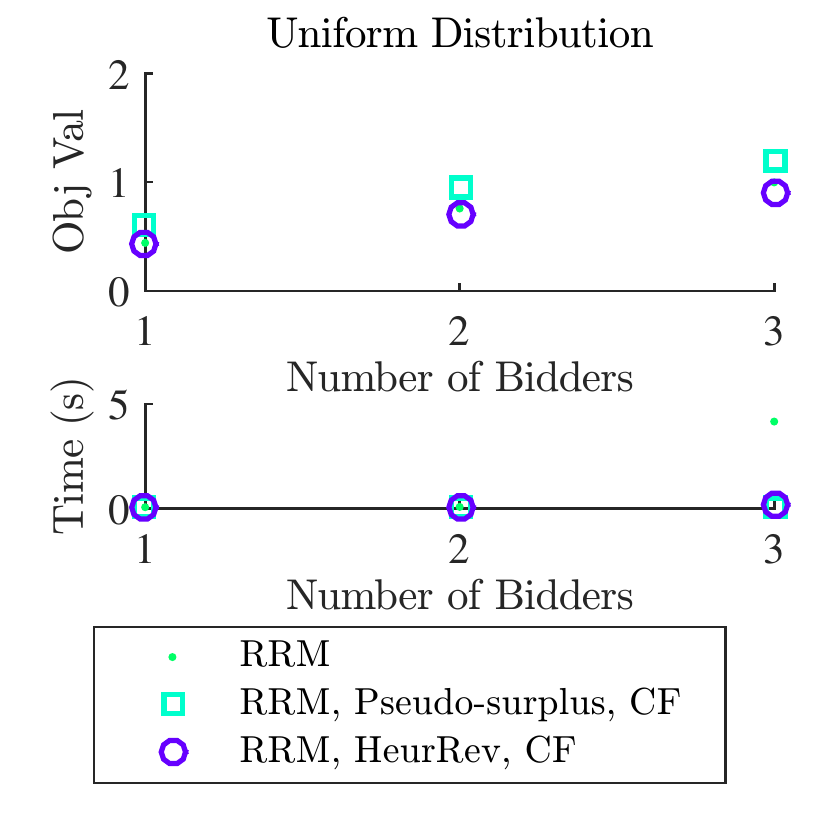}
    \caption{Uniform Distribution}
\label{fig:unifPlot_rrm_heur}
    \end{subfigure}
    \begin{subfigure}{.32\textwidth}
    \includegraphics[width=\textwidth]{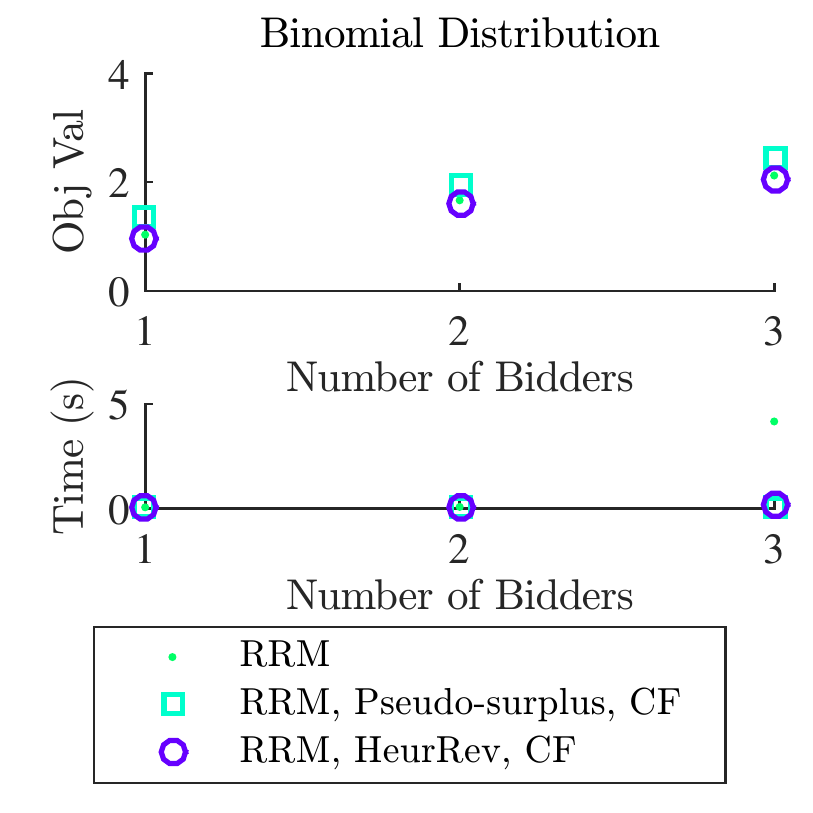}
    \caption{Binomial Distribution}
\label{fig:binomPlot_rrm_heur}
    \end{subfigure}
    \caption{RRM, Pseudo-Surplus, and Heuristic Revenue}
\end{figure}

\begin{figure}[H]
    \centering
    \begin{subfigure}{.32\textwidth}
    \includegraphics[width=\textwidth]{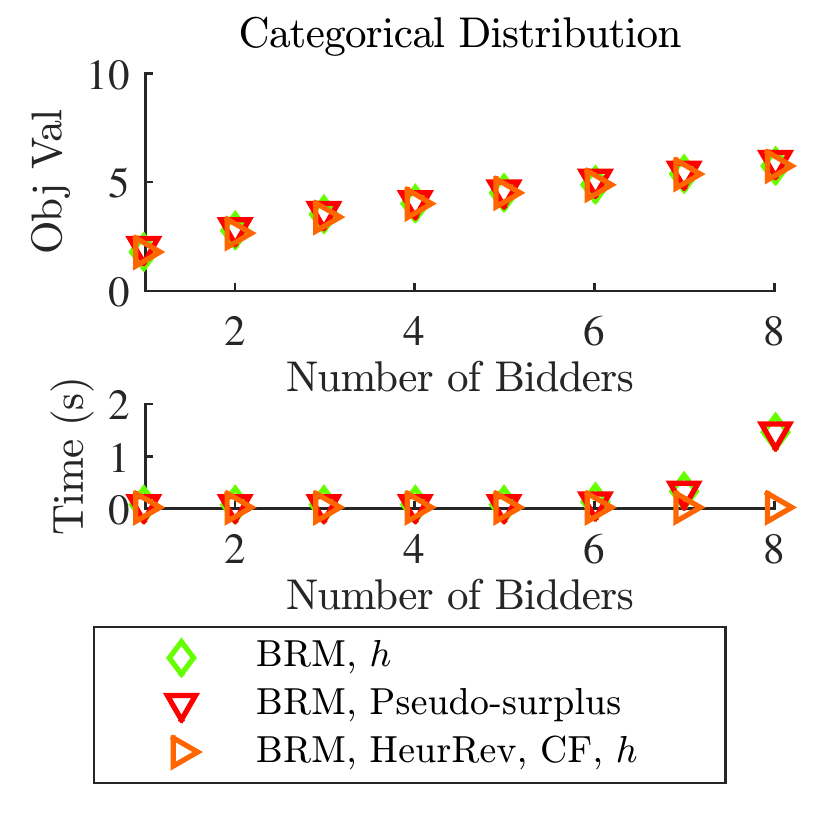}
    \caption{Categorical Distribution}
\label{fig:catPlot_brm_heur}
    \end{subfigure}
    \begin{subfigure}{.32\textwidth}
    \includegraphics[width=\textwidth]{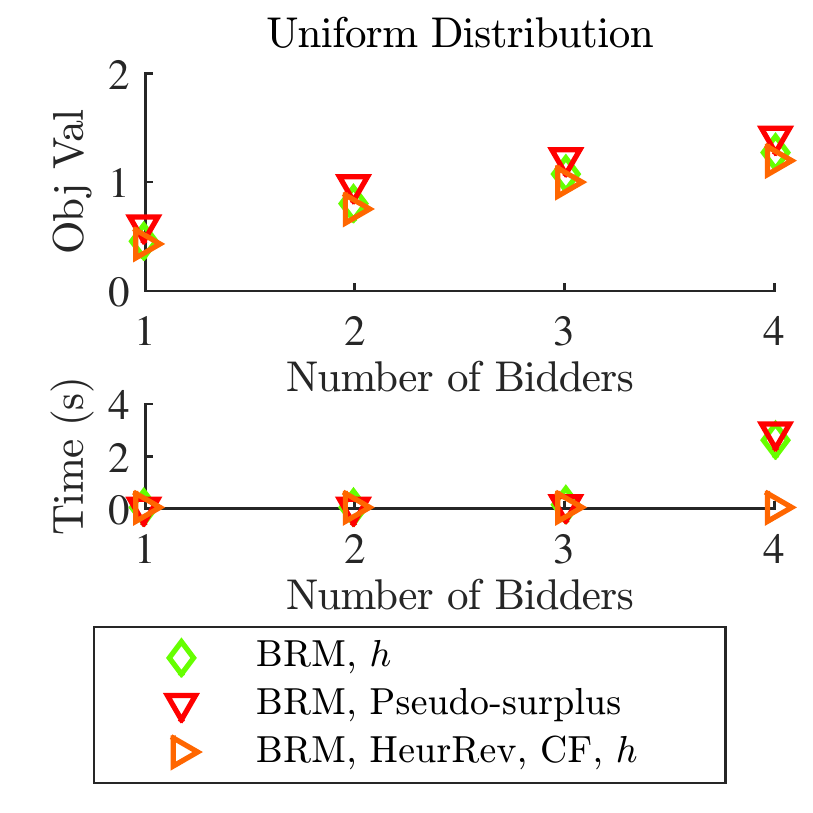}
    \caption{Uniform Distribution}
\label{fig:unifPlot_brm_heur}
    \end{subfigure}
    \begin{subfigure}{.32\textwidth}
    \includegraphics[width=\textwidth]{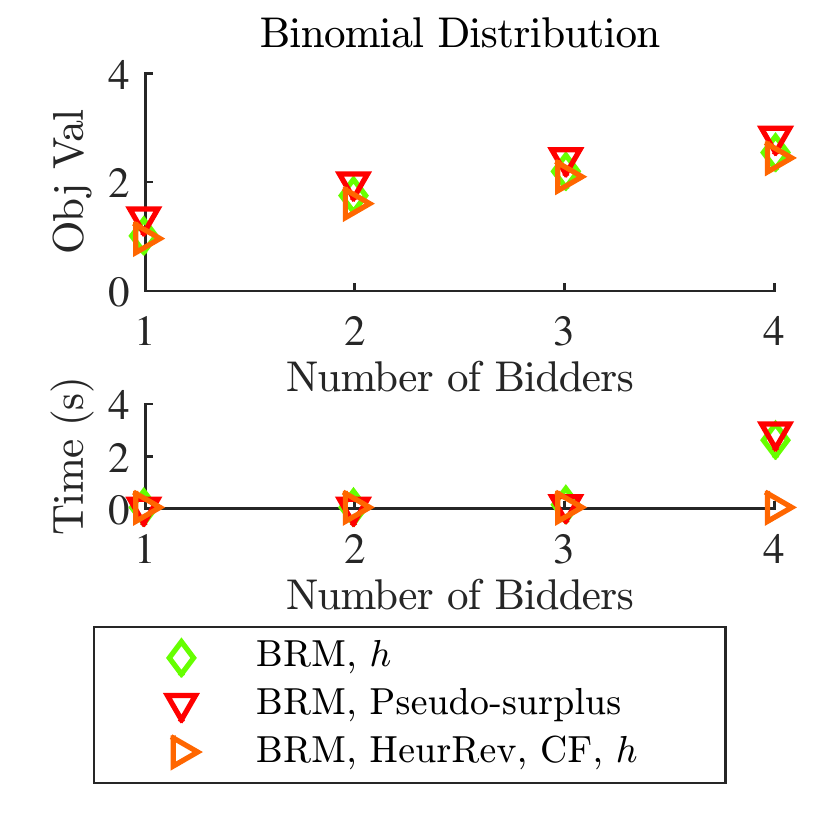}
    \caption{Binomial Distribution}
\label{fig:binomPlot_brm_heur}
    \end{subfigure}
    \caption{BRM, Pseudo-Surplus, and Heuristic Revenue}
\end{figure}

\begin{figure}[H]
    \centering
    \begin{subfigure}{.32\textwidth}
        \includegraphics[width=\textwidth]{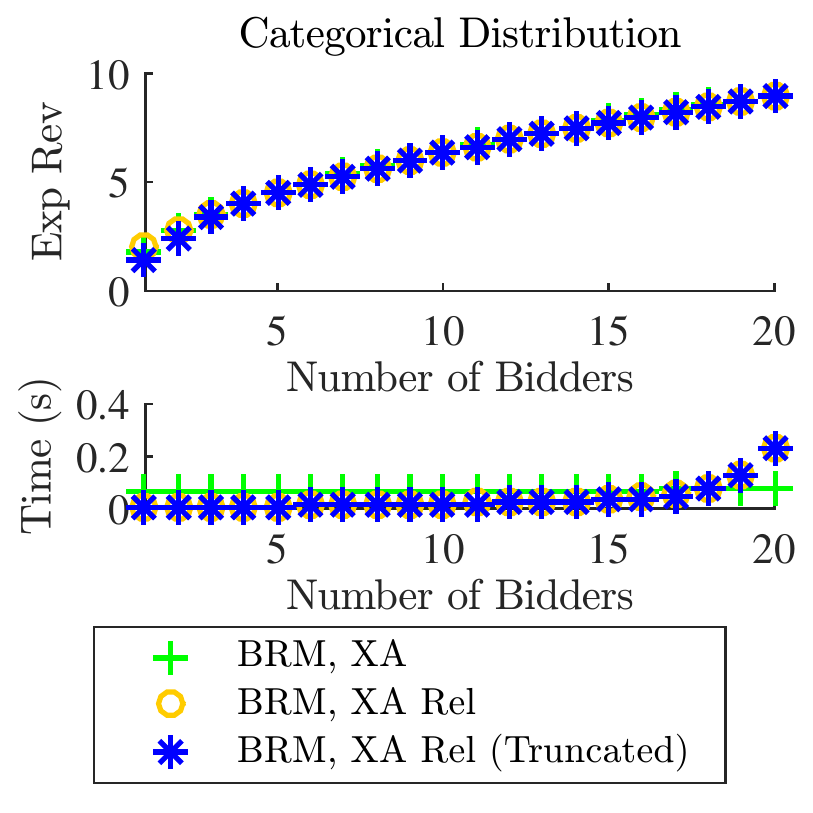}
    \caption{Categorical Distribution}
\label{fig:catPlot_bic_xa_relax}
    \end{subfigure}
    \begin{subfigure}{.32\textwidth}
    \includegraphics[width=\textwidth]{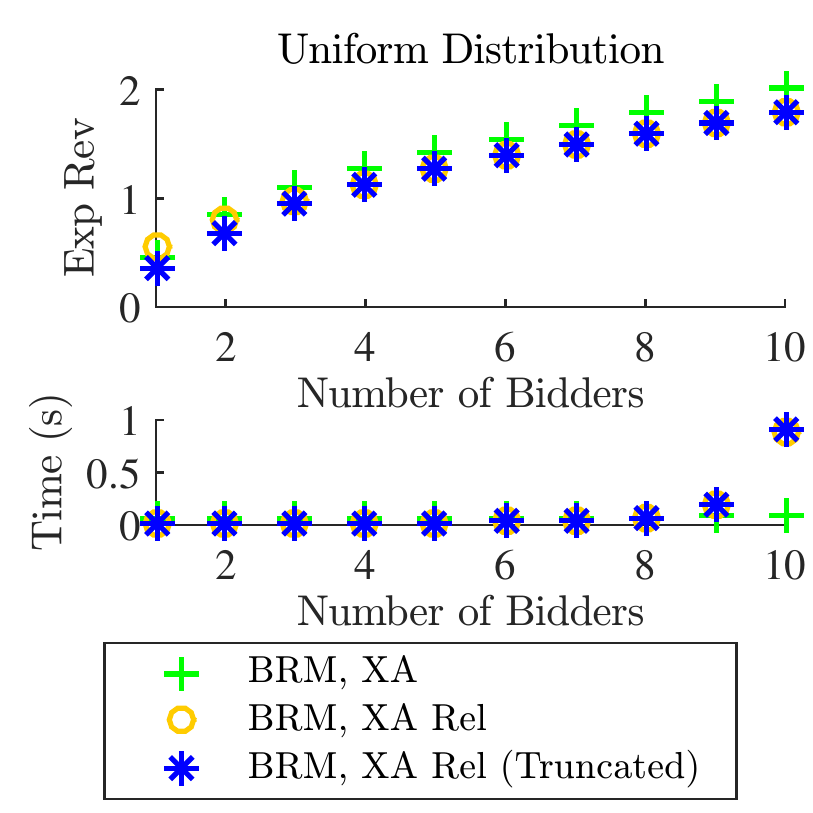}
    \caption{Uniform Distribution}
\label{fig:unifPlot_bic_xa_relax}
    \end{subfigure}
    \begin{subfigure}{.32\textwidth}
    \includegraphics[width=\textwidth]{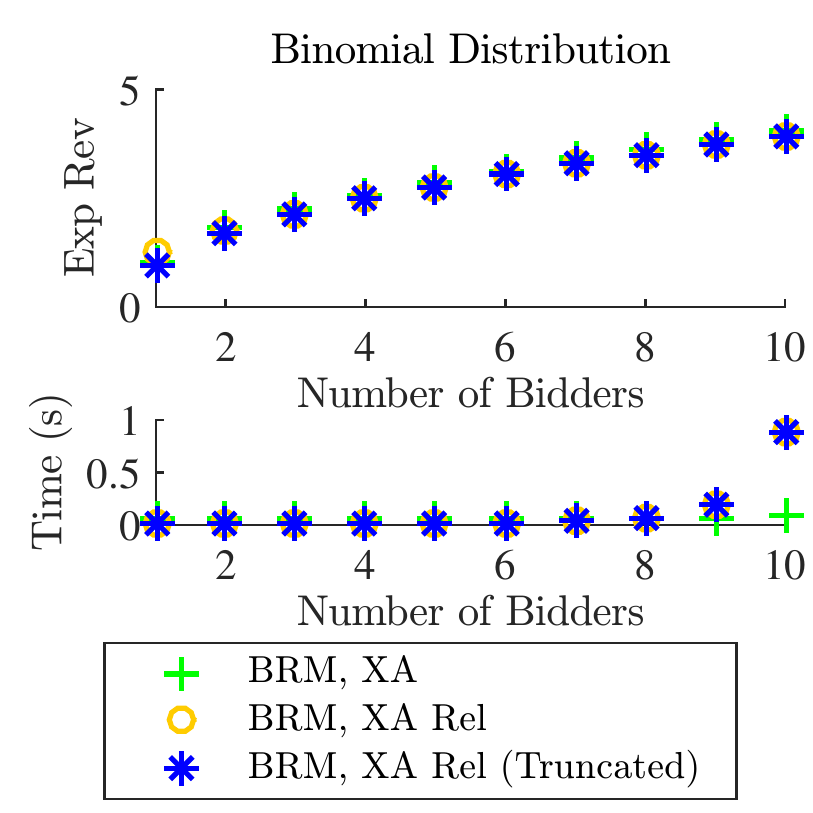}
    \caption{Binomial Distribution}
\label{fig:binomPlot_bic_xa_relax}
    \end{subfigure}
    \caption{BRM Ex-Ante Variants}
\end{figure}

Each bidder $i$ has type $\val[i] \in \valspace[i] = \{ L, H \}$, 
where $L = 3$ and $H = 10$.  
The probability mass function values of each type are 
$f_i ( L ) = 0.8$ and $f_i ( H ) = 0.2$.  
Total expected revenue and run times for this distribution are shown in Figures~%
\ref{fig:catPlot_rrm_ub_bef_aft_thm},
\ref{fig:catPlot_pseudowelfare_scaling},
\ref{fig:catPlot_rrm_heurlb_greedy_thm},
\ref{fig:catPlot_rrm_lb_bef_aft_thm},
\ref{fig:catPlot_brm_lb_bef_aft_thm}, 
\ref{fig:catPlot_rrm_brm_heur}, 
\ref{fig:catPlot_rrm_heur},
\ref{fig:catPlot_brm_heur},
and
\ref{fig:catPlot_bic_xa_relax}.

\paragraph{Uniform Distribution}

Each bidder $i$ has type $\val[i] \in \valspace[i] = \{ 0, 0.25, 0.5, 0.75, 1 \}$, 
where $f_i ( \val[i] ) = 0.2$ for all $\val[i] \in \valspace[i]$.  
Total expected revenue and run times for this distribution are shown in Figures~%
\ref{fig:unifPlot_rrm_ub_bef_aft_thm},
\ref{fig:unifPlot_pseudowelfare_scaling},
\ref{fig:unifPlot_rrm_heurlb_greedy_thm},
\ref{fig:unifPlot_rrm_lb_bef_aft_thm},
\ref{fig:unifPlot_brm_lb_bef_aft_thm}, 
\ref{fig:unifPlot_rrm_brm_heur}, 
\ref{fig:unifPlot_rrm_heur},
\ref{fig:unifPlot_brm_heur},
and
\ref{fig:unifPlot_bic_xa_relax}. 

\paragraph{Binomial Distribution}

Each bidder $i$ has type $\val[i] \in \valspace[i] = \{ k : 0 \le k \le 4, k \in \mathbb{N} \}$, 
where $f_i ( k ) = \binom{4}{k} p^k (1 - p)^{4 - k}$ and $p = 0.5$.  
Total expected revenue and run times for this distribution are shown in Figures~%
\ref{fig:binomPlot_rrm_ub_bef_aft_thm},
\ref{fig:binomPlot_pseudowelfare_scaling},
\ref{fig:binomPlot_rrm_heurlb_greedy_thm},
\ref{fig:binomPlot_rrm_lb_bef_aft_thm},
\ref{fig:binomPlot_brm_lb_bef_aft_thm}, 
\ref{fig:binomPlot_rrm_brm_heur}, 
\ref{fig:binomPlot_rrm_heur},
\ref{fig:binomPlot_brm_heur},
and
\ref{fig:binomPlot_bic_xa_relax}.

\subsection{Pseudo-surplus in the Robust Problem}

We optimize pseudo-surplus in the robust problem (see
Section~\ref{sssec:mpdesc_rrm_xp_ub_pseudowelfare}).
We report the results of the following two methods:
\begin{itemize}
\item (MATLAB) Total expected pseudo-surplus using 
Algorithm~\ref{alg:pseudo-surplus} (the greedy method), 
but without calculating payments.

\item (MATLAB) Total expected pseudo-surplus using 
Algorithm~\ref{alg:pseudo-surplusCf} (the closed form), 
but without calculating payments.
\end{itemize}

Both methods achieve the optimal pseudo-surplus, but using
Algorithm~\ref{alg:pseudo-surplusCf} (the closed form), we
achieve this value more quickly than when using
Algorithm~\ref{alg:pseudo-surplus} (the greedy method).

\subsection{Pseudo-surplus Scaling in the Robust Problem}

We optimize pseudo-surplus in the robust problem, ensuring ex-post feasibility
(see Section~\ref{sssec:mpdesc_rrm_xp_ub_pseudowelfare}),
to see how it scales with the number of bidders.
We report the results of the following:
\begin{itemize}
\item (MATLAB) Total expected pseudo-surplus using 
Algorithm~\ref{alg:pseudo-surplusCf} (the closed form), 
but without calculating payments.
\end{itemize}

While the closed-form method is nearly instantaneous when the number
of bidders is small, it cannot escape the fact that there are
exponentially-many allocation variables.  (The behavior of the closed-form
solution for the heuristic lower bound---see
Section~\ref{sec:hlb}---is identical as we scale the number of
bidders.)

\subsection{Heuristic Lower Bound in the Robust Problem}
\label{sec:hlb}

We optimize the heuristic lower bound ensuring ex-post feasibility
(see Section~\ref{sssec:mpdesc_rrm_xp_lb_h}).
We report the results of the following two methods:
\begin{itemize}
\item (MATLAB) The heuristic lower bound using 
Algorithm~\ref{alg:heurSolver} (the greedy method),
but without calculating payments.

\item (MATLAB) The heuristic lower bound using 
Algorithm~\ref{alg:heurSolverCf} (the closed form),
but without calculating payments.
\end{itemize}

Both methods achieve the optimal objective value, but using
Algorithm~\ref{alg:heurSolverCf} (the closed form), we
achieve this value more quickly than when using
Algorithm~\ref{alg:heurSolver} (the greedy method).

\subsection{Heuristic Revenue in the Robust Problem}

We first optimize our heuristic lower bound in the robust problem (see
Section~\ref{sssec:mpdesc_rrm_xp_lb_h}).  Next, we plug the resulting
allocation rule $\allocvec$ into into Myerson's formula to compute
robust payment rule $\paymentvec$ (Equation~\eqref{eq:pSquaredPayment}),
and then total expected heuristic revenue.
Varying the optimization technique,
we report the results of the following two methods:
\begin{itemize}
\item (MATLAB) Total expected heuristic revenue using Algorithm~\ref{alg:heurSolver} (the greedy method)
and robust payments (Equation~\eqref{eq:pSquaredPayment}). 

\item (MATLAB) Total expected heuristic revenue using Algorithm~\ref{alg:heurSolverCf} (the closed form)
and robust payments (Equation~\eqref{eq:pSquaredPayment}). 
\end{itemize}

Our choice of discretization factor in our greedy implementation is
sufficiently fine that both methods achieve the same objective value.
Moreover, the run times of the two methods are not noticeably
different, because both are dominated by the exponential number of
robust payment calculations.
Both these runs are noticeably slower than the pseudo-surplus and
heuristic lower bound runs, which perform no payment calculations at
all.

\subsection{Heuristic Revenue in the Bayesian Problem}

We first optimize our heuristic lower bound in the robust problem
(see Section~\ref{sssec:mpdesc_rrm_xp_lb_h}).  Next we collapse the
resulting allocation rule $\allocvec$ into
an interim allocation rule $\hatallocvec$.  Finally, we plug
$\hatallocvec$ into Myerson's formula to compute
Bayesian payment rule $\mathbf{h}$
(Equation~\eqref{eq:hSquaredPayment}), 
and then total expected heuristic revenue.
Varying the optimization technique,
we report the results of the following two methods:
\begin{itemize}
\item (MATLAB) Total expected heuristic revenue
using Algorithm~\ref{alg:heurSolverBrm} (the greedy method)
and Bayesian payments (Equation~\eqref{eq:hSquaredPayment}).

\item (MATLAB) Total expected heuristic revenue
using Algorithm~\ref{alg:heurSolverBrmCf} (the closed form)
and Bayesian payments (Equation~\eqref{eq:hSquaredPayment}).
\end{itemize}

As above, our choice of discretization factor in our greedy
implementation is sufficiently fine that both methods achieve the same
objective value.  But this time, using
Algorithm~\ref{alg:heurSolverBrmCf} (the closed form), we achieve this
value much more quickly.  This is because there are only
polynomially-many payment calculations in the Bayesian problem, so the
speed up achieved by the closed-form method is not eviscerated by
expensive payment calculations.

\subsection{Heuristic Lower Bound and Heuristic Revenue: Robust vs.\ Bayesian}

We compare the heuristic lower bound in the robust problem;
total expected heuristic revenue in the robust problem; and
total expected heuristic revenue in the Bayesian problem.
We report the results of the following three methods:
\begin{itemize}
\item (MATLAB) The heuristic lower bound in the robust problem
(see Section~\ref{sssec:mpdesc_rrm_xp_lb_h}) 
using Algorithm~\ref{alg:heurSolverCf} (the closed form),
but without calculating payments.

\item (MATLAB) Total expected heuristic revenue in the robust problem
using Algorithm~\ref{alg:heurSolverCf} (the closed form and robust payments 
using Equation~\eqref{eq:pSquaredPayment}).

\item (MATLAB) Total expected heuristic revenue in the Bayesian problem
using Algorithm~\ref{alg:heurSolverBrmCf} (the closed form and Bayesian payments
using Equation~\eqref{eq:hSquaredPayment}).
\end{itemize}

According to the theory developed in this paper, the heuristic lower
bound in the robust problem should not exceed the heuristic
revenue in the robust problem, which in turn should not exceed the
heuristic revenue in the Bayesian problem.  Our experiments are
consistent with these claims, and further provide examples where the
heuristic revenue in the robust problem strictly exceeds the heuristic
lower bound, and where the heuristic revenue in the Bayesian problem
strictly exceeds that of the robust problem.

Moreover, the Bayesian heuristic runs significantly faster than the
robust heuristic; this difference can be attributed to the computation
of polynomially- instead of exponentially-many payments.  (The
heuristic lower bound computes no payments at all.)

\subsection{Revenue, Pseudo-Surplus, and Heuristic Revenue in the Robust Problem}

We report the results of the following:
\begin{itemize}
\item (CPLEX) Total expected revenue in the robust problem
(see Section~\ref{sssec:mpdesc_rrm_xp}).

\item (MATLAB) Total expected pseudo-surplus in the robust problem
(see Section~\ref{sssec:mpdesc_rrm_xp_ub_pseudowelfare}) solved in closed form
using Algorithm~\ref{alg:pseudo-surplusCf}, 
but without calculating payments.

\item (MATLAB) Total expected heuristic revenue in the robust problem
(see Section~\ref{sssec:mpdesc_rrm_xp_lb_h}) solved in closed form,
and then plugged in to Myerson's formula for robust payments
(Algorithm~\ref{alg:heurSolverCf}).

\end{itemize}

CPLEX solves the robust problem optimally, while the other two
programs do not.  Furthermore, when there are very few bidders (e.g.,
1 or 2), CPLEX solves RRM just as fast as MATLAB solves for
pseudo-surplus or the heuristic revenue.  But as the number of bidders
grows, CPLEX does not scale.  The heuristic revenue appears to
approximate the value of RRM as determined by CPLEX quite well, and it
scales better than CPLEX.

\subsection{Revenue, Pseudo-Surplus, and Heuristic Revenue in the Bayesian problem}

We report the results of the following:
\begin{itemize}
\item (CPLEX) Total expected revenue in the Bayesian problem 
using Equation~\eqref{eq:hSquaredPayment}
(see Section~\ref{sssec:mpdesc_brm_xp}).

\item (CPLEX) Total expected pseudo-surplus in the Bayesian problem
(see Section~\ref{sssec:mpdesc_brm_xp_ub_pseudowelfare}).


\item (MATLAB) Total expected heuristic revenue in the Bayesian problem 
(see Section~\ref{sssec:mpdesc_brm_xp_ub_pseudowelfare}, Remark~\ref{rem:BRM-heuristic}),
solved in closed form, and then plugged in to Myerson's formula for Bayesian payments
(Algorithm~\ref{alg:heurSolverBrmCf}).

\end{itemize}

CPLEX solves the Bayesian problem optimally, while the other two
programs do not.  Furthermore, when there are very few bidders (e.g.,
1, 2, or 3), CPLEX solves BRM just as fast as CPLEX solves for
pseudo-surplus or MATLAB solves for the heuristic revenue.  But as the
number of bidders grows, CPLEX does not scale.  The heuristic revenue
appears to approximate the value of BRM as determined by CPLEX quite
well, and it scales better than CPLEX.

\subsection{BRM Ex-Ante}

We report the results of the following:
\begin{itemize}
\item (CPLEX) Total expected revenue of the BRM ex-ante relaxation
(see Section~\ref{sssec:mpdesc_brm_xa}).

\item (MATLAB) Total expected revenue of the BRM ex-ante relaxation
(see Section~\ref{sssec:mpdesc_brm_xa_rel}) 
solved in closed form (see Theorem~\ref{thm:ex-ante})
to find an interim allocation rule, and use it to compute Bayesian payments.

\item (MATLAB) Total expected revenue of the BRM, ex-ante relaxation
(see Section~\ref{sssec:mpdesc_brm_xa_rel}), 
solved in closed form (see Theorem~\ref{thm:ex-ante}) 
to find an interim allocation rule, which is then truncated so that
$\hatalloc[i] (\val[i]) \le 1$ for each $i \in \setofbidders$ and $\val[i] \in \valspace[i]$, 
before using it to compute Bayesian payments.
\end{itemize}

For the BRM ex-ante relaxation, a problem with only polynomially-many variables and constraints,
even CPLEX, which is optimal, appears to scale just fine with the number of bidders.
Indeed, while the approximations are faster than CPLEX on small problems,
their run time eventually grows, while CPLEX does not.

\section{Conclusion and Future Work}
\label{sec:conc}

We analyzed optimal auctions where bidder utilities are quasi-linear, but defined in terms of convex, instead of the usual linear, payments.
We adapted Myerson's analysis to this setting when it is required that constraints always hold (i.e., the robust problem) and when it is required only that they hold in expectation (i.e., the Bayesian problem).
We showed that total expected revenue in the Bayesian problem can exceed the total expected revenue in the robust problem, and consequently, we analyzed each problem in turn.

In the robust setting, we developed upper and lower bounds on total expected revenue that can be computed easily using the equi-marginal principle.  
In the Bayesian setting, we derived a payment formula that lends itself to a mathematical program that involves fewer variables than a straightforward implementation of the optimal auction problem.
Additionally, we derived closed-form solutions to a relaxation of the Bayesian ex-ante problem, as well as problems that seek robust upper and lower bounds.

Based on our analysis, we derived heuristics that approximate solutions to both the robust and Bayesian revenue maximization problems.
With three different distributions, we saw through experiments that our heuristics produce solutions that are close to optimal, and scale better than the mathematical programming solver we used.
However, our experiments thus far were restricted to quasi-linear utility functions with quadratic perceived payments.
In the future, we hope to run experiments with a wider class of payment functions, beyond quadratic, and possibly even develop theory that goes beyond convex entirely.
Additionally, we hope to prove guarantees about our heuristics, to further characterize their quality.

\section{Acknowledgments}

This research was supported by NSF Grant \#1217761 and Microsoft
Research.  Amy and Takehiro would also like to thank Jason Hartline
and Tim Roughgarden for always being willing to help as we grappled
with the details of Myerson's seminal work.

\appendix

\section{The Discretization Effect}
\label{sec:discretization}

Let $\allocvec^*$ be an optimal allocation rule in an arbitrary auction design problem, and
let $\allocvec$ be an optimal allocation rule in the corresponding discretized problem,
with discretization factor $\Delta > 0$.
For all bidders $i \in \setofbidders$, 
let $0 \le k_i \in \mathbb{Z}$ be such that
\begin{equation}
\alloc[i] (\val[i], \valvec[-i]) = k_i \Delta
.\end{equation}
The optimal values $\alloc[i]^* (\val[i], \valvec[-i])$ and our
approximation $\alloc[i] (\val[i], \valvec[-i])$ are related by a residual vector $\boldsymbol{\rho}$,
each entry of which can be be positive, negative, or zero:
\begin{align}
\alloc[i]^* (\val[i], \valvec[-i]) 
&= \alloc[i] (\val[i], \valvec[-i]) + \rho_i \\
&= k_i \Delta + \rho_i,
\end{align}

To determine $k_i$s, we use the method of least squares.
Specifically, we minimize the square of the residual $L^2$-norm:
\begin{equation}
\norm{\boldsymbol{\rho}}_2^2
= \sum_{i=1}^{\numbidders} \left( \alloc[i]^* (\val[i], \valvec[-i]) - k_i \Delta \right)^2
.\end{equation}
To do so, we take derivatives with respect to $k_i$:
\begin{equation}
\frac{\partial \sum_{i=1}^{\numbidders} \left(\alloc[i]^* (\val[i], \valvec[-i]) - k_i \Delta \right)^2}{\partial k_i}
= 2 \left( \alloc[i]^* (\val[i], \valvec[-i]) - k_i \Delta \right) (-\Delta)
.\end{equation}
Setting the right-hand side of this equation to 0 yields the following intuitive result:
\begin{equation}
k_i = \frac{\alloc[i]^* (\val[i], \valvec[-i])}{\Delta}
.\end{equation}
Enforcing the constraint that $k_i$ must be integral leaves two
possible candidate $k_i$s:
\begin{align}
k_i = \left\lfloor{\frac{\alloc[i]^* (\val[i], \valvec[-i])}{\Delta}}\right\rfloor
\quad \text{or} \quad
k_i = \left\lceil{\frac{\alloc[i]^* (\val[i], \valvec[-i])}{\Delta}}\right\rceil.
\end{align}
In either case, 
\begin{equation}
\rho_i
= \alloc[i]^* (\val[i], \valvec[-i]) - \alloc[i] (\val[i], \valvec[-i])
\le \abs{\alloc[i]^* (\val[i], \valvec[-i]) - \alloc[i] (\val[i], \valvec[-i])}
\le \Delta.
\end{equation}
Therefore, the residual from bidder $i$ is bounded above by $O (\Delta)$.
Likewise, the total residual is bounded above by $O (n \Delta)$.

%
%

Having established a method of determining $\allocvec$, we now
describe how using $\allocvec$ can impact the total expected perceived
payment.  Let $\paymentvecq^*$ and $\paymentvecq$ be the perceived
payments resulting from the allocation rules $\allocvec^*$ and
$\allocvec$, respectively.
First, we show that the difference in expected the perceived payment
from bidder $i$ when using $\allocvec$ instead of $\allocvec^*$ is
$O (\Delta)$.  Then, we conclude that the total expected perceived
payment is $O (\numbidders \Delta)$.

\begin{lemma}
\label{lem:discDiffPercPayOneBidder}
The difference in bidder $i$'s expected perceived payment
when using $\allocvec$ instead of $\allocvec^*$ is $O (\Delta)$: i.e.,
%
$\abs{\Exp_{\valz[i,\ell] \sim F_i} \left[ \paymentq[i]^* ( \valz[i,\ell], \valvec[-i] ) \right]
- \Exp_{\valz[i,\ell] \sim F_i} \left[ \paymentq[i] ( \valz[i,\ell], \valvec[-i] ) \right]}
\in O (\Delta)$.
\end{lemma}

\begin{proof}
We can express the difference between 
$\Exp_{\valz[i,\ell] \sim F_i} \left[ \paymentq[i]^* ( \valz[i,\ell], \valvec[-i] ) \right]$
and $\Exp_{\valz[i,\ell] \sim F_i} \left[ \paymentq[i] ( \valz[i,\ell], \valvec[-i] ) \right]$
in terms of virtual values:
\begin{align}
\Exp_{\valz[i,\ell] \sim F_i} \left[ \paymentq[i]^* ( \valz[i,\ell], \valvec[-i] ) \right]
- \Exp_{\valz[i,\ell] \sim F_i} \left[ \paymentq[i] ( \valz[i,\ell], \valvec[-i] ) \right]
&= \Exp_{\valz[i,\ell] \sim F_i} \left[ \vvald[i] (\valz[i,\ell]) \, \alloc[i]^* ( \valz[i,\ell], \valvec[-i] ) 
- \vvald[i] (\valz[i,\ell]) \, \alloc[i] ( \valz[i,\ell], \valvec[-i] ) \right] \\
&= \Exp_{\valz[i,\ell] \sim F_i} \left[ \vvald[i] (\valz[i,\ell]) \, \left( \alloc[i]^* ( \valz[i,\ell], \valvec[-i] ) - \alloc[i] ( \valz[i,\ell], \valvec[-i] ) \right) \right] \\
&\le \Exp_{\valz[i,\ell] \sim F_i} \left[ \vvald[i] (\valz[i,\ell]) \, \Delta \right] \\
&\in O (\Delta)
.\end{align}
\end{proof}

\begin{corollary}
The difference in total expected perceived payment 
when using $\allocvec$ instead of $\allocvec^*$
is $O (\numbidders \Delta)$: i.e.,
%
$\abs{\sum_{i=1}^{\numbidders} \Exp_{\valvec} \left[ \paymentq[i]^* ( \valz[i,\ell], \valvec[-i] )
- \paymentq[i] ( \valz[i,\ell], \valvec[-i] ) \right]}
\in O (\numbidders \Delta)$.
\end{corollary}

%
%

En route to deriving the difference in \emph{total\/} expected revenue
from using allocation rule $\allocvec$ instead of $\allocvec^*$, we
present the following lemma, which argues that the maximum difference
in perceived payment from bidder $i$ is $O (\Delta)$.

\begin{lemma}
\label{lem:maxErrDiscq}
The difference in bidder $i$'s perceived payment when using allocation
rule $\allocvec$ instead of $\allocvec^*$ is $O (\Delta)$: i.e.,
%
$\abs{\paymentq[i]^* (\valz[i,\ell], \valvec[-i]) - \paymentq[i] (\valz[i,\ell], \valvec[-i])}
\in O (\Delta)$.
\end{lemma}

\begin{proof}
Without loss of generality, assume 
$\paymentq[i]^* (\valz[i,\ell], \valvec[-i]) \ge \paymentq[i] (\valz[i,\ell], \valvec[-i])$.
Allocation variables $\alloc[i]^* (\valz[i,\ell], \valvec[-i])$ and 
$\alloc[i] (\valz[i,\ell], \valvec[-i])$ can differ by at most $\Delta$,  
so the difference between $\paymentq[i]^* (\valz[i,\ell], \valvec[-i])$ and 
$\paymentq[i] (\valz[i,\ell], \valvec[-i])$ 
is maximized when
$\alloc[i]^* (\valz[i,\ell], \valvec[-i]) = \alloc[i] (\valz[i,\ell], \valvec[-i]) + \Delta$,
for an arbitrary value $\ell$,
and $\alloc[i]^* (\valz[i,j], \valvec[-i]) = \alloc[i] (\valz[i,j], \valvec[-i]) - \Delta$,
for $j < \ell$. Now
\begin{align}
\paymentq[i]^* (\valz[i,\ell], \valvec[-i]) 
&= \left( \valz[i,\ell] \alloc[i]^* ( \valz[i,\ell], \valvec[-i] ) - 
\sum_{j=1}^{\ell - 1} ( \valz[i,j+1] - \valz[i,j] ) \alloc[i]^* ( \valz[i,j], \valvec[-i] ) \right) \\
&= \left( \valz[i,\ell] (\alloc[i] ( \valz[i,\ell], \valvec[-i] ) + \Delta) - 
\sum_{j=1}^{\ell - 1} ( \valz[i,j+1] - \valz[i,j] ) (\alloc[i] ( \valz[i,j], \valvec[-i] ) - \Delta) \right) \\
&= \valz[i,\ell] (\alloc[i] ( \valz[i,\ell], \valvec[-i] ) - 
\sum_{j=1}^{\ell - 1} ( \valz[i,j+1] - \valz[i,j] ) \alloc[i] ( \valz[i,j], \valvec[-i] ) 
+ \valz[i,\ell] \Delta + \sum_{j=1}^{\ell - 1} ( \valz[i,j+1] - \valz[i,j] ) \Delta \\
&= \paymentq[i] (\valz[i,\ell], \valvec[-i]) 
+ \valz[i,\ell] \Delta + \sum_{j=1}^{\ell - 1} ( \valz[i,j+1] - \valz[i,j] ) \Delta \\
&= \paymentq[i] (\valz[i,\ell], \valvec[-i])
+ \valz[i,\ell] \Delta + \left( \valz[i,\ell] - \valz[i,1] \right) \Delta \\
&= \paymentq[i] (\valz[i,\ell], \valvec[-i])
+ \Delta \left( 2 \valz[i,\ell] - \valz[i,1] \right) \\
&= \paymentq[i] (\valz[i,\ell], \valvec[-i]) + O (\Delta)
.\end{align}
Therefore, the difference in perceived payments is $O (\Delta)$.
\end{proof}

%
%

Having established Lemma~\ref{lem:maxErrDiscq},
we now show that discretization affects the expected revenue
collected from bidder $i$ by $O \left( \sqrt{\Delta} \right)$,
and total expected revenue by $O \left( \numbidders \sqrt{\Delta} \right)$.

\begin{lemma}
\label{lem:discDiffRevOneBidder}
The difference in expected revenue from bidder $i$
when using $\allocvec$ instead of $\allocvec^*$
is $O (\sqrt{\Delta})$: i.e.,
$\abs{\Exp_{\valz[i,\ell] \sim F_i} \left[ \payment[i]^* ( \valz[i,\ell], \valvec[-i] )
- \payment[i] ( \valz[i,\ell], \valvec[-i] ) \right]}
\in O \left( \sqrt{\Delta} \right).$
\end{lemma}

\begin{proof}
First, we construct payment rules 
$\barpaymentvec \in \mathbb{R}^{\numbidders}$ and
$\tildepaymentvec \in \mathbb{R}^{\numbidders}$.
We choose $\barpayment[i] (\valz[i,\ell], \valvec[-i])$ and
$\tildepayment[i] (\valz[i,\ell], \valvec[-i])$,
for all $i \in \setofbidders$ and $\valvec \in \valspace$, so that
\begin{align}
\label{eq:barpayment}
\barpayment[i] (\valz[i,\ell], \valvec[-i])
&= \min \left\{ \payment[i]^* (\valz[i,\ell], \valvec[-i]), \payment[i] (\valz[i,\ell], \valvec[-i]) \right\} \\
\label{eq:tildepayment}
\tildepayment[i] (\valz[i,\ell], \valvec[-i])
&= \max \left\{ \payment[i]^* (\valz[i,\ell], \valvec[-i]), \payment[i] (\valz[i,\ell], \valvec[-i]) \right\}
.\end{align}
Likewise, for $\barpaymentvecq, \tildepaymentvecq \in \mathbb{R}^{\numbidders}$.

By Jensen's inequality, we can write the difference in 
expected revenue from bidder $i$ as
\begin{align}
& \Exp_{\valz[i,\ell] \sim F_i} \left[ \payment[i]^* ( \valz[i,\ell], \valvec[-i] )
- \payment[i] ( \valz[i,\ell], \valvec[-i] ) \right] \\ \nonumber
&\quad \le \left( \Exp_{\valz[i,\ell] \sim F_i} \left[ \left( \payment[i]^* ( \valz[i,\ell], \valvec[-i] ) - \payment[i] ( \valz[i,\ell], \valvec[-i] ) \right)^2 \right] \right)^{1/2} \\
&\quad = \left( \Exp_{\valz[i,\ell] \sim F_i} \left[ 
\left( \payment[i]^* ( \valz[i,\ell], \valvec[-i] ) \right)^2  
- 2 \payment[i]^* ( \valz[i,\ell], \valvec[-i] ) \payment[i] ( \valz[i,\ell], \valvec[-i] )
+ \left( \payment[i] ( \valz[i,\ell], \valvec[-i] ) \right)^2
\right] \right)^{1/2} \\
\label{eq:discBoundLoosenWithConstruct}
&\quad \le \Big( 2
\Exp_{\valz[i,\ell] \sim F_i} \left[ 
\left( \tildepayment[i] ( \valz[i,\ell], \valvec[-i] ) \right)^2  
- \left( \barpayment[i] ( \valz[i,\ell], \valvec[-i] ) \right)^2
\right] \Big)^{1/2}
,\end{align}
where Equation~\eqref{eq:discBoundLoosenWithConstruct} follows from 
Equations~\eqref{eq:barpayment} and \eqref{eq:tildepayment} and
\begin{align}
\Exp_{\valz[i,\ell] \sim F_i} \left[ \barpayment[i] (\valz[i,\ell], \valvec[-i])^2 \right]
& \le \Exp_{\valz[i,\ell] \sim F_i} \left[ \payment[i]^* ( \valz[i,\ell], \valvec[-i] ) \payment[i] ( \valz[i,\ell], \valvec[-i] ) \right] \\
\Exp_{\valz[i,\ell] \sim F_i} \left[ \tildepayment[i] (\valz[i,\ell], \valvec[-i])^2 \right]
& \ge \Exp_{\valz[i,\ell] \sim F_i} \left[ \left( \payment[i]^* ( \valz[i,\ell], \valvec[-i] ) \right)^2 \right] \\
\Exp_{\valz[i,\ell] \sim F_i} \left[ \tildepayment[i] (\valz[i,\ell], \valvec[-i])^2 \right]
& \ge \Exp_{\valz[i,\ell] \sim F_i} \left[ \left( \payment[i] ( \valz[i,\ell], \valvec[-i] ) \right)^2 \right]
.\end{align}
Using Lemma~\ref{lem:maxErrDiscq}, we arrive at our result:
\begin{align}
\left( 2 \Exp_{\valz[i,\ell] \sim F_i} \left[ 
\left( \tildepayment[i] ( \valz[i,\ell], \valvec[-i] ) \right)^2  
- \left( \barpayment[i] ( \valz[i,\ell], \valvec[-i] ) \right)^2
\right] \right)^{1/2}
&= \left( 2 \Exp_{\valz[i,\ell] \sim F_i} \left[ 
\tildepaymentq[i] ( \valz[i,\ell], \valvec[-i] )
- \barpaymentq[i] ( \valz[i,\ell], \valvec[-i] )
\right] \right)^{1/2} \\
&\le \left( 2 \Exp_{\valz[i,\ell] \sim F_i} \left[ 
\abs{\paymentq[i]^* ( \valz[i,\ell], \valvec[-i] )
- \paymentq[i] ( \valz[i,\ell], \valvec[-i] )}
\right] \right)^{1/2} \\
&\le \left( 2 \ \Exp_{\valz[i,\ell] \sim F_i} \left[ O(\Delta) \right] \right)^{1/2} \\
&\in O \left( \sqrt{\Delta} \right)
.\end{align}
\end{proof}

\begin{corollary}
The difference in total expected revenue
when using $\allocvec$ instead of $\allocvec^*$
is $O \left( \numbidders \sqrt{\Delta} \right)$: i.e.,
%
$\abs{\sum_{i=1}^{\numbidders} \Exp_{\valvec} \left[ \payment[i]^* ( \valz[i,\ell], \valvec[-i] )
- \payment[i] ( \valz[i,\ell], \valvec[-i] ) \right]}
\in O \left( \numbidders \sqrt{\Delta} \right)$.
\end{corollary}

To summarize, in our setting, where a budget $B$ is finitely divisible by a factor $\Delta_B$,
so that allocations are multiplicative factors of $\Delta = \Delta_B / B$,
in the usual linear case, where $\paymentq[i] = \payment[i]$,
total expected revenue is within $O \left( \numbidders \Delta_B / B \right)$ of optimal.
Furthermore, in the case of quadratic perceived payments,
total revenue is within $O \left( \numbidders \sqrt{\Delta_B / B} \right)$ of optimal.

\section{Another Upper Bound}
\label{sec:upperBound}

While non-operational as of yet, 
we present an easy-to-derive upper bound on revenue for completeness.

\begin{lemma}
\label{lemma:upperBound}
Expected payments, when bidders have quasi-linear 
utility functions as described by Equation~\eqref{eq:util_with_q} and 
$\paymentq[i] ( \payment[i] ( \valz[i,k], \valvec[-i] ) ) = 
\left( \payment[i] ( \valz[i,k], \valvec[-i] ) \right)^2$,
can be upper-bounded as follows:
\begin{equation}
\label{eq:upperBound}
\sqrt{\Exp_{\valz[i,k] \sim F_i} \left[ \vvald[i,k] \, \alloc[i] ( \valz[i,k], \valvec[-i] ) \right]}
\ge \Exp_{\valz[i,k] \sim F_i} \left[ \payment[i] ( \valz[i,k], \valvec[-i] ) \right] %
.\end{equation}
\end{lemma}

\begin{proof}
Applying Theorem~\ref{thm:ExpRevVirVal} to concave utilities with
$\paymentq[i] ( \valz[i,k], \valvec[-i] ) = ( \payment[i] ( \valz[i,k], \valvec[-i] ) )^2$
yields:
\begin{equation*}
\Exp_{\valz[i,k] \sim F_i} \left[ ( \payment[i] ( \valz[i,k], \valvec[-i] ) )^2 \right] = 
\Exp_{\valz[i,k] \sim F_i} \left[ \vvald[i,k] \, \alloc[i] ( \valz[i,k], \valvec[-i] ) \right] %
.\end{equation*}
By Jensen's inequality (since squaring is a convex function):
\begin{equation*}
\Exp_{\valz[i,k] \sim F_i} \left[ ( \payment[i] ( \valz[i,k], \valvec[-i] ) )^2 \right] 
\ge \left( \Exp_{\valz[i,k] \sim F_i} \left[ \payment[i] ( \valz[i,k], \valvec[-i] ) \right] \right)^2
.\end{equation*}
Combining the two equations and taking square roots completes the proof.
\end{proof}

\begin{example}
Following up on Example~\ref{ex:tight3},
we again assume $\valspace[i] = \{ 1 \}$, for all bidders $i \in \setofbidders$, so that $\vvald[i,1] = 1$.
In this case, the upper bound (Equation~\eqref{eq:upperBound}) is tight:
\begin{equation*}
\sqrt{\vvald[i,1] \, \alloc[i] (\val[i], \valvec[-i])}
= \sqrt{(1) \left( \frac{1}{n} \right)}
= \sqrt{\frac{1}{n}}
= \payment[i] (\val[i], \valvec[-i])%
.\end{equation*}
\end{example}

\section{Program Descriptions}
\label{sec:programs}

We describe the programs implemented, 
including the number of variables and constraints.  
Let $k = \max_{i} \{ \valspacesize[i] : i \in \setofbidders \}$, 
so that $k \ge \valspacesize[i]$, for all bidders $i \in \setofbidders$.

\subsection{RRM Mathematical Programs}

\subsubsection{RRM}
\label{sssec:mpdesc_rrm_xp}

\paragraph{Mathematical Program}

\begin{align}
\max_{\allocvec} \,
& \sum_{\valvec} \sum_{i=1}^{n} f (\valvec) \payment[i](\valvec) \\
\text{subject to }
& \sum_{i=1}^{n} \alloc[i] (\valvec) \le 1, 
&& \forall \valvec \in \valspace \\
& 0 \le \alloc[i] (\valvec), 
&& \forall i \in \setofbidders, \forall \valvec \in \valspace \\
& \alloc[i] (\valvec) \le 1, 
&& \forall i \in \setofbidders, \forall \valvec \in \valspace \\
& \alloc[i] (\valz[i,\ell], \valvec[-i]) \ge \alloc[i] (\valz[i,\ell-1], \valvec[-i]), 
&& \forall i \in \setofbidders, \forall \valz[i,\ell] > \valz[i,\ell-1] \in \valspace[i], \forall \valvec[-i] \in \valspace[-i] \\
& \left( \payment[i] (\valz[i,\ell], \valvec[-i]) \right)^2 = \valz[i,\ell] \alloc[i] (\valz[i,\ell], \valvec[-i]) - 
\\ \nonumber
& \qquad 
\sum_{j=1}^{\ell-1} (\valz[i,j+1] - \valz[i,j]) \hatalloc[i] (\valz[i,j], \valvec[-i]), %
&& \forall i \in \setofbidders, \forall \valz[i,\ell] \in \valspace[i], \forall \valvec[-i] \in \valspace[-i]
\end{align}

\paragraph{Variables}

For each bidder $i \in \setofbidders$,
and for each $\valvec \in \valspace$, 
there are variables
$\alloc[i] (\val[i], \valvec[-i])$
and $\payment[i] (\val[i], \valvec[-i])$.  
The total number of variables is $O (n k^n)$.

\paragraph{Constraints}

The total number of constraints is $O(n k^n)$.  

\begin{itemize}
\item Ex-post feasibility.  This requires $O(k^n)$ equations.
\item Lower and upper bounds on $\alloc[i] (\val[i], \valvec[-i])$.  This requires $O(n k^n)$ equations.
\item Monotonicity.  There are $O(k^{n-1})$ $\valvec[-i]$ vectors and $O(k)$ comparisons per bidder; so $O(k^n)$ equations per bidder, and $O(n k^n)$ equations in all.
\item Payment formula.  This requires $O(n k^n)$ equations.
\end{itemize}


\subsubsection{RRM, Upper Bound (Pseudo-Surplus)}
\label{sssec:mpdesc_rrm_xp_ub_pseudowelfare}

\paragraph{Mathematical Program}

\begin{align}
\max_{\allocvec} \, 
& \sum_{\valvec} \sum_{i=1}^{n} f (\valvec) \sqrt{\val[i] \alloc[i] (\valvec)} \\
\text{subject to }
& \sum_{i=1}^{n} \alloc[i] (\valvec) \le 1, 
&& \forall \valvec \in \valspace \\
& 0 \le \alloc[i] (\valvec), 
&& \forall i \in \setofbidders, \forall \valvec \in \valspace \\
& \alloc[i] (\valvec) \le 1, 
&& \forall i \in \setofbidders, \forall \valvec \in \valspace \\
& \alloc[i] (\valz[i,\ell], \valvec[-i]) \ge \alloc[i] (\valz[i,\ell-1], \valvec[-i]), 
&& \forall i \in \setofbidders, \forall \valz[i,\ell] > \valz[i,\ell-1] \in \valspace[i], \forall \valvec[-i] \in \valspace[-i]
\end{align}

\paragraph{Variables}

For each bidder $i \in \setofbidders$ 
and for each $\valvec \in \valspace$, 
there are variables $\alloc[i] (\val[i], \valvec[-i])$.
The total number of variables is $O(n k^n)$.

\paragraph{Constraints}

The total number of constraints is $O(n k^n)$.  

\begin{itemize}
\item Ex-post feasibility.  This requires $O(k^n)$ equations.
\item Lower and upper bounds on $\alloc[i] (\val[i], \valvec[-i])$.  This requires $O(n k^n)$ equations.
\item Monotonicity.  There are $O(k^{n-1})$ $\valvec[-i]$ vectors and $O(k)$ comparisons per bidder; so $O(k^n)$ equations per bidder, and $O(n k^n)$ equations in all.
\end{itemize}


\subsubsection{RRM, Lower Bound (Heuristic)}
\label{sssec:mpdesc_rrm_xp_lb_h}

\paragraph{Mathematical Program}

\begin{align}
\max_{\allocvec} \, 
& \sum_{\valvec} \sum_{i=1}^{\numbidders} f (\valvec) \sqrt{\vvald[i]^+ (\val[i]) \, \alloc[i] (\valvec)} \\
\text{subject to }
& \sum_{i=1}^{\numbidders} \alloc[i] (\valvec) \le 1, 
&& \forall \valvec \in \valspace \\
& 0 \le \alloc[i] (\valvec), 
&& \forall i \in \setofbidders, \forall \valvec \in \valspace \\
& \alloc[i] (\valvec) \le 1, 
&& \forall i \in \setofbidders, \forall \valvec \in \valspace \\
& \alloc[i] (\valz[i,\ell], \valvec[-i]) \ge \alloc[i] (\valz[i,\ell-1], \valvec[-i]), 
&& \forall i \in \setofbidders, \forall \valz[i,\ell] > \valz[i,\ell-1] \in \valspace[i], \forall \valvec[-i] \in \valspace[-i]
\end{align}

\paragraph{Variables}

For each bidder $i \in \setofbidders$ 
and for each $\valvec \in \valspace$, 
there are variables $\alloc[i] (\val[i], \valvec[-i])$.
The total number of variables is $O(n k^n)$.

\paragraph{Constraints}

The total number of constraints is $O(n k^n)$.  

\begin{itemize}
\item Ex-post feasibility.  This requires $O(k^n)$ equations.
\item Lower and upper bounds on $\alloc[i] (\val[i], \valvec[-i])$.  This requires $O(n k^n)$ equations.
\item Monotonicity.  There are $O(k^{n-1})$ $\valvec[-i]$ vectors and $O(k)$ comparisons per bidder; so $O(k^n)$ equations per bidder, and $O(n k^n)$ equations in all.
\end{itemize}

\subsection{BRM, Ex-post Mathematical Programs}

\subsubsection{BRM, Ex-post}
\label{sssec:mpdesc_brm_xp_naive}

This program does not use Equation~\eqref{eq:hSquaredPayment} to compute payments.  
Instead, it attempts to maximize revenue by relating 
$\hatpaymentq$ and $\payment$ terms.

\paragraph{Mathematical Program}

\begin{align}
\max_{\allocvec} \,
& \sum_{i=1}^{n} \sum_{\val[i] \in \valspace[i]} f_i (\val[i]) \hatpayment[i](\val[i]) \\
\text{subject to }
& \sum_{i=1}^{n} \alloc[i] (\valvec) \le 1, 
&& \forall \valvec \in \valspace \\
& 0 \le \alloc[i] (\valvec), 
&& \forall i \in \setofbidders, \forall \valvec \in \valspace \\
& \alloc[i] (\valvec) \le 1, 
&& \forall i \in \setofbidders, \forall \valvec \in \valspace \\
& \hatalloc[i] (\val[i]) = \sum_{\valvec[-i]} f (\valvec[-i]) \alloc[i] (\val[i], \valvec[-i]), 
&& \forall i \in \setofbidders, \forall \val[i] \in \valspace[i] \\
& \hatpayment[i] (\val[i]) = \sum_{\valvec[-i]} f (\valvec[-i]) \payment[i] (\val[i], \valvec[-i]), 
&& \forall i \in \setofbidders, \forall \val[i] \in \valspace[i] \\
& \hatpaymentq[i] (\val[i]) = \sum_{\valvec[-i]} f (\valvec[-i]) \left( \payment[i] (\val[i], \valvec[-i]) \right)^2,
&& \forall i \in \setofbidders, \forall \val[i] \in \valspace[i] \\
& \hatalloc[i] (\valz[i,\ell]) \ge \hatalloc[i] (\valz[i,\ell-1]), 
&& \forall i \in \setofbidders, \forall \valz[i,\ell] > \valz[i,\ell-1] \in \valspace[i] \\
& \hatpaymentq[i] (\valz[i,\ell]) = \valz[i,\ell] \hatalloc[i] (\valz[i,\ell]) - \\ \nonumber
& \qquad \sum_{j=1}^{\ell-1} (\valz[i,j+1] - \valz[i,j]) \hatalloc[i] (\valz[i,j]), %
&& \forall i \in \setofbidders, \forall \valz[i,\ell] \in \valspace[i]
\end{align}

\paragraph{Variables}

For each bidder $i \in \setofbidders$,
and for each $\val[i] \in \valspace[i]$
and $\valvec[-i] \in \valspace[-i]$, 
there are variables
$\payment[i] (\val[i], \valvec[-i])$ 
and 
$\alloc[i] (\val[i], \valvec[-i])$.
The total number of variables is $O (n k^n)$.

\paragraph{Constraints}

The total number of constraints is $O(n k^n)$.  

\begin{itemize}
\item Ex-post feasibility.  This requires $O(k^n)$ equations.
\item Lower and upper bounds on $\alloc[i] (\val[i], \valvec[-i])$.  This requires $O(n k^n)$ equations.
\item Relating $\hatalloc[i] (\val[i])$ with $\alloc[i] (\val[i], \valvec[-i])$.  This requires $O(nk)$ equations.
\item Relating $\hatpayment[i] (\val[i])$ with $\payment[i] (\val[i], \valvec[-i])$.  This requires $O(nk)$ equations.
\item Relating $\hatpaymentq[i] (\val[i])$ with $\left( \payment[i] (\val[i], \valvec[-i]) \right)^2$.  This requires $O(nk)$ equations.
\item Monotonicity.  This requires $O(nk)$ equations.
\item Payment formula.  This requires $O(nk)$ equations.
\end{itemize}


\subsubsection{BRM, Ex-post, Simplified}
\label{sssec:mpdesc_brm_xp}

This program uses Equation~\eqref{eq:hSquaredPayment} to compute payments.

\paragraph{Mathematical Program}

\begin{align}
\max_{\allocvec} \,
& \sum_{i=1}^{n} \sum_{\val[i] \in \valspace[i]} f_i (\val[i]) h_i(\val[i]) \\
\text{subject to }
& \sum_{i=1}^{n} \alloc[i] (\valvec) \le 1, 
&& \forall \valvec \in \valspace \\
& 0 \le \alloc[i] (\valvec), 
&& \forall i \in \setofbidders, \forall \valvec \in \valspace \\
& \alloc[i] (\valvec) \le 1, 
&& \forall i \in \setofbidders, \forall \valvec \in \valspace \\
& \hatalloc[i] (\val[i]) = \sum_{\valvec[-i]} f (\valvec[-i]) \alloc[i] (\val[i], \valvec[-i]), 
&& \forall i \in \setofbidders, \forall \val[i] \in \valspace[i] \\
& \hatalloc[i] (\valz[i,\ell]) \ge \hatalloc[i] (\valz[i,\ell-1]), 
&& \forall i \in \setofbidders, \forall \valz[i,\ell] > \valz[i,\ell-1] \in \valspace[i] \\
& \left( h_i (\valz[i,\ell]) \right)^2 = \valz[i,\ell] \hatalloc[i] (\valz[i,\ell]) - 
\sum_{j=1}^{\ell-1} (\valz[i,j+1] - \valz[i,j]) \hatalloc[i] (\valz[i,j]), 
&& \forall i \in \setofbidders, \forall \valz[i,\ell] \in \valspace[i]
\end{align}

\paragraph{Variables}

For each bidder $i \in \setofbidders$,
and for each $\val[i] \in \valspace[i]$
and $\valvec[-i] \in \valspace[-i]$, 
there are variables $\alloc[i] (\val[i], \valvec[-i])$.
The total number of variables is $O (n k^n)$.

\paragraph{Constraints}

The total number of constraints is $O(n k^n)$.  

\begin{itemize}
\item Ex-post feasibility.  This requires $O(k^n)$ equations.
\item Lower and upper bounds on $\alloc[i] (\val[i], \valvec[-i])$.  This requires $O(n k^n)$ equations.
\item Relating $\hatalloc[i] (\val[i])$ with $\alloc[i] (\val[i], \valvec[-i])$.  This requires $O(nk)$ equations.
\item Monotonicity.  This requires $O(nk)$ equations.
\item Payment formula.  This requires $O(nk)$ equations.
\end{itemize}


\subsubsection{BRM, Ex-post, Upper Bound (Pseudo-Surplus)}
\label{sssec:mpdesc_brm_xp_ub_pseudowelfare}

\paragraph{Mathematical Program}

\begin{align}
\max_{\allocvec} \, 
& \sum_{i=1}^{n} \sum_{\val[i] \in \valspace[i]} f_i (\val[i]) \sqrt{\val[i] \hatalloc[i] (\val[i])} \\
\text{subject to }
& \sum_{i=1}^{\numbidders} \alloc[i] (\valvec) \le 1, 
&& \forall \valvec \in \valspace \\
& 0 \le \alloc[i] (\valvec), 
&& \forall i \in \setofbidders, \forall \valvec \in \valspace \\
& \alloc[i] (\valvec) \le 1, 
&& \forall i \in \setofbidders, \forall \valvec \in \valspace \\
& \hatalloc[i] (\val[i]) = \sum_{\valvec[-i]} f_{-i} (\valvec[-i]) \alloc[i] (\val[i], \valvec[-i]),
&& \forall i \in \setofbidders, \forall \val[i] \in \valspace[i] \\
& \hatalloc[i] (\valz[i,\ell]) \ge \hatalloc[i] (\valz[i,\ell-1]), 
&& \forall i \in \setofbidders, \forall \valz[i,\ell] > \valz[i,\ell-1] \in \valspace[i]
\end{align}

\paragraph{Variables}

For each bidder $i \in \setofbidders$ 
and for each $\val[i] \in \valspace[i]$
and $\valvec[-i] \in \valspace[-i]$, 
there are variables $\alloc[i] (\val[i], \valvec[-i])$.
The total number of variables is $O(n k^n)$.

\paragraph{Constraints}

The total number of constraints is $O(n k^n)$.  

\begin{itemize}
\item Ex-post feasibility.  This requires $O(k^n)$ equations.
\item Lower and upper bounds on $\alloc[i] (\val[i], \valvec[-i])$.  This requires $O(n k^n)$ equations.
\item Relating $\hatalloc[i] (\val[i])$ with $\alloc[i] (\val[i], \valvec[-i])$.  This requires $O (nk)$ equations.
\item Monotonicity.  This requires $O(n k)$ equations.
\end{itemize}

\begin{remark}
\label{rem:BRM-heuristic}
The BRM, ex-post, lower bound (heuristic) mathematical program has
constraints that are identical to these, but differs in its objective
function, which involves virtual values instead of values:
\begin{align}
\max_{\allocvec} \, 
& \sum_{i=1}^{\numbidders} \sum_{\val[i] \in \valspace[i]} f_i (\val[i]) \sqrt{\vvald[i]^+ (\val[i]) \, \hatalloc[i] (\val[i])}
\end{align}
\end{remark}


\subsection{BRM Ex-ante Mathematical Programs}

\subsubsection{BRM, Ex-ante}
\label{sssec:mpdesc_brm_xa}

\paragraph{Mathematical Program}

\begin{align}
\max_{\allocvec} \,
& \sum_{i=1}^{n} \sum_{\val[i] \in \valspace[i]} f_i (\val[i]) h_i (\val[i]) \\
\text{subject to }
& \sum_{i=1}^{n} f (\val[i]) \hatalloc[i] (\val[i]) \le 1
&& \\
& 0 \le \hatalloc[i] (\val[i]), 
&& \forall i \in \setofbidders, \forall \val[i] \in \valspace[i] \\
& \hatalloc[i] (\val[i]) \le 1, 
&& \forall i \in \setofbidders, \forall \val[i] \in \valspace[i] \\
& \hatalloc[i] (\valz[i,\ell]) \ge \hatalloc[i] (\valz[i,\ell-1]), 
&& \forall i \in \setofbidders, \forall \valz[i,\ell] > \valz[i,\ell-1] \in \valspace[i] \\
& \left( h_i (\valz[i,\ell]) \right)^2 = \valz[i,\ell] \hatalloc[i] (\valz[i,\ell]) - 
\sum_{j=1}^{\ell-1} (\valz[i,j+1] - \valz[i,j]) \hatalloc[i] (\valz[i,j]), 
&& \forall i \in \setofbidders, \forall \valz[i,\ell] \in \valspace[i]
\end{align}

\paragraph{Variables}

For each bidder $i \in \setofbidders$,
and for each $\val[i] \in \valspace[i]$, 
there are variables
$\hatalloc[i] (\val[i])$ and $h_i (\val[i])$.  
The total number of variables is $O (nk)$.

\paragraph{Constraints}

The total number of constraints is $O(nk)$.    

\begin{itemize}
\item Ex-ante feasibility.  This requires $O(1)$ equations.
\item Lower and upper bounds on $\hatalloc[i] (\val[i])$.  This requires $O(nk)$ equations.
\item Monotonicity.  This requires $O(nk)$ equations.
\item Payment formula.  This requires $O(nk)$ equations.
\end{itemize}


\subsubsection{BRM, Ex-ante Relaxation}
\label{sssec:mpdesc_brm_xa_rel}

\paragraph{Mathematical Program}

\begin{align}
\max_{\allocvec} \,
& \sum_{i=1}^{\numbidders} \sum_{\val[i] \in \valspace[i]} f_i (\val[i]) \sqrt{\vvald[i]^+ (\val[i]) \, \hatalloc[i] (\val[i])} \\
\text{subject to }
& \sum_{i=1}^{\numbidders} f (\val[i]) \hatalloc[i] (\val[i]) \le 1
&& \\
& 0 \le \hatalloc[i] (\val[i]), 
&& \forall i \in \setofbidders, \forall \val[i] \in \valspace[i] \\
& \hatalloc[i] (\valz[i,\ell]) \ge \hatalloc[i] (\valz[i,\ell-1]), 
&& \forall i \in \setofbidders, \forall \valz[i,\ell] > \valz[i,\ell-1] \in \valspace[i]
\end{align}

\paragraph{Variables}

For each bidder $i \in \setofbidders$,
and for each $\val[i] \in \valspace[i]$, 
there are variables $\hatalloc[i] (\val[i])$.
The total number of variables is $O (nk)$.

\paragraph{Constraints}

The total number of constraints is $O(nk)$.  

\begin{itemize}
\item Ex-ante feasibility.  This requires $O(1)$ equations.
\item Lower bounds on $\hatalloc[i] (\val[i])$.  This requires $O(nk)$ equations.
\item Monotonicity.  This requires $O(nk)$ equations.
\end{itemize}



\bibliographystyle{plainnat}
\bibliography{bibliography}

\end{document}